# Lasing in Group-IV materials

V. Reboud[1*], D. Buca[2], H. Sigg[3], J. M. Hartmann[1], Z. Ikonic[4], N. Pauc[5], V. Calvo[5], P. Rodriguez[1], and A. Chelnokov[1]

[1] Univ. Grenoble Alpes, CEA, LETI, 38054 Grenoble, France.
[2] Jülich Aachen Research Alliance (JARA)-Institute Green IT, RWTH Aachen, 52074 Aachen, Germany.
[3] Laboratory for Micro- and Nanotechnology, Paul Scherrer Institut, 5232 Villigen, Switzerland.
[4] Pollard Institute, School of Electronic and Electrical Engineering, University of Leeds, Leeds, United Kingdom
[5] Univ. Grenoble Alpes, CEA, IRIG-DePhy, 38054 Grenoble, France.

[*] Corresponding author: Vincent Reboud, e-mail address: vincent.reboud@cea.fr

## Abstract

Silicon photonics in the near-Infra-Red, up to 1.6 µm, is already one of key technologies in optical data communications, particularly short-range. It is also being prospected for applications in quantum computing, artificial intelligence, optical signal processing, where complex photonic integration is to be combined with large-volume fabrication. However, silicon photonics does not yet cover a large portion of applications in the mid-IR. In the 2 to 5 µm wavelength range, environmental sensing, life sensing, and security all rely on optical signatures of molecular vibrations to identify complex individual chemical species. The markets for such analysis are huge and constantly growing, with a push for sensitivity, specificity, compactness, low-power operation and low cost. An all-group-IV, CMOS-compatible mid-IR integrated photonic platform would be a key enabler in this wavelength range. As for other wavelengths, such a platform should be complete with low-loss guided interconnects, detectors, modulators, eventually, and most importantly efficient and integrated light sources. This chapter reviews recent developments in the fields of mid-IR silicon-compatible optically and electrically pumped lasers, light emitting diodes and photodetectors based on Ge, GeSn and SiGeSn alloys. It contains insights into the fundamentals of these developments, including band structure modelling, material growth and processing techniques.

# Table of contents:

## Abbreviations:

| | |
|---|---|
| AFM: Atomic Force Microscopy<br>APT: Atom Probe Tomography<br>AP: Atmospheric Pressure<br>APB: Anti-Phase Boundaries<br>ASE: Amplified spontaneous emission<br>CMOS: Complementary metal–oxide–semiconductor<br>CVD: Chemical Vapor Deposition<br>CW : continuous Wave<br>DBR: Distributed Bragg reflector<br>DFB: Distributed feedback<br>DHS: Double heterostructure<br>DOS: Density of states<br>EL: Electroluminescence<br>FCA: Free carrier absorption<br>FLP: Fermi-level pinning<br>F-P: Fabry-Perot<br>GBs: Grain boundaries<br>Ge: Germanium<br>GeOI: Germanium On Insulator<br>GeSn: Germanium Tin<br>GR: Growth rate<br>HH: Heavy-holes<br>IVBA: Inter-valence band absorption<br>LED: Light emitting device<br>LH: Light-holes | L-L: Light in-Light out<br>MD: Micro disc<br>MIGS: Metal-induced gap states<br>MQW: Multi quantum well<br>MSM: Metal-semiconductor-metal<br>PAI: Pre-amorphized by implantation<br>PC: Photonic crystals<br>PD: Photodiode<br>PMMA: Polymethyl methacrylate<br>QDs: Quantum dots<br>rms: (Surface) root mean square (roughness)<br>RP-CVD: Reduced Pressure – Chemical Vapor Deposition<br>RSM: Reciprocal space map<br>Si: Silicon<br>SiGeSn: Silicon germanium tin<br>Sn: Tin<br>SO: Spin orbit<br>SRB: Strain-Relaxed Buffers<br>SSR: Solid-state reaction<br>TDD: Threading Dislocations Density<br>TEM: Transmission Electron Microscope<br>TMAH: Tetramethylammonium hydroxide<br>VSL: Variable stripe length<br>VS: Virtual substrate<br>WG: Waveguide<br>XRD: X-Ray Diffraction |

# Introduction

Silicon photonics in the near-IR, up to 1.6 µm, is already one of key technologies in optical data communications, particularly short-range. It also is being prospected for applications in quantum computing, artificial intelligence, optical signal processing, where complex photonic integration is to be combined with large-volume fabrication. However, silicon photonics does not yet cover a large portion of applications in the mid-IR. In the 2 to 5 µm wavelength range, environmental sensing, life sensing and security, all rely on optical signatures of molecular vibrations to identify complex individual chemical species. The markets for such analysis are huge and constantly growing, with a push for sensitivity, specificity, compactness, low-power operation and low cost.

An all-group-IV, CMOS-compatible mid-IR integrated photonic platform would be a key enabler in this wavelength range. As for other wavelengths, such a platform should be complete with low-loss guided interconnects, detectors, modulators, eventually, and most importantly efficient and integrated light sources. Since the first demonstration of lasing in a GeSn optical cavity in 2015 [1], remarkable progress has been achieved in the development of Silicon Germanium Tin (SiGeSn) group-IV semiconductors as laser materials, including near room temperature laser operation [2], lasing at ultra-low threshold power densities [3] and near unity wall plug efficiency [4]. Most recently, an electrically pumped Germanium Tin (GeSn) laser operating at temperatures up to 100 K was reported [5]. We will review here recent developments concerning optically pumped lasers and other optoelectronic devices such as GeSn-based Light Emitting Diodes and photodetectors. We will also present insights into the fundamentals of these developments, e.g. the material growth and processing techniques as well as the band structure modelling. We hope that this will give the reader the opportunity to learn from the past and find his own best strategy for the future.

Over the years, numerous concepts had been put forward such as SiGe superlattice zone-folding [6], confinement induced reciprocal space engineering [7] as well as hexagonal core shell nanowires [8], to name a few. These are concepts that “by-pass” the problem of the indirect band structure of silicon and, to a lesser extent, germanium. Defect doping [9], as described in the “Stimulated emission in the near infrared from disordered Ge dots” chapter by Moritz Brehm, Johannes Kepler University, Linz, in this book, is another way of bypassing this limitation [10]. Here, instead of working around it, we solve the problem from scratch. Namely, a group-IV material is developed that, by its very nature, is a direct band gap system. The community is currently exploring two directions. The first one consists in transforming the indirect Ge band-gap into a direct one using high amounts of tensile strain [4,11]. The second one relies on the alloying of Ge with Sn and the control of the GeSn layer’s strain [12,13]. Both methods improve dramatically the Ge light emission properties in the short mid-infrared domain (in the 2 to 5 µm wavelength range) and, as shown in the following, enables the fabrication of lasing devices.

The current research is focused on (i) the reduction of the laser threshold and (ii) the increase of the maximum lasing temperature. The lasing threshold density is closely related to loss mechanisms such as non-radiative recombination via defects at the surface, at the interfaces and also in the bulk of the optically active material. Such defects may be minimized by dedicated growth and etching strategies, as detailed below. Carrier confinement is another route to lower the laser threshold. This however requires appropriate barrier layers and the growth of hetero- or (multi)-quantum well structures. Maximal temperature operation is – as will be shown – in the first place related to the amount of Sn incorporated into the active layer. As expected, most of the strategies come with trade-offs. For instance, the defect density is higher in higher Sn content, direct bandgap GeSn layers, resulting in non-radiative recombination channels which jeopardize the threshold density. Meanwhile, carrier confinement in wells induce quantization effects that reduce bandgap directness.

This review is meant to reflect ongoing discussions on material growth, device manufacturing and device concepts. As often in a review, we sometimes had to choose between going into details but not cover all possible options, or the opposite, mention all options but only cursorily. For the technical part, the growth and processing, we chose the detailed approach, which necessarily reflects our personal experiences. This was to give the reader a sense of what is needed, but we of course acknowledge the many good works done elsewhere and thoroughly cite them. The review also gives an overview of recent experimental achievements.

Our review is organized as follows: Section 1 describes recent Ge and SiGeSn epitaxy developments and processes to fabricate devices in group-IV indirect and direct band gap materials. The impact of strain and of Sn incorporation on the band structures of Ge, to transform it from an indirect into a direct bandgap material, is detailed in Section 2. In Section 3, the latest advances on lasing in Ge-based materials are reviewed. A state of the art of Ge-based optoelectronic devices such as light emitting devices (LEDs), photodetectors and lasers, is provided in section 4. In Section 5, we will conclude by describing the opportunities Ge and GeSn offer to a future monolithic integration on a Si photonic platform, and discuss future developments needed to reach this goal.

# 1. Fabrication: Ge-based epitaxy and processes for group-IV indirect and direct band gap material

Germanium is a strategic material for microelectronic and photonic applications. It can be grown on Si(100): it is widely used for (i) near Infra-Red photo-detectors thanks to its indirect band gap of 0.77 eV at room temperature and (ii) Metal Oxide Semiconductor Field Effect Transistors (MOSFETs), with higher hole mobilities than in Si. We will review in this section Ge, GeSn and SiGeSn epitaxial growth and some of the processes, such as etching, which are specific to those materials. Metallization, in order to benefit from efficient contacts on GeSn-based layers and fabricate electrically pumped group-IV devices, will also be discussed, with a focus on thermal stability.

## 1.1 Germanium growth

Thick Ge Strain-Relaxed Buffers (SRBs) on Si(001) can be used for a variety of purposes in microelectronics and optoelectronics. The top part of microns-thick Ge layers can be peeled-off from the Ge/Si stack underneath and bonded on oxidized Si using the SmartCut$^{TM}$ approach [14], resulting in Germanium-On-Insulator (GeOI) substrates. The latter can be used as templates for the fabrication of high mobility p-type Metal Oxide Semiconductor Field Effect Transistors [15]. The slight tensile strain present in thick Ge films on top of buried oxides can be exploited for the fabrication of highly strained suspended micro-bridges and micro-crosses [16,17,18]. Because of their lattice parameter, which is 4.2% higher than that of Si (5.65785 Å ⇔ 5.43105 Å) and close to that of GaAs (5.653 Å), Ge SRBs can be used as templates for the epitaxy of anti-phase boundaries-free, superior quality GaAs and InAlAs-based buffers [19,20]. The later can be called upon for the fabrication of high electron mobility n-type MOSFETs on III-V-On-Insulator substrates [21]. Thick Ge layers grown selectively at the end of Silicon-On-Insulator (SOI) waveguides can act as the active cores of superior performance near Infra-Red photo-detectors [18], [22,23]. Large area, high quality graphene 2D layers were recently deposited on Ge SRBs [24,25]. Finally, Ge SRBs are handy for the deposition of high Sn content GeSn layers which are typically used for the fabrication of mid-IR optically pumped lasers [1,2], as we will see in the following Sections.

Such SRBs are typically grown on Si(001) using a Low Temperature / High Temperature approach [26], with a short duration Thermal Cycling or anneal afterwards [27] to minimize the Threading Dislocations Density (TDD). This TDD is typically around $10^7$ $cm^{-2}$ for 2.5 µm thick Ge layers which are really smooth given the large lattice parameter mismatch between Si and Ge (surface root mean square roughness typically around 0.8 nm for 20 µm x 20 µm Atomic Force Microscopy images). The TDD monotonously decreases with the Ge thickness, as shown for layers thicknesses up to 2.5 µm [28]. The universality of such a behavior was conclusively demonstrated in the Ge/Si system and the GaAs/Si system in Ref. [29]. These findings were extended, for the Ge/Si system, to larger thicknesses (up to 5 µm) in Ref. [30] by CEA-LETI and recently confirmed by Leibniz-Institut für innovative Mikroelektronik (IHP) in Ref. [31].

The impact of thickness, in the 0.12 – 1.56 µm range, thermal cycling and substrate nature (nominal or slightly vicinal) on the properties of Ge layers grown on Si(001) was recently assessed in a 300mm Reduced Pressure – Chemical Vapor Deposition (RP-CVD) chamber from Applied Materials. The original results presented in the following Subsections are typical of Ge SRBs grown in industrial epitaxy tools whatever their brand and the wafer diameters used.

### 1.1.1 Ge growth protocol

The flow of $H_2$ carrier gas was set at a fixed value of a few tens of standard litres per minute. Germane ($GeH_4$) diluted at 2% in $H_2$ was used as the source of Ge. The $F(GeH_4)/F(H_2)$ mass-flow ratio was always equal to $10^{-3}$. The slightly p-type doped 300 mm Si(001) substrates used were either nominal (± 0.25°) or slightly vicinal, with a 0.5° misorientation towards one of the <110> directions, to facilitate Anti-Phase Boundaries (APB) - free GaAs growth on top [20]. During growth, the wafer laid horizontally on top of a circular SiC - coated susceptor plate that rotated to improve the spatial thickness uniformity of the films. It was heated by 76 tungsten – halogen lamps located above and below the susceptor assembly. Temperature monitoring and control was ensured through the lower pyrometer, i.e. the one looking at the backside of the susceptor plate on which the wafer laid.

The 0.12 µm – 1.56 µm thick Ge layers were grown on nominal or 0.5° off-axis Si(001) substrates in three steps. After a 1100°C, 2 minutes $H_2$ bake (to get rid of chemical oxide through the following reaction: $SiO_2(s) + 2H_2(g) \rightarrow Si(s) + 2H_2O(g)$), a 120 nm thick Ge layer was grown in 460s at 400°C, 100 Torr, in order to start from a rather flat, nearly fully relaxed Ge “seed” layer. The Ge growth rate was then equal to 16 nm $min.^{-1}$. In the second step, the temperature was ramped from 400°C up to 750°C (2.5°C/s) and the growth pressure from 100 Torr down to 20 Torr while still having germane flowing into the growth chamber. Around 80 nm of Ge were deposited during the 2$^{nd}$ step. In the third step, a Ge layer was grown at 750°C, 20 Torr, with a 45 nm $min.^{-1}$ growth rate, in order to obtain the desired thickness. A 3 x (875°C, 10s / 750°C,10s) Thermal Cycling under $H_2$ was used on the thickest Ge layers to minimize the Threading Dislocations Density.

### 1.1.2 Surface morphology

2 µm x 2 µm and 20 µm x 20 µm Atomic Force Microscopy (AFM) images of the surfaces of 1.56 µm thick, cyclically annealed Ge layers after growth on nominal and 0.5° off Si(001) substrates can be found in Figure 1. They are characterized by the super-position of a cross-hatch along the <110> directions on top of holes and small hills bordered by bi-atomic step edges. This <110> cross-hatch is the surface signature of the propagation on (111) planes of the threading arms of 60° misfit dislocations in the Ge layers, notably during Thermal Cycling. Those threading arms left in their wake <110> “plough” lines.

The main difference between the two types of surfaces is the bi-atomic step orientation and spacing. Those steps are randomly oriented, with a spacing between steps which fluctuates when starting from a nominal Si(001) substrate (Figure 1 left column images). Meanwhile, they are parallel to one of the <110> directions (from the top left down to the bottom right of pictures) and are more closely spaced with a 0.5° off-axis Si(001) template (Figure 1 right column images), which should be favourable for the growth of APB-free GaAs layers on top.

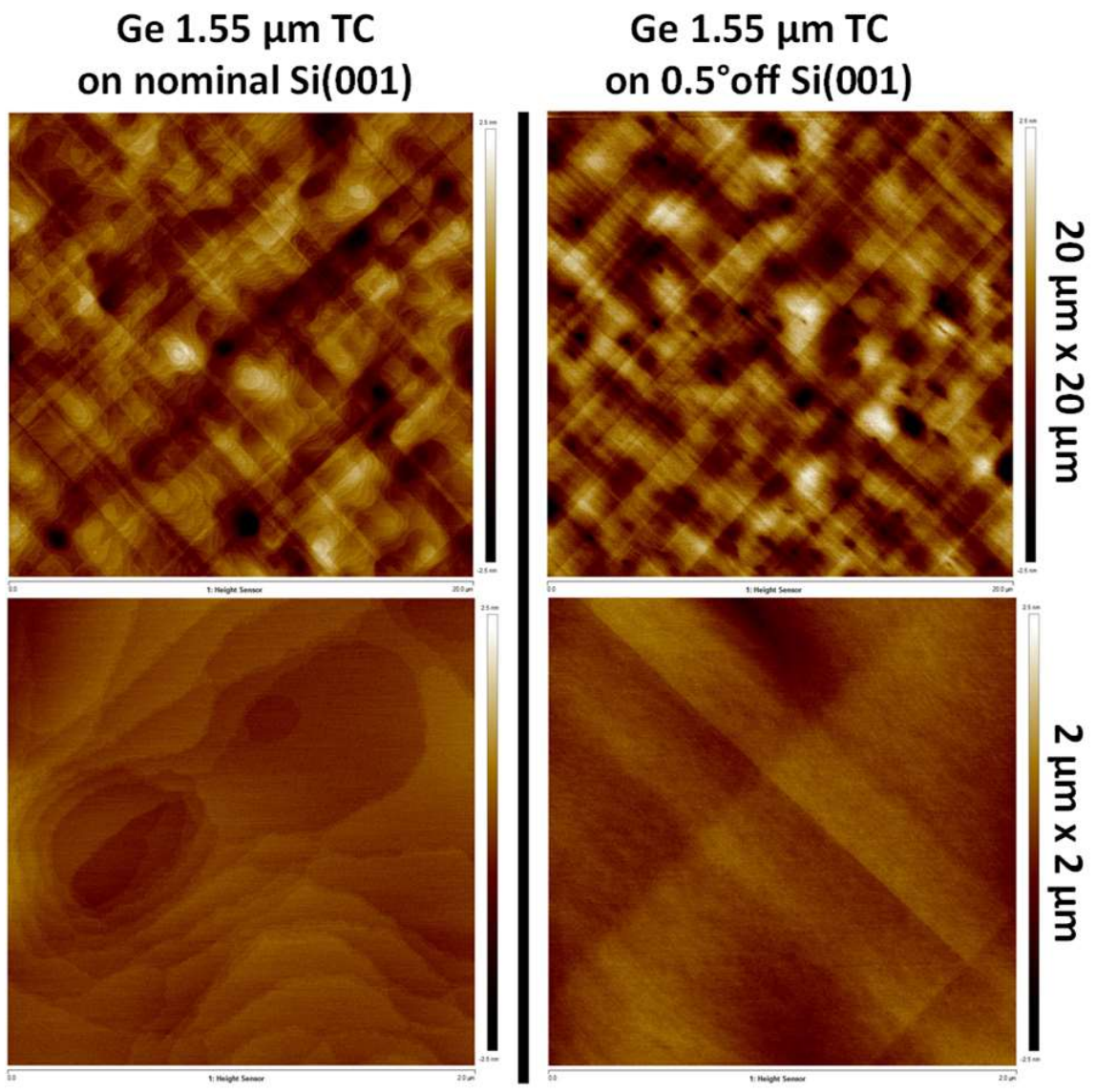

Figure 1: 20µm x 20 µm AFM images and 2 µm x 2 µm close-ups of the surface of 1.56 µm thick Ge layers grown on nominal and 0.5°off-axis Si(001) substrates (left and right images, respectively). Image sides are along the <100> directions, i.e. at 45° to the wafer notch and the cleaving directions.

The surface root mean square (rms) roughness and the Z ranges (= $Z_{max.} - Z_{min.}$) of those layers were extracted from those AFM images. They were rather small, with values for growth on a nominal surface 30% less, on average, than that for growth on a 0.5° off-axis substrate: 0.2 nm and 1.9 nm, to be compared with 0.3 and 2.5 nm (2 µm x 2 µm images) and 0.55 nm and 6.2 nm, to be compared with 0.7 nm and 7.1 nm (20 µm x 20 µm images).

### 1.1.3 Macroscopic degree of strain relaxation and crystalline quality from X-Ray Diffraction

High resolution X-Ray Diffraction (XRD) was used to quantify the macroscopic degree of strain relaxation $R = (a_{Ge}^{//} - a_{Si}) / (a_{Ge} - a_{Si})$ as a function of Ge layer thickness and cyclic anneal (or not). $a_{Ge}^{//}$, $a_{Si}$ and $a_{Ge}$ are the in-plane lattice parameter of the Ge layer, the lattice parameter of the Si substrate (5.43105 Å) and the lattice parameter of bulk Ge (5.65785 Å), respectively. Omega-2Theta profiles around the (004) XRD order are plotted in Figure 2a for various thickness Ge layers on Si. Apart from the Si substrate peak (located at 34.564°), there is, at slightly more than 33°, the Ge layer peak, whose intensity drastically increases as the Ge layer thickness increases. The Ge peak is slightly asymmetric. The high incidence angle component is due to the interfacial GeSi alloy formed during the thermal cycling [27]. From the angular position $\omega_{Ge}$ of this peak, we can gain access to $a_{Ge}^{\perp}$, the lattice parameter of the Ge layer in the growth direction, thanks to Bragg's Law : $2\left(a_{Ge}^{\perp}/4\right)\sin\omega_{Ge}=\lambda$, λ being the Cu Kα1 wavelength (1.5406 Å). $a_{Ge}^{//}$ and then $R$ can then be determined thanks to the formalism detailed in Ref. [32].

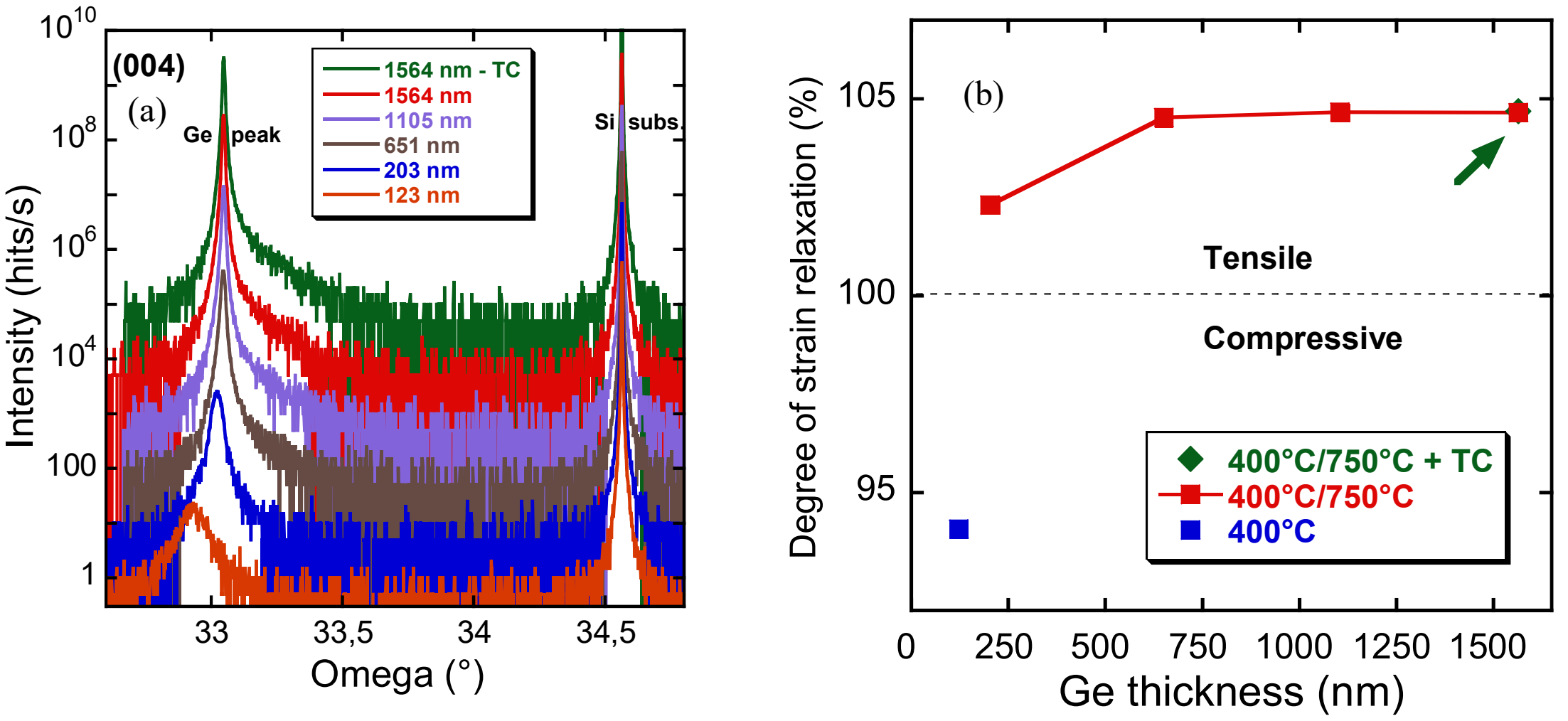

Figure 2 : (a) Omega-2Theta scans around the (004) XRD order on various thickness Ge layers grown on Si(001). TC stands for Thermal Cycling, (b) Macroscopic degree of strain relaxation as a function of Ge deposited thickness for layers grown at 400°C, at 400°C/750°C or at 400°C/750°C followed by a short duration Thermal Cycling between 750°C and 875°C.

Almost all Ge layers were slightly tensily-strained, see Figure 2b, with R values which monotonously increased then stabilized (at a 104.5% value) as the thickness increased. Such a strain state was due to differences in thermal expansion coefficients between Ge and Si (see Ref. [32] for numerical values). The lattice parameter of Ge thick layers on Si substrates, which were almost fully relaxed at 750°C (the upper growth temperature) or 875°C (the upper temperature of thermal cycles), shrank differently from that of a bulk Ge crystal. $a_{Ge}^{\perp}$ contracted with the thermal dilatation coefficient of Ge, whereas $a_{Ge}^{//}$ shrank with the smaller one of the much thicker Si substrate underneath. We thus ended up with tensile-strained Ge layers as soon as there was a ramping-up to 750°C or a thickening at 750°C, with R values in the 102.3% - 104.7% range. Using a short duration thermal cycling had no clear impact on R, which was exactly the same (104.7%) for a 1.56 µm cyclically annealed Ge layer than for the same thickness un-annealed one. The only layer which was still compressively strained was the 120 nm thick Ge layer grown at 400°C, with an R value, 94%, which was very close to 100% (full strain relaxation). Prestrain values as obtained by alternative methods are summarized in 2.1.5, Table 2.

### 1.1.4 Threading dislocations density from X-Ray Diffraction

The Ayers' theory [33] was then used in order to convert the Full Widths at Half Maximum β of Ge peaks from omega scans around the (004) XRD order, such as the ones provided in Figure 3a for various thickness Ge SRBs, into Threading Dislocation Densities (TDDs). Ayers *et al.* have indeed shown that the TDD was related to β by TDD = $(\beta/3b)^2$, b being the length of the Burger's vector if the rocking curve width is determined solely by threading dislocations. For pure Ge layers with 60° misfit dislocations, b = $0.707a^{bulk}_{(Ge)}$ = 4.00 Å. The TDD (in $cm^{-2}$) is then linked to β (in arc seconds) by TDD = $1632\beta^2$ for pure Ge.

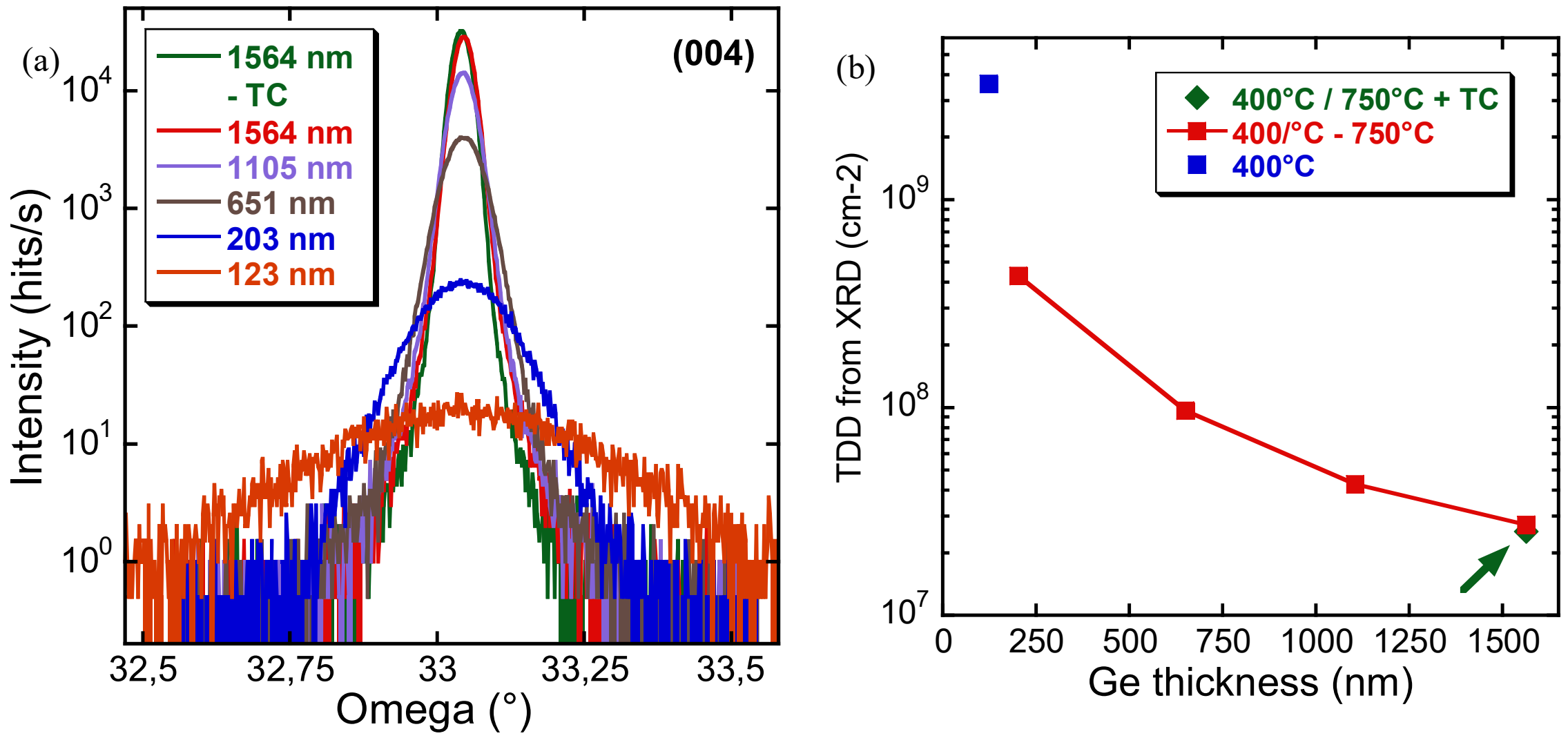

Figure 3: (a) Omega scans around the (004) XRD order on the various thickness Ge layers grown on Si(001). TC stands for Thermal Cycling, (b) Threading Dislocations Densities ($cm^{-2}$) inferred from Omega scans around the (004) Ge XRD peak and the use of Ayers" formula.

TDDs coming from XRD and the use of Ayer's formula are plotted in Figure 3b. Using thick Ge layers and high growth temperatures is advantageous: the TDD indeed drops from $3.6\times10^9$ $cm^{-2}$ (123 nm @ 400°C) down to $2.5\times10^7$ $cm^{-2}$ (1564 nm @ 400°C/750°C + thermal cycling) as the Ge thickness increases.

One might wonder about the significance of such TDD values, which are indirect. We have shown in Ref. [34] that, for 2.5 µm thick cyclically annealed Ge SRBs, we had after (i) the removal of one micron of Ge with Secco and chromium-free wet etching solutions and (ii) plane view transmission electron microscopy, a TDD of ~ $10^7$ $cm^{-2}$, to be compared with $8\times10^6$ $cm^{-2}$ after HCl defect etching and $1.8\times10^7$ $cm^{-2}$ from XRD in Ref. [30] layers. Figure 3b data thus seem trustworthy.

The reason as to why TDDs from HCl defect etching were on average 2.5 times lower than from XRD in Ref. [30] was tentatively due to the following: we had after HCl defect etching a TDD value which was a "surface" one, as only 130 nm of Ge was etched, while the XRD one came from the bulk of the Ge layer, with therefore a TDD mean value which might be higher than in reality. This might also be the reason why differences were deceptively small, in XRD, between as-grown and cyclically annealed 1.56 µm thick Ge SRBs (Figure 3b). We actually showed, for instance in Refs. [27], [28], [35], that the TDD, from Secco or HCl defect decoration, was 3 to 5 times lower in cyclically annealed than in as-grown Ge SRBs.

Finally, we have used cross-sectional Transmission Electron Microscopy (TEM) to image, in Ref. [35], a 2.8 µm thick cyclically annealed Ge layer. Although the volume probed in the TEM lamella was small, we showed that the 0.15 µm thick zone close to the Ge/Si interface had a dislocation density of the order of $10^{10}$ $cm^{-2}$. The 1.3 µm Ge layer on top had a measured dislocation density around $1–2\times10^{9}$ $cm^{-2}$. Finally, the topmost 1.35 µm thick Ge layer was characterized by a dislocation density of the order of $3–6\times10^{7}$ $cm^{-2}$ (see Figure 4). Such a value is actually close to the ~ $2x10^{7}$ $cm^{-2}$ TDD from XRD in similar thickness Ge layers, validating our use of Ayers' theory.

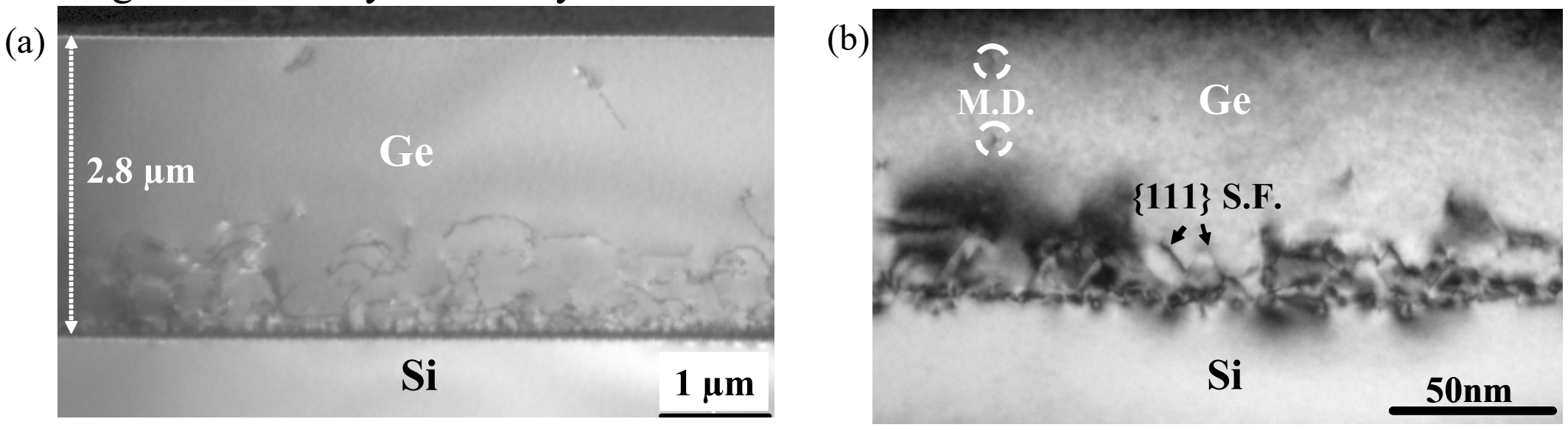

Figure 4 : (a) cross-sectional Weak Beam Dark Field image of a 2.8 µm thick, annealed Ge layer grown on a nominal Si(001) substrate. (b) Dark Field with close to Bragg conditions image (taken with g = 220) of the interface between Ge and Si. The {111} Stacking Faults (SF) observed in Figure 4b are the result of the dissociation of perfect Misfit Dislocations (MD) into partial dislocations [35].

## 1.2 Germanium tin growth

Silicon (Si) remains the backbone of the semiconductor industry due to the unique features offered by this semiconductor. Because of its indirect bandgap, it is however unable to efficiently emit light. The heterogeneous integration of direct bandgap III-V semiconductors on Si, which is the mainstream approach used to obtain near infra-red lasers onto Si [36,37], has also some drawbacks, including a limited compatibility with CMOS fabrication techniques. Meanwhile, Ge, an almost direct bandgap group-IV semiconductor, gained considerable attention in recent years for the monolithic integration of lasers on Si platforms. Ge, which is fully CMOS compatible, can indeed be turned into a direct bandgap material by applying high amounts of tensile stress [4, 16].

Germanium can otherwise be alloyed with tin [38], a semi-metal, in order to fabricate optically pumped mid infra-red lasers [1, 2, 3, 39]. GeSn alloys can also be used in photo-diodes [40,41], p-i-n Light Emitting Devices and electrically pumped lasers [5, 42], Fin-type Field Effect Transistors [43], vertical gate-all-around FETs [44] and so on. They indeed offer the possibility of engineering the energy bandgap by changing the Sn content and the built-in strain. They have a direct bandgap above a threshold Sn concentration around 8% (in unstrained layers). Their growth, which has to be conducted at low temperatures (because of Sn segregation and precipitation) can be an advantage for monolithic 3D integration. A GeSn alloy crystallizes in a diamond structure, enabling its epitaxial growth on Si(001) substrates. Its lattice parameter is much higher, however, making strain management complex ($a_{Sn}$ = 6.489 Å ⇔ $a_{Si}$ = 5.43105 Å and $a_{Ge}$ = 5.65785 Å). These features triggered the interest of researchers and significant efforts have been made since 2010 to improve the epitaxial quality of GeSn, notably in Chemical Vapor Deposition (CVD). However, the low solid solubility (<1%) of Sn in Ge and the low thermal stability of $Ge_{1-x}Sn_x$ alloys make their epitaxy quite challenging. This is particularly true for high Sn contents *x*. Thick Ge Strain-Relaxed Buffers (SRBs) are typically used to mitigate the deleterious impact of the large lattice parameter mismatch between Si and $Ge_{1-x}Sn_x$. Growth conditions far from equilibrium must otherwise be used, with a combination of high growth rates and low growth temperatures (below 350°C, typically). This can be achieved by selecting the proper CVD precursors and growth parameters.

B. Vincent *et al.* from imec, were the first to use a mixture of digermane ($Ge_2H_6$) and tin tetrachloride ($SnCl_4$) to grow, at Atmospheric Pressure (760 Torr), pseudomorphic GeSn layers on Ge SRBs [45]. This chemistry was later on used by the ForschungsZentrum Juelich to build the first GeSn laser [1]. Since then, it has been adopted by for instance ForschungsZentrum Juelich, CEA-LETI or KTH Royal Institute of Technology in Stockholm to fabricate different types of devices. Typical growth features can be found in Refs. [46,47,48,49,50].

Digermane has however the reputation of being unstable, notably for high concentrations in the bottle (10%, typically). It is otherwise costly and difficult to order. This spurred ASM America and the University of Arkansas to evaluate mainstream germane ($GeH_4$). They reproductively showed that it yielded, together with $SnCl_4$, the same optical quality layers as $Ge_2H_6$ [51,52]. They, together with other entities like imec or the University of Warwick, benchmarked $GeH_4$ and $Ge_2H_6$ for the growth of GeSn [53,54,55]. Process details were scarce, however, especially for $GeH_4$. Temperatures were provided but pressure data were at best vague or missing altogether. The only thorough comparison was that J. Margetis *et al.*, from ASM America, who systematically explored the impact of the $F(SnCl_4)/F(GeH_4)$ Mass-Flow Ratio and the temperature on the GeSn growth rate and Sn content. There was no data on chamber pressure, however [55].

In the following, a one-to-one comparison of $GeH_4$ and $Ge_2H_6$ in a 200 mm RP-CVD cluster tool from Applied Materials, which can operate up to several hundreds of Torrs, will be presented. Growth temperature, pressure, precursor flows and $H_2$ carrier flows were changed in order to grow high quality GeSn layers [56].

### 1.2.1 GeSn growth protocol

The growth of Ge and $Ge_{1-x}Sn_x$ layers was performed on nominal Si(001) substrates. 2.5 µm thick Ge Strain-Relaxed Buffers were grown in a regular temperature epitaxy chamber to accommodate the lattice mismatch between $Ge_{1-x}Sn_x$ and Si. Germane was then used as a source of Ge. The Ge SRBs were grown using a low temperature / high temperature approach followed by a short duration thermal cycling in order to reduce the threading dislocations density, which was close to $10^7$ $cm^{-2}$. Samples were then kept at 20 Torr under ultra-pure $N_2$ in the load-locks of the tool to avoid surface oxidation. Afterwards, the wafers were loaded in the dedicated epitaxy chamber of the cluster tool equipped with low temperature infra-red pyrometers. Prior to $Ge_{1-x}Sn_x$ growth, the samples were annealed under $H_2$ at 800ºC for 2 minutes, i.e. a temperature significantly lower than the highest temperature used during the thermal cycling (875°C). Surfaces had therefore, after such 800°C bakes, the same cross-hatched morphology and roughness parameters as that of as-grown Ge buffers.

$GeH_4$ or $Ge_2H_6$ diluted at 2% in $H_2$ were used as Ge sources. As $SnCl_4$ is a liquid precursor, a bubbler was used to feed tin atoms into the growth chamber. The temperature and the pressure inside the bubbler were set to deliver $SnCl_4$ 1% in $H_2$ when hydrogen was flown in. In the following, the $GeH_4$ and the $Ge_2H_6$ flows were constant and such that $F(GeH_4) = 4xF(Ge_2H_6)$. This meant that the Ge atomic flow from germane was twice that from digermane. The growth pressures were always (i) 400 Torr for $GeH_4 + SnCl_4$ and (ii) 100 Torr for $Ge_2H_6 + SnCl_4$. The overall $H_2$ carrier flow was constant and the same for both chemistries. Finally, the $SnCl_4$ flow was either the same or half with $GeH_4$ than with $Ge_2H_6$.

### 1.2.2 Impact of temperature and germanium precursor on GeSn growth kinetics

As shown in the left part of Figure 5, there was, over the 301°C-349°C range, an exponential increase of the GeSn growth rate with the temperature, with similar activation energies for both chemistries. The activation energy increased with the $SnCl_4$ flow, from 8.6 kcal mol.$^{-1}$ for $F(SnCl_4)/F(H_2) = 2x10^{-5}$ up to 11.8 kcal mol.$^{-1}$ for $F(SnCl_4)/F(H_2) = 4x10^{-5}$. The latter is close indeed to the activation energy with $Ge_2H_6$ and the same $SnCl_4$ flow, 10.4 kcal. mol.$^{-1}$. Those values are very similar to those obtained a few years ago with $Ge_2H_6$ in the same epitaxy chamber: 10.4 kcal. mol.$^{-1}$ [48] and 10.6 kcal. mol.$^{-1}$ [50]. They are also close to those reported in Ref. [55] for $GeH_4$ + $SnCl_4$, although the chamber design was different and the pressure not disclosed: 9.0 - 12.0 kcal. mol.$^{-1}$.

As shown in the right part of Figure 5, the Sn content decreased linearly as the growth temperature increased, with a slope that was less for $GeH_4$ (- 1.0 or -1.1 % / 10°C) than for $Ge_2H_6$ (- 1.7% / 10°C). Those slopes are once again close to literature values. Slopes of - 1.85% / 10°C [48] and -1.8% / 10°C [50] were indeed found for $Ge_2H_6$ in the same epitaxy chamber, in full agreement with the -1.8% / 10°C value obtained, also for $Ge_2H_6$, in an ASM Epsilon 2000 epitaxy chamber [49]. Slopes of -1.1% / 10°C were associated, for $GeH_4$, with $F(SnCl_4)/F(GeH_4)$ MFRs of 0.0085 and 0.012 in Ref. [55]. The Sn content dropped with a -1.3% / 10°C slope in pseudomorphic GeSn layers grown with a $GeH_4$ + $SnCl_4$ + $H_2$ chemistry in a home-made Low Pressure CVD reactor [57].

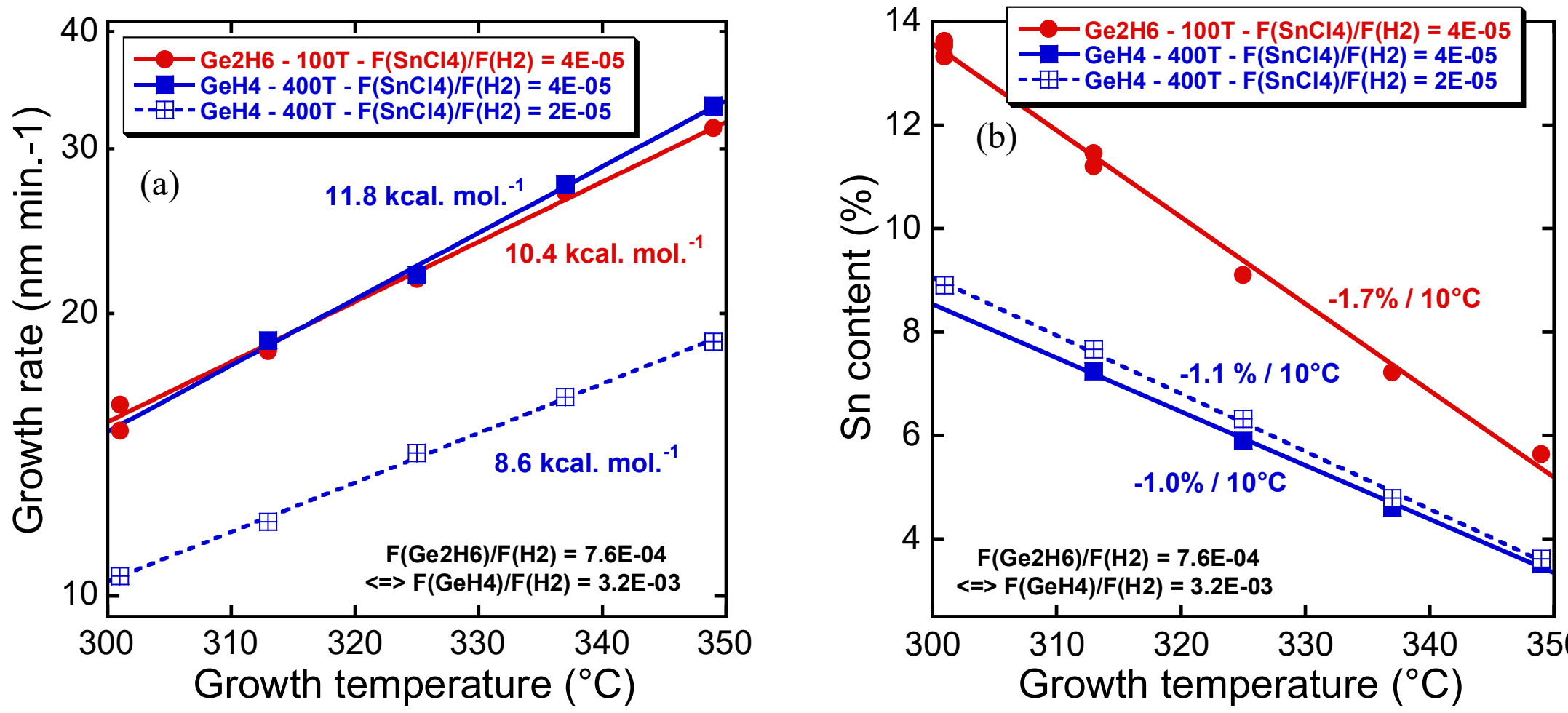

Figure 5 : Evolution with the temperature of GeSn growth rate and Sn content in GeSn layers grown at (a) 400 Torr with $GeH_4$ + $SnCl_4$ and (b) 100 Torr with $Ge_2H_6$ + $SnCl_4$ [56].

In order to gain more insight about the growth mechanisms, we have extracted from Figure 5 data the elemental Sn and Ge growth rate components, by multiplying the GeSn Growth Rate (GR) by the Sn fraction $x$ or the Ge fraction (1-$x$) (i.e. Sn GR component = $x$*GeSn GR and Ge GR component = (1-$x$)*GeSn GR)). Over the 301°C-349°C range, the Ge growth rate component increased exponentially with the temperature, with similar activation energies for both chemistries (left part of Figure 6). Ge GR components were almost the same for a given $SnCl_4$ flow, although the growth pressure was 4 times higher and the Ge atomic flow twice higher with $GeH_4$ than with $Ge_2H_6$. Halving, for the *same $GeH_4$ flow*, the $SnCl_4$ flow resulted in Ge GR components which were, *counterintuitively*, 40% lower.

Meanwhile, the Sn GR component monotonously decreased as the temperature increased, whatever the Ge precursor and the $SnCl_4$ flow probed (right part of Figure 6). This was likely due to Sn sublimation. The Sn GR component was 50% higher, for the same $SnCl_4$ flow, with $Ge_2H_6$ than with $GeH_4$ (although the Ge atomic flow was half of it). Halving, for the same $GeH_4$ flow, the $SnCl_4$ flow otherwise resulted in Sn GR components 35% lower.

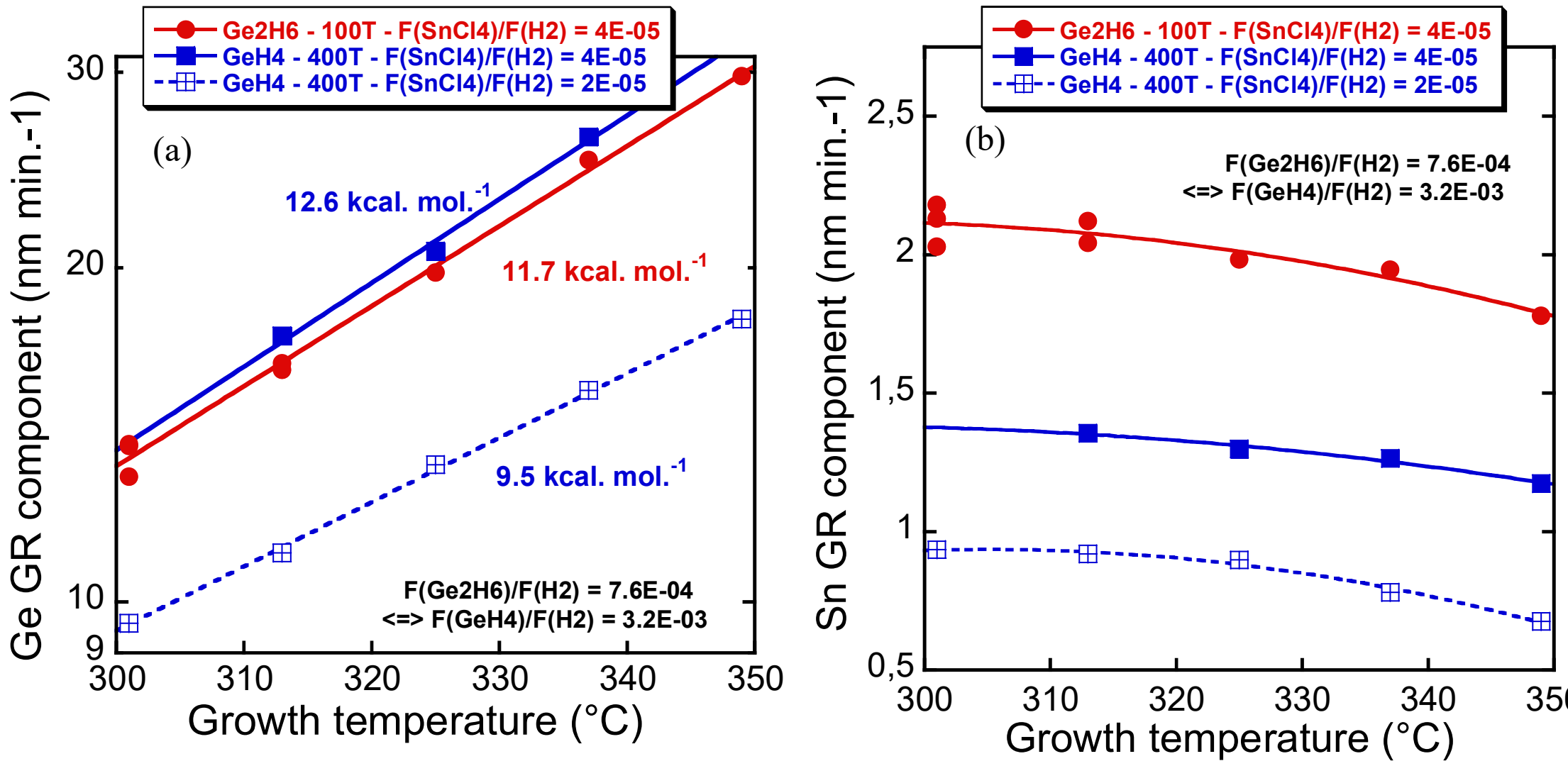

Figure 6: Evolution with the temperature of Ge and Sn growth rate components for GeSn layers grown at (a) 400 Torr with $GeH_4$ + $SnCl_4$ or (b) 100 Torr with $Ge_2H_6$ + $SnCl_4$ [56].

Complex interplays between Ge precursors and $SnCl_4$ thus governed the GeSn growth kinetics, with most likely the formation of reactive intermediaries. J. Margetis *et al.* [55] indeed suggested that, for the $GeH_4$ + $SnCl_4$ system, Ge incorporation mainly occurred through dissociative adsorption ($GeH_4$(g) + 3_ -> $GeH_2$ + 2H, then $GeH_2$ -> Ge + $H_2$(g)). Meanwhile, Sn incorporation likely happened through the formation of dichloro-products in the gaseous phase (i.e. $GeH_4$(g) + $SnCl_4$(g) -> $GeH_2Cl_2$(g) + $SnH_2Cl_2$(g), followed by $SnH_2Cl_2$(g) -> SnHCl(g) + HCl(g), SnHCl(g) + _ -> SnHCl and finally SnHCl -> Sn + HCl(g)). This might be one of the reasons why Sn concentrations were, in Figure 6, almost the same for $F(SnCl_4)/F(H_2)$ = $2x10^{-5}$ and $4x10^{-5}$: Sn incorporation might have been hampered by the $GeH_4$ flow, which was the same and not high enough to result in higher Sn contents for twice higher $SnCl_4$ flows.

### 1.2.3 Structural properties of thin, compressively strained GeSn layers

X-Ray Diffraction profiles highlighting the structural quality of thin, pseudomorphic GeSn layers grown at 400 Torr with $GeH_4$ + $SnCl_4$ in the 301°C – 349°C temperature range can be found in Figure 7a. As the growth temperature decreased, the GeSn peak moved away from the Ge SRB peak, which was a clear sign that the Sn content increased. GeSn peaks were well defined and intense, with thickness fringes on both sides. They were properly reproduced by simulations, which was another sign that layers were fully monocrystalline.

The root mean square (rms) surface roughness and the Z range (= Zmax. – Zmin.) associated with 5 µm x 5 µm AFM images of those GeSn layers can be found in Figure 7b. Surfaces were cross-hatched, as the Ge SRBs underneath, with little or no impact of the Ge precursor and growth temperature on roughness. The mean rms roughness and Z range of those pseudomorphic, tens of nm thick GeSn layers were equal to 0.54 nm and 4.84 nm, respectively.

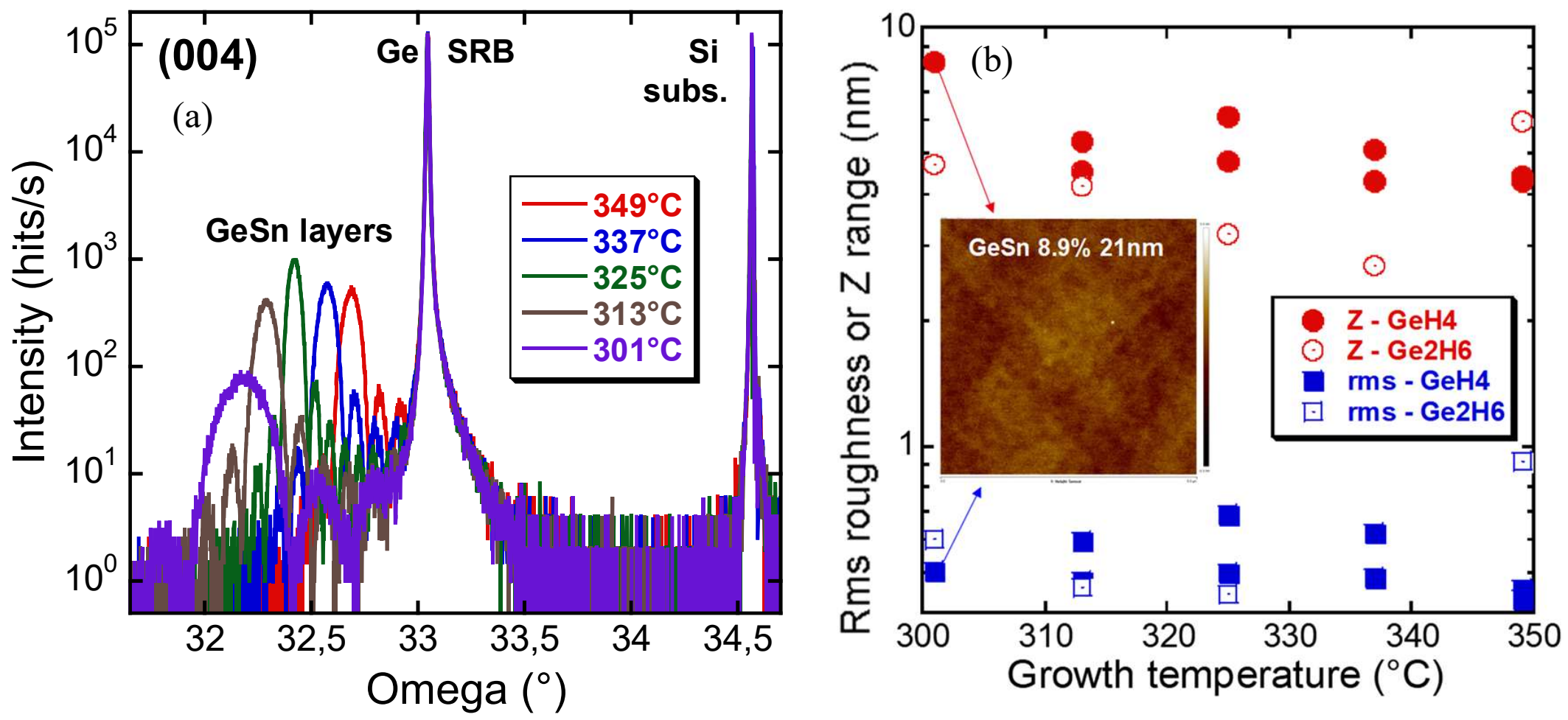

Figure 7 : (a) ω-2θ scans around the (004) XRD order associated with thin, pseudomorphic GeSn layers grown at 400 Torr with $GeH_4$ + $SnCl_4$ in the 301°C – 349°C temperature range, (b) rms roughness and Z range associated with 5 µm x 5 µm images of the surface of GeSn layers grown with $GeH_4$ + $SnCl_4$ or $Ge_2H_6$ + $SnCl_4$ on Ge SRBs, over the 301°C – 349°C temperature range. <100> scan directions. Inset: AFM image of the highest Sn content layer grown with $GeH_4$ [56].

### 1.2.4 SiGeSn epitaxy: some specificities

Before ending section 1.2 with a discussion on the properties of thick, relaxed GeSn layers on Ge SRBs which are typically used for optically pumped lasing, we will say a few words about SiGeSn epitaxy. Such ternary alloys can indeed be of use as energy barriers in SiGeSn / GeSn Multi-Quantum Wells or Double Hetero-Structures [58]. Indeed, the energy bandgap of Si, 1.12 eV, is higher than that of Ge, 0.77 eV and Sn, which is actually a semimetal (-0.41 eV).

It was shown in Ref. [59] that, at 349°C, 100 Torr and with a $Ge_2H_6$ + $SnCl_4$ + $Si_2H_6$ chemistry, there was, a *simultaneous increase of the Si and Sn contents with the $Si_2H_6$ flow* in thin, pseudomorphic SiGeSn layers on Ge SRBs, although the $SnCl_4$ flow was constant. Meanwhile, the SiGeSn growth rate decreased substantially with the $Si_2H_6$ flow, with some degradation, above a given threshold, of the surface morphology and the formation of Sn droplets. Those findings were confirmed at 337°C and 325°C in Ref. [50]. For given precursor flows, we also showed, in Ref. [50], that a growth temperature increase, in the 313°C – 363°C range, resulted in the following: (i) an exponential increase of the SiGeSn growth rate, with an activation energy, 9.5 kcal mol.$^{-1}$, close to that of GeSn, 10.6 kcal mol.$^{-1}$ (although growth rates were always lower for SiGeSn than for GeSn) and (ii) a Si content which increased, from 4% up to 12% (+ 1.31% / 10°C) and a Sn content which decreased, from 12% down to 4% (-1.25% / 10°C), *while the Ge content in the ternary alloy stayed constant at 84%.*

Results summarized above are definitely in line with ForschungsZentrum Juelich data in Refs. [60,61], although the carrier gas and the precursor injection were different ($H_2$ and laminar flow above the wafer surface in Refs. [50] and [59], to be compared with $N_2$ and a showerhead in Refs. [60,61]), strengthening those findings. The epitaxy of SiGeSn / GeSn stacks is not straightforward, because of requirements very much at odds with each other. High Sn content and thus definitely direct bandgap GeSn layers should for instance be grown at very low temperature to avoid Sn precipitation/surface segregation, while high Si contents and thus significant energy barriers would favour the high temperature growth of SiGeSn barriers. A $SnCl_4$ flow which is lower than in binaries should also be selected to have, in the end, the targeted Sn content in SiGeSn. Indeed, the addition of $Si_2H_6$ to the gaseous mixture simultaneously increases the Si and Sn contents, reduces the growth rate and so on.

### 1.2.5 Epitaxy of thick, relaxed GeSn layers

As we will see in the following, Sn content and strain have a major impact on the bandgap directness of GeSn: the lower the compressive strain and the higher the Sn content in GeSn layers, the more direct their bandgap will be and the better their light emission properties will be. This is typically achieved, in GeSn laser devices, by growing GeSn layers on Ge SRB with a thickness well above the critical thickness for plastic relaxation. Layers will then relax, with the emission of vast amounts of strain-relieving dislocations.

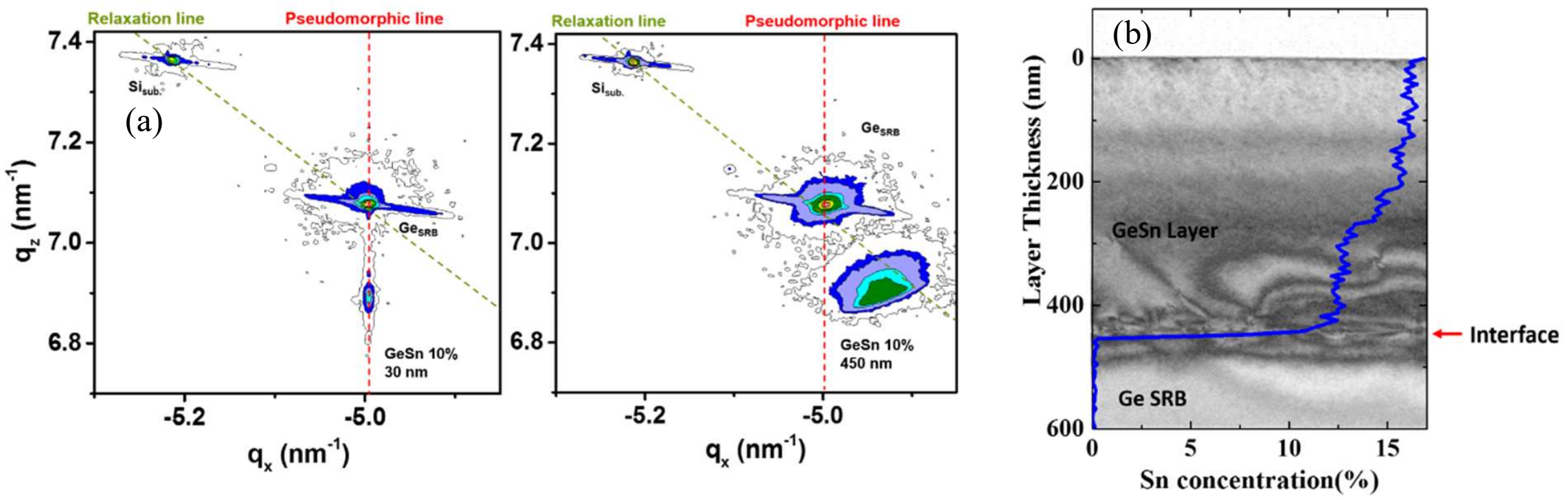

Figure 8 : (a) Reciprocal Space Maps around the (224) XRD order showing the plastic relaxation occurring when switching from 30 nm to 450 nm thick GeSn 10% layers grown at 325°C. The thinner layer has the same in-plane lattice parameter and thus $q_X$ value as the Ge SRB underneath, while the top part of the thicker layer is 66% relaxed. (b) Sn concentration profile (from Energy Dispersive X-ray analysis) superimposed on a Bright Field TEM image of a 465 nm thick GeSn layer grown at 313°C with nominally 12% of Sn [62].

We have explored, in Ref. [62], the impact of thickness on the properties of GeSn 10%, 12% and 15% layers grown at 325°C, 313°C and 301°C with the same $Ge_2H_6$ and $SnCl_4$ mass-flows and growth pressure (100 Torr) as in Figure 5. Above a given threshold that depended on the built-in strain and thus the Sn content, layers plastically relaxed, with, on top, the appearance of less heavily dislocated, higher Sn content layers, as shown in Figure 8. This Sn enrichment was shown by ASM America and the University of Arkansas [63] or by Ecole Polytechnique de Montréal [64] to be due to a gradual dissipation of the compressive strain and therefore an increase of the GeSn in-plane constant, improving the incorporation of larger size Sn atoms. Such an involuntary Sn grading is actually advantageous for the fabrication of high performance lasers, as optical recombination will occur in the thick, higher Sn content and crystalline quality layers on top and not in the bottom, lower Sn content and therefore higher bandgap bottom layers, which are more defective [65].

The feasibility of growing GeSn step-graded heterostructures, in order to gradually relax the built-in compressive strain in a controlled manner and confine the misfit dislocations in the bottom layers, was evaluated in Ref. [66]. To that end, Figure 5 data points were used, with a gradual decrease of the growth temperature from 349°C down to 313°C in 12°C steps. We succeeded in obtaining, as shown in Figure 9, superior quality stacks that were used to obtain optically pumped lasers in the mid infra-red with record high performances [2, 39,182].

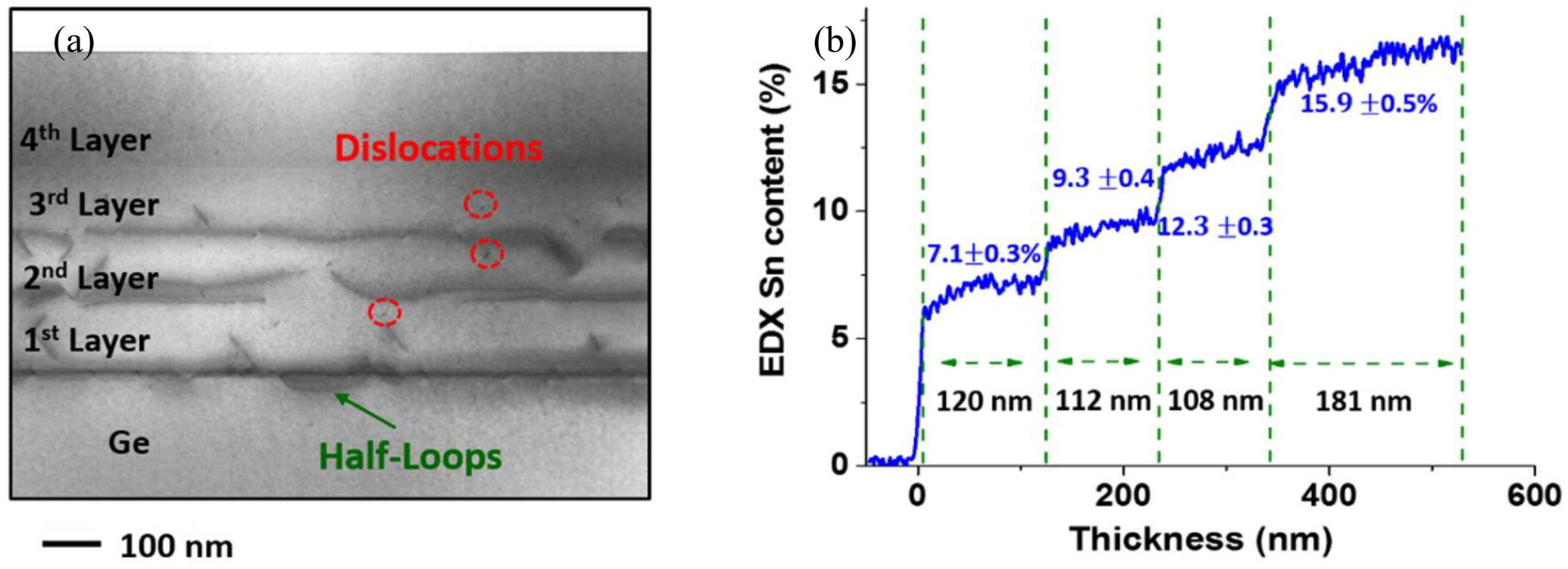

Figure 9 : (a) Scanning transmission electron microscopy (STEM) images of a GeSn step-graded sample (top). Some dislocations are highlighted. (b) A Sn concentration profile extracted from TEM–EDX measurements is provided in the right part of the figure. Layers 1, 2, 3 and 4, with 7%, 9%, 12% and 16% of Sn were 95%, 92%, 78% and 64% relaxed, respectively [66].

## 1.3 Ge-based materials processing

Lithographic steps used to fabricate Ge-GeSn optoelectronic devices must take into consideration the overall fabrication process. Indeed, many parameters such as the minimum pattern size, the layer thickness to be etched, the compatibility of some metals with the developer chemistry and so on have a direct impact on the lithography technique along with the resist type to be employed. For instance, diluting the resist with the proper solvent results in a thinner resist after spin coating, helping to reach smaller pattern sizes.

Optical lithography has been successfully used to fabricate waveguides [52]. However, electron beam lithography is the most versatile and spatially resolved tool found in most research papers. It enables to fabricate all sets of optical resonators, going from microdisks to photonic crystals [39, 67, 68]. For processes employing the lift-off approach commonly used in academic research laboratories, polymethyl methacrylate (PMMA) is the resist of choice. Unfortunately, PMMA suffers from very poor selectivities when exposed to dry etching plasmas. Doing so requires resists which are more tolerant to plasma etching, such as the ZEP 520a (positive tone) or the ma-N 2400 series (negative tone). Both have a very good pattern resolution, particularly for infrared resonators where constraint on resolution is less demanding than for shorter wavelength resonators. These resists can be easily stripped in an oxygen plasma after the dry etching step.

### 1.3.1 Anisotropic etching

Etching is a critical step in the fabrication of semiconducting optical resonators. Control of the selectivity between the layer to be etched and the mask or the buried layer is of prime importance. In addition, the resonance properties of the light are strongly influenced by the shape of the etched sidewalls (verticality, roughness). Etching recipes must also address these aspects. Ge can be anisotropically dry etched in fluorine type plasmas ($SF_6$ [69], mainly, or a mixture of $C_4F_8$ and $SF_6$ [70]). $Cl_2$ based chemistries can be used for Ge [71] and GeSn [72]. The latter approach was shown to give very smooth etched surfaces and vertical sidewalls. Table 1 gives a representative recipe for the anisotropic etching of Ge and GeSn in an Inductively Coupled Plasma Reactive Ion Etching (ICP-RIE) reactor with a $Cl_2$ chemistry. Typical etching rates are of the order of 450-600 nm/min for Ge and GeSn, while the selectivity (semiconductor to resist etched thickness ratio) is in the range of 2.5-4 with ma-N 2400 or ZEP 520a resists.

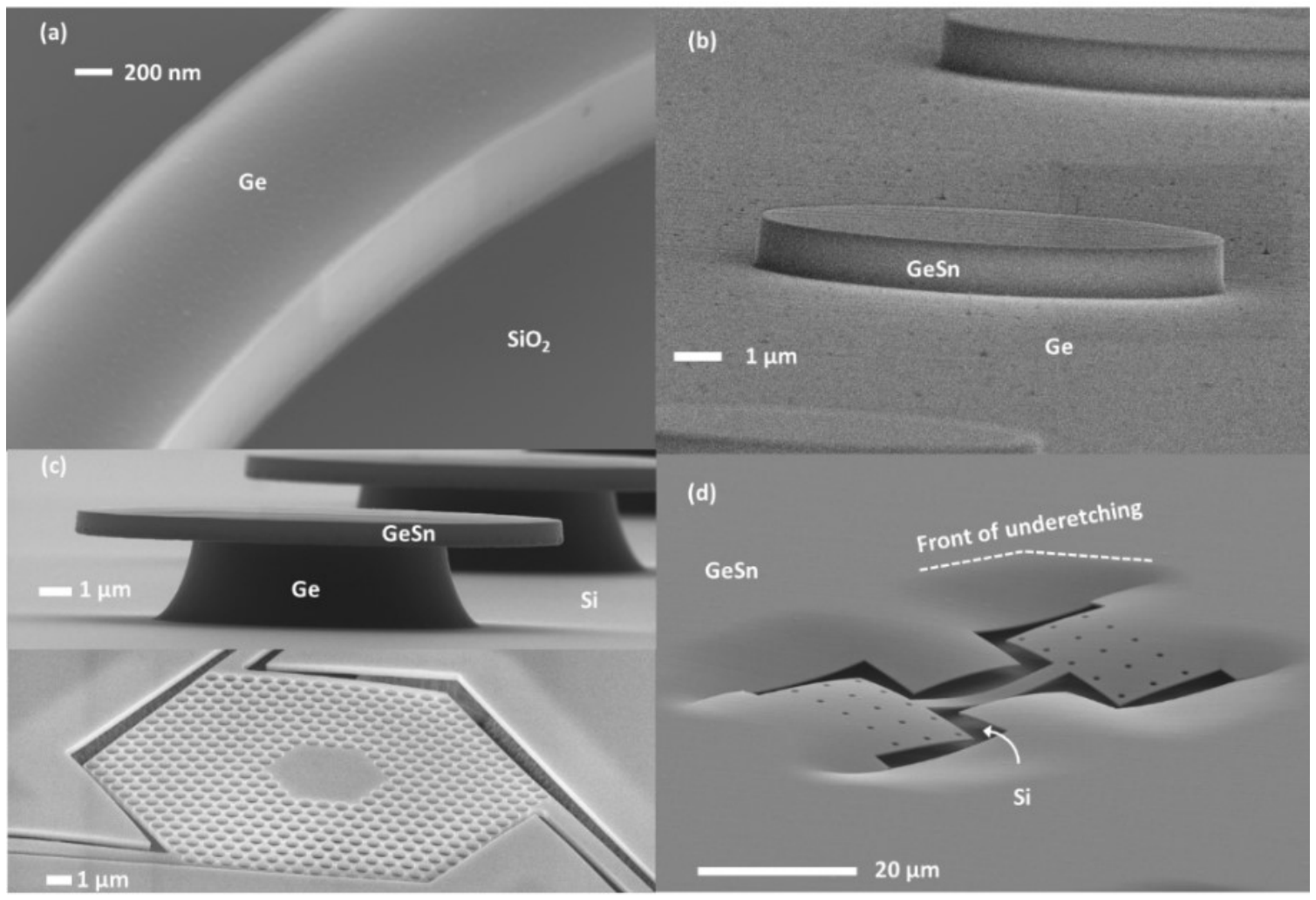

Figure 10: (a) Ge waveguide made from a Germanium On Insulator (GOI) substrate etched with the anisotropic etching recipe given in Table 1; (b): GeSn microdisk on the Ge buffer, etched with the same recipe as in (a); (c) GeSn microdisk (top) and H4 cavity (bottom) obtained after the anisotropic etching of GeSn and the isotropic and selective etching of Ge with respect to GeSn (see Table 1) and (d): released GeSn microstructure with two anchored sides, showing a dramatic membrane bending (note the presence of through holes in the GeSn membrane which can be used to initiate the underetching reaction in areas far from the trench). Underetched zones appear as bright areas on these SEM pictures.

| | Pressure (mT) | Gas flow (sccm) | | | Coil power (W) | Platten power (W) |
|---|---|---|---|---|---|---|
| Anisotropic etching of Ge and GeSn | 10 | $Cl_2$ | $N_2$ | $O_2$ | 200 | 100 |
| | | 100 | 25 | 10 | | |
| Isotropic etching of Ge vs GeSn | 50 | $CF_4$ | $N_2$ | $O_2$ | 500 | 0 |
| | | 30 | 40 | 50 | | |

Table 1: anisotropic etching recipe of Ge and GeSn and isotropic etching recipe of Ge versus GeSn in an ICP-RIE reactor.

Aluminium can be a very efficient hard mask for the anisotropic etching of Ge and GeSn. The selectivity is very high, reaching about 100 on Ge [72] with the recipe given in Table 1. A 10 nm thick Al mask is thus enough to etch a 1 micron thick Ge layer. The mask can be selectively removed afterwards in a hot and concentrated tetramethylammonium hydroxide (TMAH) solution, leaving the Ge and GeSn layers unattacked. The Al mask definition can also be done by locally etching the top 10 nm thick Al layer through a soft mask, and with an oxygen free Cl plasma [72]. After pattern transfer in the hard mask, oxygen is introduced in the recipe, leading to a continuous regeneration of the sputtered alumina via aluminium oxidation.

### 1.3.2 Isotropic etching

Having GeSn layers grown on thick Ge strain relaxed buffers, as in most of the works, is very attractive. Indeed, underetching the Ge SRB results in a better optical confinement and dissipates the (residual) compressive strain present in as-grown GeSn layers. This in turn improves the semiconductor directness. Gupta *et al.* noticed that low power $CF_4$ Radio-Frequency (RF) plasmas selectively etched Ge with respect to GeSn [73]. Such a selectivity was likely due to the formation of a thin surface passivation layer of tin fluoride on GeSn, because of Ge surface depletion and reaction of F with Sn atoms. To maximize the etching selectivity, the sputtering component of etching has then to be kept as low as possible to preserve the passivation layer. Working in an ICP reactor, in a pure « chemical » etching mode, therefore offers the highest selectivity. A typical recipe for the selective isotropic etching of Ge with respect to GeSn (etching rates around 500 nm/min) is given in Table 1. Figure 10c) shows a GeSn microdisk on a Ge pedestal (top) and a freestanding photonic crystal H4 hexagonal cavity (bottom) obtained with Table 1 recipes. A $SF_6$ chemistry can also be used, at zero RF power for the isotropic and selective etching of Ge with respect to GeSn. However, the selectivity is poorer than with $CF_4$. Indeed, etching of low Sn content layers will occur.

Underetching a part or the whole resonator area can have a profound impact on the overall device since this step comes with a substantial displacement of matter at the micron scale (thick Ge films on Si are tensile strained and GeSn films on Ge are compressively strained). Special care must then be given to the total underetching length or to the device design in order to benefit from strain release (see the strategies of strain redistribution in [4]), without generating undesired layer cracks or bending (see for instance Figure 10d).

A wet chemical etching of GeSn was also used, for instance in [74]. Al-Kabi *et al.* developed a low temperature (0°C) etching of GeSn, with a mixture of $HCl:H_2O_2:H_2O$ 1:1:10. The authors reported the fabrication of smooth surfaces for GeSn waveguides on Ge, with lasing upon optical pumping. This section is far from being exhaustive and does not report all the different chemistries used for etching Ge and its alloys. Nevertheless it seems at this stage interesting to note that hot TMAH does not attack Ge nor GeSn. This opens new perspectives and makes possible underetching strategies involving Si as a sacrificial material.

### 1.3.3 Surface passivation of (SiGeSn) based materials

Parasitic and non-radiative recombination of free carriers in semiconductors is highly undesirable as it degrades the performances of light emitters or photodetectors (high threshold values and inherent heating problems in lasers, non radiative « loss » of carriers in photodetectors...). With the scaling down of components, the ratio of surface to volume atoms significantly increases. Recombination on surface traps might then influence the recombination dynamics of the whole device. Finding a way to keep the density of surface traps at reasonably low levels in Ge is the topic of recurring works (see for instance [75]). Nevertheless, the rise of SiGeSn alloys renewed research around surface passivation. Indeed, the growth of SiGeSn barriers is nowadays possible on GeSn [42]. Stange *et al.* demonstrated lasing in GeSn/SiGeSn multi quantum wells with a threshold much lower than in bulk materials at cryogenic temperatures [58], indicating that these systems were suitable for carrier confinement and that the interface between the barriers and the wells were of good quality. Getting high barrier heights for both electrons and holes is however necessary for room temperature operation. Current barrier materials are limited to the low Sn and Si content range for $Ge_{1-x-y}Si_xSn_y$ compounds, requiring additional epitaxy developments to cover a more complete barrier materials spectrum. GeSn layers with a Sn content grading can also provide a way to confine carriers. J. Chrétien *et al.* demonstrated lasing in $Ge_{0.84}Sn_{0.16}$ layers embedded into $Ge_{0.87}Sn_{0.13}$ layers, with then a moderate electronic confinement [2].

Various wet cleaning and dielectric capping steps were evaluated in [76] for the passivation of GeSn layers. Mahjoub *et al.* quantified the density of interface traps $D_{it}$ in Metal Oxide Semiconductor (MOS) capacitors with the capacitance voltage technique (an alumina layer was deposited beforehands thanks to Atomic Layer Deposition) [77]. Sn droplets were found on the surface of $Ge_{0.9}Sn_{0.1}$ layers and removed by dipping in a HF:HCl mixture. The authors reported a retardation of the reoxidation prior to alumina deposition with the use of a $(NH4)_2S$ - based wet treatment. They estimated the density of interface traps to be approximately $9x10^{11}$ $cm^{-2}.eV^{-1}$.

## 1.4 Electrical contacts on Ge-based materials: a focus on GeSn

Lasing, upon optical pumping, has been demonstrated in 2015 in GeSn alloys [1]. Electrically pumped GeSn-based photonics devices, such as photodetectors, LEDs or lasers remain elusive, however [5]. Fermi-level pinning at the metal / Ge(Sn) interface, doping and thermal stability of GeSn layers are indeed some of the technological hurdles that have to be overcome to obtain optimum electrical contacts.

The strong Fermi-level pinning (FLP) at the metal / germanium interface is well documented in the literature [78]; it is predominantly governed by metal-induced gap states (MIGS) and defects-induced gap states (DIGS). GeSn, in particular n-type doped layers, are not spared by the FLP issue. GeSn surface passivation by plasma treatment (typically $O_2$ treatment) or thin dielectric deposition ($Al_2O_3$) have been shown to alleviate FLP [79,80] as well as the formation of stanogermanide or high doping in GeSn layers [81,82].

Having high levels of electrically active dopants in GeSn layers, and therefore low contact resistivities, is challenging. Ion implantation and in-situ doping have both been reported in the literature [83,84]. One of the limiting factors for ion implantation is the activation step that can lead to a degradation of the junction and of the GeSn properties because of the poor thermal stability of GeSn layers (mainly due to Sn segregation [85]). The in-situ Ga doping of GeSn layers that results in ultra-low p-type specific contact resistivity seems promising [86].

Beyond the above-mentioned topics, a comprehensive understanding of the solid-state reaction during the formation of stanogermanides is of utmost importance to precisely characterize the system and benefit from reliable contacts on GeSn. In this section, we will provide some data about the metallization of GeSn layers; more specifically, we will focus on Ni stanogermanide formation. In a second part, technological levers to enhance the thermal stability of stanogermanides will be proposed to improve GeSn devices performances.

### 1.4.1 Metallization of GeSn layers

A complete analysis of the literature concerning the metallization of GeSn layers showed that many metals were used to contact these layers. Al and TaN [79], Ti [87], Zr [88] and even Yb [81] were evaluated. Nevertheless, Ni [83] is by far the most used metal. Indeed, Ni enables to obtain ohmic contacts with low sheet and specific contact resistances at relatively low temperatures [89, 90, 91,92]. Several teams have investigated the solid-state reaction of the Ni / GeSn system. However, there were still open questions concerning the system evolution, the phases present, the impact of Sn content and the behavior of Sn atoms during the solid-state reaction.

The Ni / $Ge_{0.9}Sn_{0.1}$ solid-state reaction can be monitored by in-situ X-ray diffraction (XRD), in-plane reciprocal space map (RSM) measurements and in-plane pole figures. A sequential growth was evidenced using in-situ XRD (Figure 11a). The solid-state reaction between a Ni thin film and $Ge_{0.9}Sn_{0.1}$ layer can be summarized as follows. First, after the total consumption of Ni at 120 °C, a Ni-rich phase is formed. Then, at 245 °C, the Ni-rich phase disappears and the growth of the mono-stanogermanide phase $Ni(Ge_{0.9}Sn_{0.1})$ is initiated. This phase is stable up to 600 °C [93]. Based on advanced in-plane RSM and pole figure measurements (Figure 11b), it was demonstrated that the Ni-rich phase was actually the hexagonal ε-$Ni_5(Ge_{0.9}Sn_{0.1})_3$ metastable phase [93].

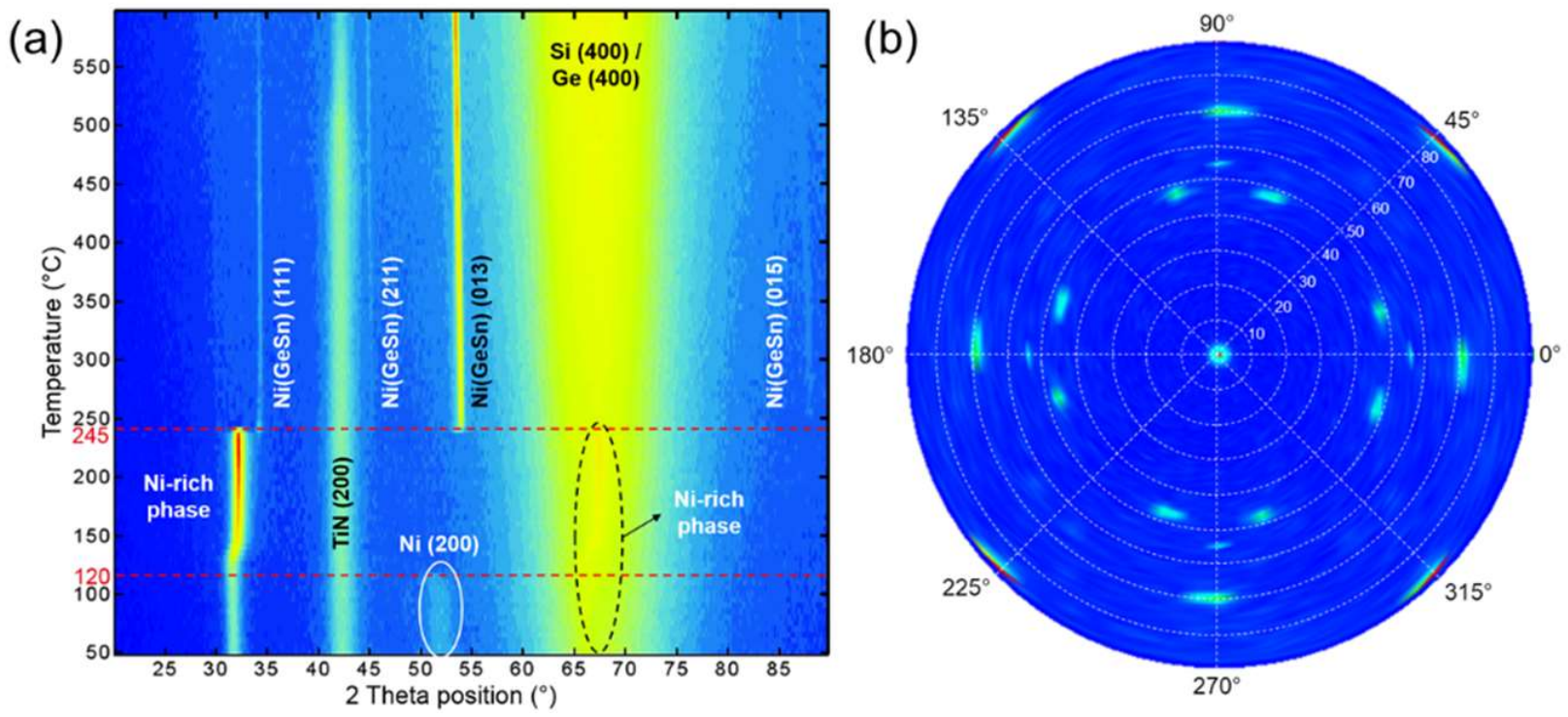

Figure 11: (a): In-situ X-ray diffraction measurement of the reaction between Ni and $Ge_{0.9}Sn_{0.1}$, adapted from reference [94], (b): in-plane pole figure recorded at 2θ = 32° for Ni/ $Ge_{0.9}Sn_{0.1}$ sample annealed at 180 °C, adapted from reference [94].

The role Sn might have during the Ni/GeSn solid-state reaction (SSR) was, until recently not fully understood. A comprehensive analysis focused on Sn segregation during the Ni/GeSn SSR was carried out. In situ X-ray diffraction and cross-section transmission electron microscopy measurements coupled with energy-dispersive X-ray spectrometry and electron energy-loss spectroscopy atomic mappings were performed to follow the phase sequence, Sn distribution and segregation [94,95]. The results showed that, during the SSR, Sn incorporated into intermetallic phases. As the temperature increased, Sn segregation happened first around grain boundaries (GBs) and then towards the surface. At higher temperatures and when the Ni(GeSn) film agglomerated, Sn easily migrated towards the top surface (see Figure 12).

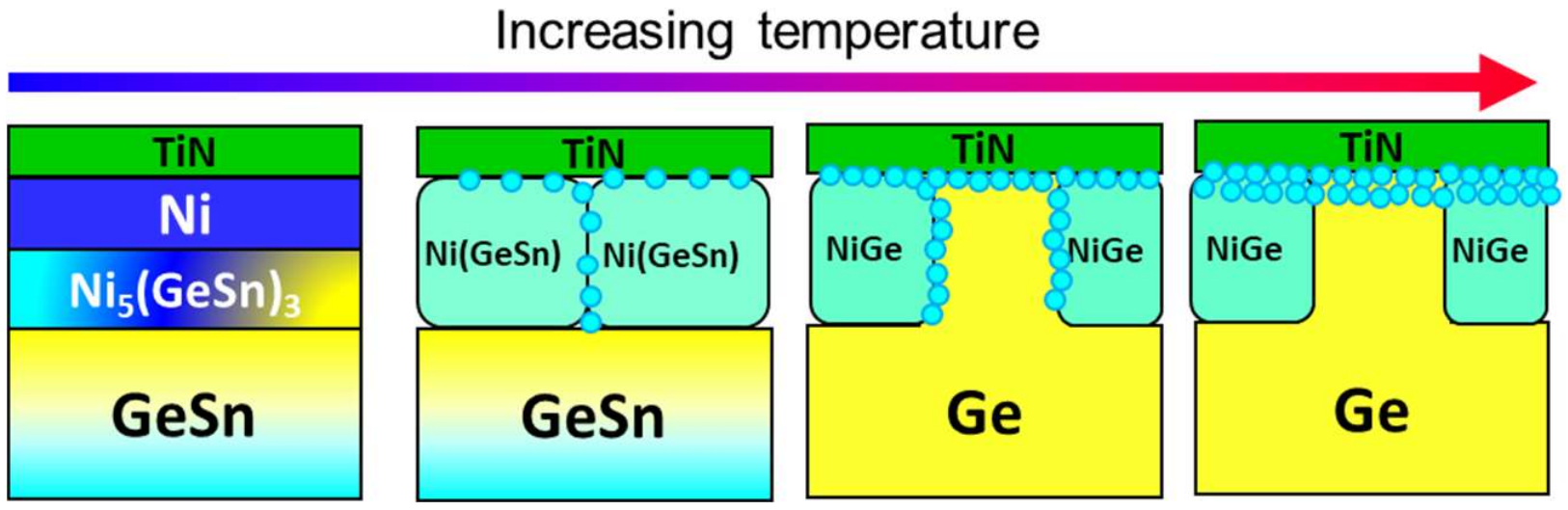

Figure 12: Schematics of Sn behavior during Ni / GeSn solid-state reaction taken from reference [100].

Sn accumulation around GBs hampered atom diffusion, delaying the growth of the Ni(GeSn) phase. Higher thermal budgets will thus be mandatory for the formation of contacts in high-Sn-content photonic devices, which could be detrimental for thermal stability. This last point is crucial. Indeed, it has clearly been shown that the Ni / GeSn system suffered from a lack of thermal and electrical stability, mainly due to the agglomeration of the Ni(GeSn) phase and to Sn segregation. It was thus of utmost importance to propose various means of extending the thermal stability of Ni-based stanogermanides.

### 1.4.2 Technological levers to extend the thermal stability of Ni-based stanogermanides

The metastability of GeSn alloys and surface Sn segregation phenomena led international research teams to propose alternative solutions to extend the thermal stability of Ni-based stanogermanides. Technological levers can be used at various steps of the contact module. The GeSn surface can be pre-amorphized by implantation (PAI) prior to Ni metallization. C implantation was successfully used to extend the thermal stability of Ni(GeSn) by about 100 °C [96]. The use of Pt interlayer between Ni and GeSn [97] or the co-sputtering of Pt and Ni [98] were shown to enhance the thermal stability of the Ni / GeSn system. More recently, the classical rapid thermal annealing was compared to the much shorter and thus more metastable pulsed laser thermal annealing. Results were promising [99].

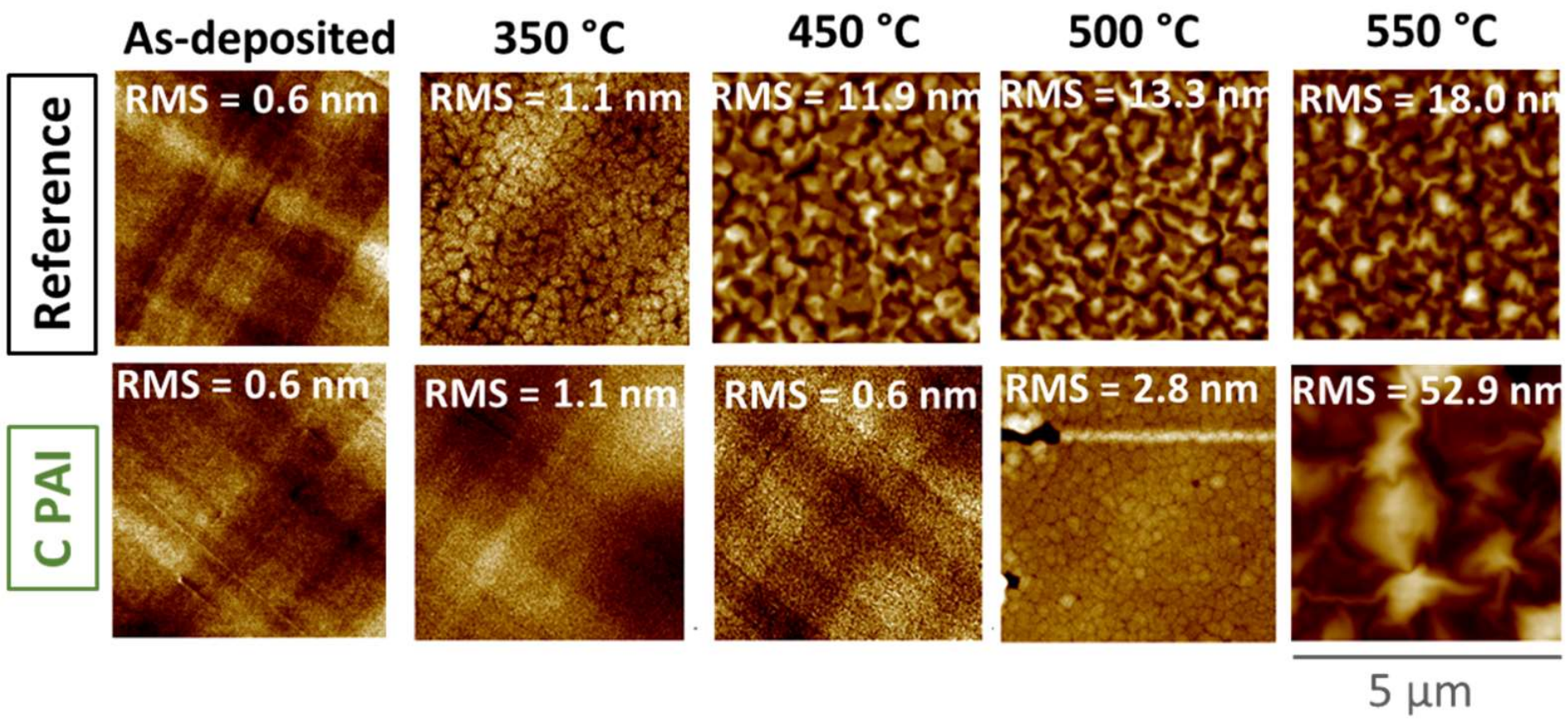

Figure 13: AFM images (scan 5 µm x 5 µm) of the surface of samples annealed at various temperatures for the Ni / GeSn system without PAI (top) and with C PAI (bottom) taken from reference [100].

Various pre-amorphization processes using Si, Ge, C or C + Ge implantation were implemented to prepare the GeSn surface prior to metallization. The use of C PAI not only extended the thermal stability of the Ni(GeSn) by delaying the agglomeration process as shown in Figure 13, it also postponed Sn segregation towards higher temperatures (delay of 150-200 °C); the electrical properties were thus strongly improved [100]. Such results are very promising.

The impact of Pt or Co as alloying elements for Ni-based metallization of GeSn layers was also investigated [101]. As far as the solid-state reaction is concerned, the overall phase sequence is the same for all metallizations: at low temperature, a Ni-rich phase is obtained; it is then consumed to form the low resistivity mono-stanogermanide phase. Nevertheless, the addition of Pt or Co as alloying elements has an impact on Ni consumption, Ni-rich and monostanogermanide phases' formation temperatures. It also leads to the formation of $PtSn_x$ or $CoSn_y$ compounds. Moreover, the addition of Co or Pt positively influences Sn segregation by delaying this phenomenon. Co has a weak influence on morphological and electrical properties. On the other hand, Pt improves the surface morphology by delaying the Ni(GeSn) phase agglomeration and enhancing the process window in which the sheet resistance remains low [102,103]. Figure 14 illustrates the impact of alloying elements on the electrical properties of the Ni / GeSn system.

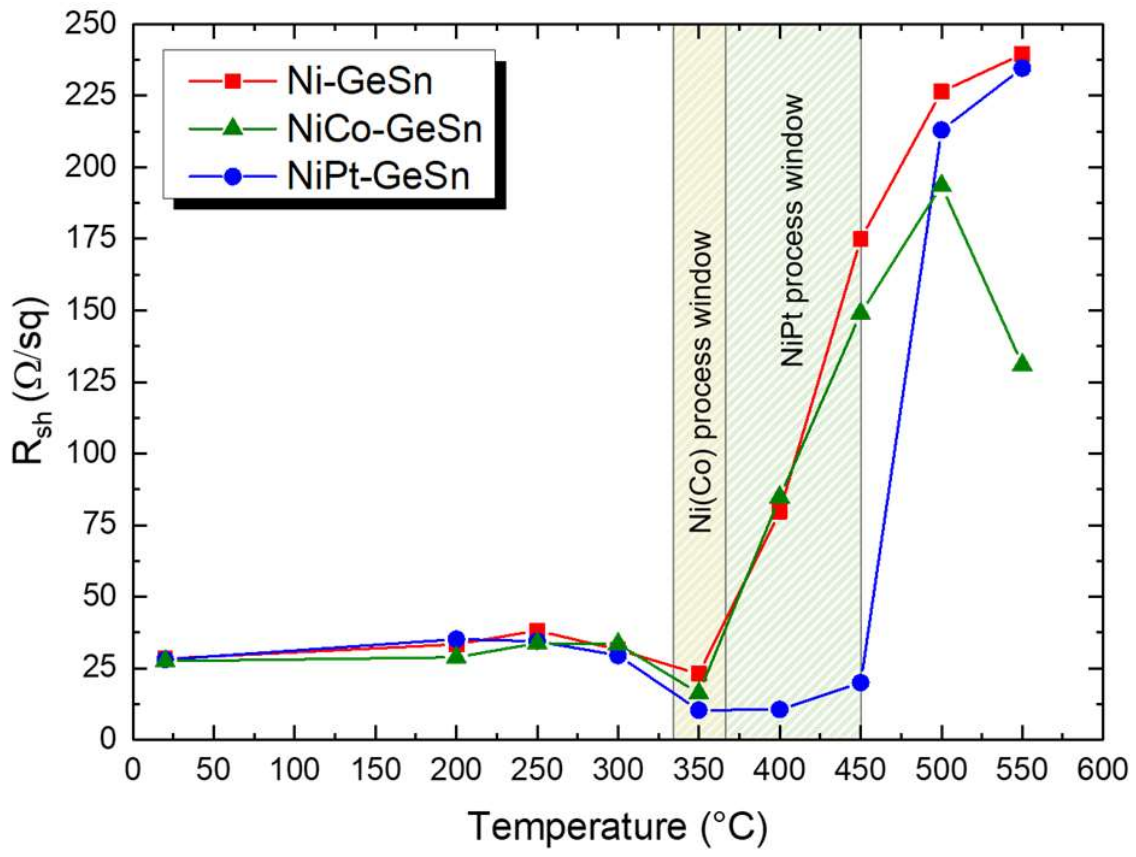

Figure 14: Evolution of the sheet resistance as a function of the rapid thermal annealing temperature for the Ni, $Ni_{0.9}Co_{0.1}$ and $Ni_{0.9}Pt_{0.1}/Ge_{0.9}Sn_{0.1}$ systems taken from reference [100].

The impact of annealing Ni / GeSn layers thanks to laser pulses (ultra-violet nanosecond laser or UV-NLA) was also analyzed. The overall evolution of the phase sequence with the thermal budget was comparable to that in RTA samples: the $Ni_5(GeSn)_3$ phase was obtained first, followed by the mono-stanogermanide (Ni(GeSn)) phase. A texture change was however evidenced; indeed, the size of crystallites in the Ni(GeSn) phase seemed to be reduced with UV-NLA, which could result in thermal stability improvements. Yet, agglomeration or Sn segregation (island formation on the surface) has not been noticed for laser-annealed samples, at variance with RTA samples. Laser annealing therefore seems to be promising to form, with low thermal budgets, optimum electrical performance Ni contacts on GeSn layers [100].

To conclude, the development of a complete contact module on GeSn is mandatory to have performant electrically activated devices. A better knowledge and understanding of the physico-chemical properties of the metal / GeSn systems was essential to propose technological and process levers to overcome the metastability of GeSn alloys and Sn segregation. The use of pre-amorphization by implantation, the addition of alloying elements such as Pt to Ni, the use of UV-nanosecond laser annealing or the combination of such technological levers are promising tracks towards a stable and reliable contact module on GeSn devices.

# 2 Effects of strain and Sn for band structure manipulation

The direct bandgap in intrinsic Ge has to be carefully estimated as the deformation potentials model [104] under-estimates the crossover between the L- and Γ- valleys. A good experimental agreement of this crossover in intrinsic Ge has recently been achieved using the tight-binding model [105]. The crossing to the direct band is predicted to be close to 6% at 20 K for a [100] uniaxial stress [4] instead of previous predictions at around 4.4 % [104,106,107]. The tight-binding model forecasts a crossover at around 2.1% at 20K for (001) biaxial stress [4] instead of 1.8% predicted by the deformation potential model [108].

We present here the different stress induction schemes used in recent years to reach such high amounts of strain in Ge and GeSn layers. In the second part, we describe the principle of the bandstructure modifications due to strain, alloying, confinement and combinations of those. Where possible, the bandstructure effects are given in analytical form to express the general trends. However, for more detailed considerations we refer our readers to the more specialized literature.

## 2.1 Strain control in Germanium based materials

### 2.1.1 Strained Ge growth

The growth of pseudomorphic Ge on larger lattice constant buffers such as InGaAs [109,110,111,112,113] or GeSn [114,115,116] can result in high biaxial tensile strain. Direct bandgap behaviour has been observed using photoluminescence analysis versus temperature in Ge layers and Ge quantum dots [117]. Strains up respectively to 2.3 % and 2.4% have been measured in Ge layers and Ge quantum dots, respectively [118]. This technique is limited to thin Ge layers below the critical thickness for plastic relaxation (typically around 50 nm [119, 120]). Similarly, III-V semiconductors – as virtual substrates [121] – are not acceptable in a Si-based platform. To the best of our knowledge, no tensily strained epitaxial GeSn layers have been reported yet in the literature.

### 2.1.2 Thermal strain

As presented in Part I – Fabrication: Ge-based epitaxy and processes for group-IV direct band gap material - thick Ge layers epitaxially grown on Si are quasi-relaxed and present only slight tensile strains at room temperature. This residual strain is induced by the thermal expansion coefficients difference between Ge and Si coming into play during the cooling-down after epitaxy or annealing [32, 122]. The strain thus depends on the process temperature [123,124] and may lead to thermally induced strains up to 0.3 % [125] and 0.4% in patterned structures [126].

However, to achieve biaxial strain Ge with a direct bandgap, we need methods that provide much higher strain values than those given just above. Currently, the main techniques used are strain redistribution and external stress transducers or stressor layers. We are going to describe those technologies in the following. As source materials, we used GeOI wafers. Ge layers on top showed a homogeneously distributed residual strain of 0.16 % [127] and were produced in a 200 mm wafer production line [128]. In the case of GeSn layers grown on thick Ge strain relaxed buffer (SRBs), Ge has once again a residual tensile strain of 0.16%, while GeSn is, because of the larger lattice constant parameter of Sn than of Ge, in a compressive strain state of – 0.5 %, typically.

### 2.1.3 Micro-bridges and membranes

#### 2.1.3.1 In Germanium

The strain redistribution concept was first developed in Si using 0.6% pre-strained SOI [129]. The generalisation of the method and its potential as photonic platform has been described in [130]. M. J. Suess *et al.* redistributed successfully the residual strain in Ge using micro-bridges up to uniaxial strain values of 3.1 % [16]. Micro-bridges were patterned along the [100] crystallographic direction of Ge. The pre-strained Ge layer (Figure 15a) grown on Si or SOI was first etched into a Ge micro-bridge (Figure 15b). Then, the substrate or the buried oxide under the Ge layer was under-etched (Figure 15c). The liberated structure relaxed, focusing the longitudinal strain of the two arms in the structure centre. The strain enhancement at the centre of the suspended micro-bridge can be analytically calculated from the four dimensions listed in d and the under-etch length below the Ge. The length $A$ defines the central micro-bridge length, $B$ is the overall length of the micro-bridge, $a$ the microbridge width and $b$ the microbridge arm width.

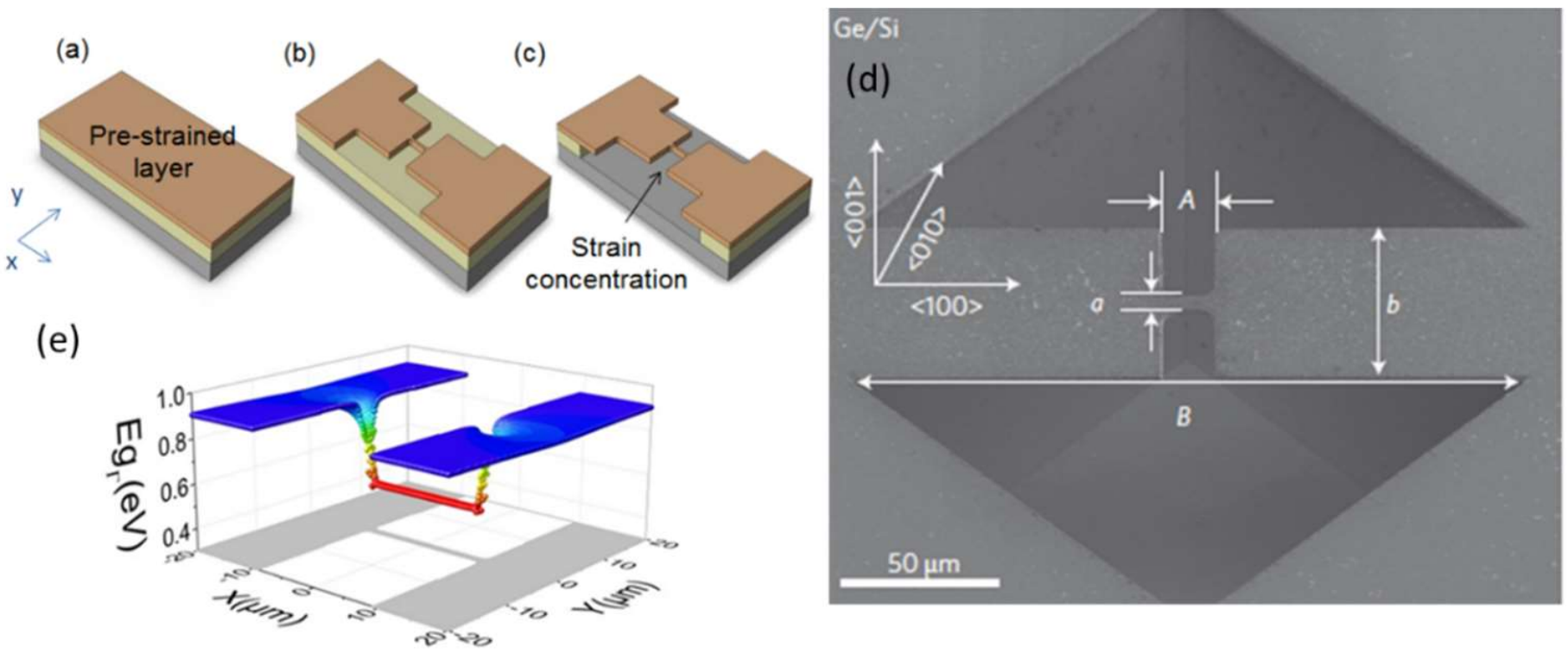

Figure 15 : Schematic of the process steps used to fabricate strained Ge micro-bridges from (a) a residual tensile strained Ge layer, (b) which is patterned and (c) under-etched. (d) Tilted SEM image of a Ge/Si structure showing the critical dimensions A, a, B and b of the micro-bridge (figure from [16]), (e) Tight binding modelling of the bandgap energy of the strained micro-bridge. Note that strain is loaded uniaxially along <100>, and the transition regime is narrow, i.e. less than 1 μm wide.

The strain enhancement factor $f = \varepsilon_{xx}/\varepsilon_0$ in the narrow constriction of the bridge is defined by the ratio of the $\varepsilon_{xx}$ micro-bridge strain divided by the $\varepsilon_0$ residual biaxial strain. It follows, for Ge deposited on a SOI substrate, that $f$ is approximately given by equation 1 where $E_i$ are the Young's moduli in the [100] direction of Si and Ge and $t_i$ are the corresponding layer thicknesses,

$$f = \eta \frac{2L+B}{B}\left[1+\frac{A}{B-A}\right] / \left[\frac{a}{b}+\frac{A}{B-A}\right] \quad \text{with} \quad \eta = \frac{t_{Ge}E_{Ge}}{t_{Si}E_{Si}+t_{Ge}E_{Ge}}$$

Above 3.1 % of uniaxial strain, corresponding to a 4.8 $cm^{-1}$ Raman frequency shift, the Ge micro-bridges grown on SOI broke due to the presence of misfit dislocations at the Ge/Si interface [16]. GeOI substrates with a significantly better crystallographic quality than Ge directly grown on Si, enabled the fabrication of highly strained micro-bridges. Record-breaking strain values of 4.9% at room temperature, corresponding to a Raman shift of 9.9 $cm^{-1}$, were reported [17]. Higher strains were achieved by lowering the temperature. At 20 K, strain values as large as 6% were obtained [4, 131].

#### 2.1.3.2 In Germanium Tin

The situation is different for GeSn, as the GeSn layers grown on Ge SRBs on Si are compressively strained, as the Sn lattice constant is higher than the Ge one (6.489 Å, versus 5.65785 Å). When the strategy described above is used on GeSn, the compressive strain will inevitably lead to a bowing of the central part of the microbridge. A two-step lithography strategy has therefore been developed to use the residual strain in the Ge SRBs [2]. In Figure 16, the two liberated Ge arms' length dictates the level of strain in the suspended GeSn microbridges. The residual tensile strain of the Ge SRBs is here the driving force to tensile strain GeSn micro-bridges, which are in a compressive strain state just after epitaxial growth.

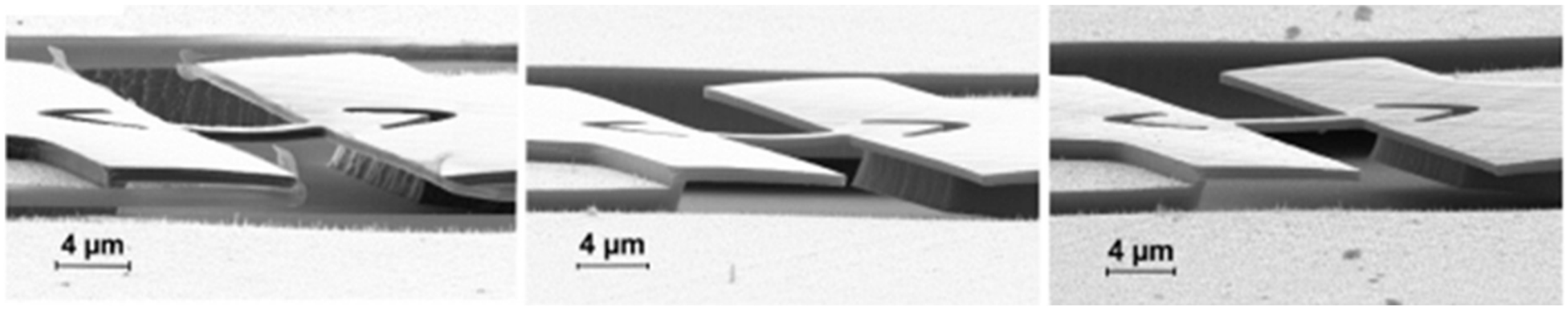

Figure 16: SEM images of the GeSn microbridge (left) before $SiO_2$ etching (GeSn micro-bridge bending downward due the release of its built-in compressive strain), (middle) at the beginning of the $SiO_2$ etching, (right) at the final stage of $SiO_2$ etching when the GeSn is tensile strained (figures from [2]).

The GeSn micro-bridge bends downward due the release of its built-in compressive strain after the etching of the GeSn/Ge/SOI and before the HF etching (Figure 16 left). The $SiO_2$ etching allows the Ge arm to relax and slowly strains the GeSn micro-bridge (Figure 16 middle). Using this method, GeSn micro-bridges with strain up to 2.2% were obtained [2] (Figure 16 right).

### 2.1.4 Biaxially strained micro-bridges

#### 2.1.4.1 Suspended membranes

Bi-axial stress or, equivalently, strain in Ge and GeSn can be induced the same way as described previously by using a micro-cross design instead of a micro-bridge one. Strain relaxation in the large stretching arms during the $SiO_2$ under-etching results in a biaxial tensile strain in the central region. Figure 17a shows the strain in a Ge micro-cross modelled by the finite element method (FEM) along x= [110], y = [110] and z = [001] crystalline directions. The full strain tensor and the lattice orientation distribution in highly strained Ge micro-crosses were evaluated at the sub-micrometer scale using Laue micro-diffraction and simultaneous rainbow-filtered micro-diffraction [132]. The complete strain tensor maps extracted from standard Laue micro-diffraction measurements on micro-crosses showed a very good agreement with FEM modelling (Figure 17a). This agreement is confirmed in Figure 17b by comparing the strain profile measured along the x= [110] direction with the modelled one.

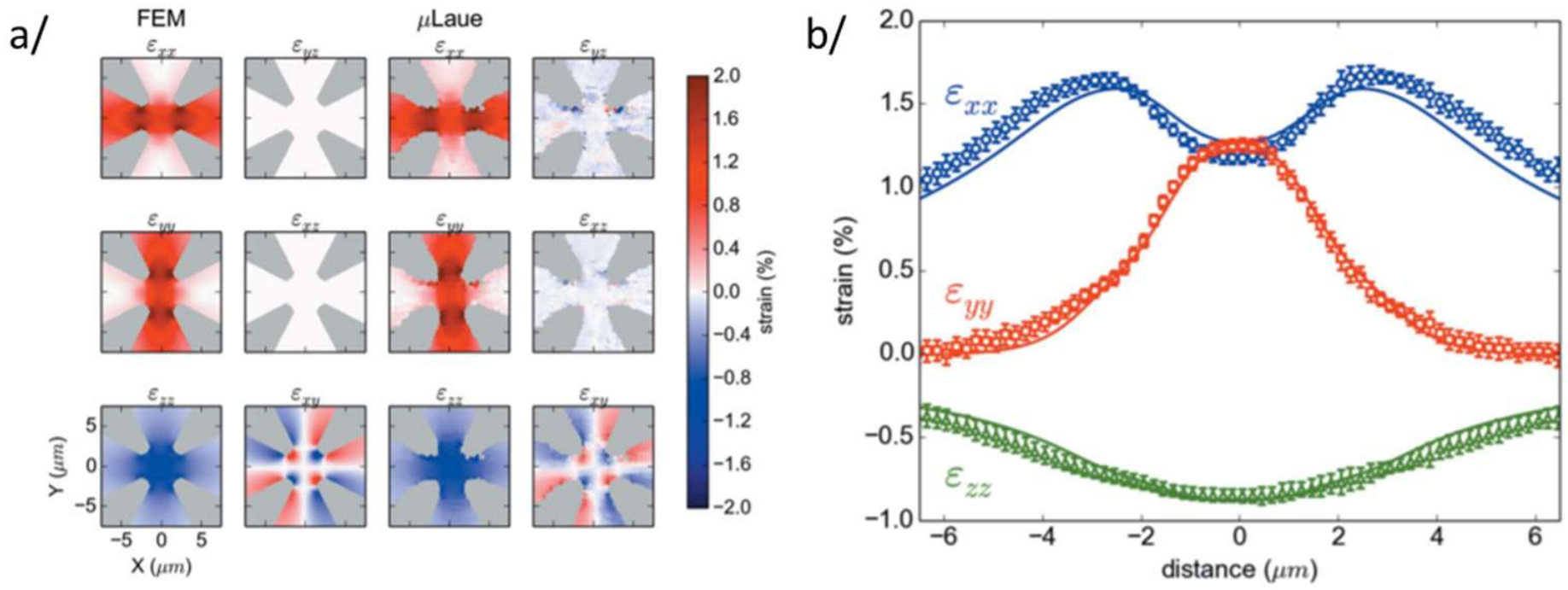

Figure 17: a/ Strain distribution in a Ge micro-cross along x= [110], y = [110] and z = [001] crystalline directions from FEM modelling (left) and by rainbow-filtered Laue micro-diffraction measurements (right), b/ Profile of strain along the x= [110] direction of the micro-cross calculated by FEM (solid line) and measured by Laue micro-diffraction (symbols) (figures from [132]).

The largest in-plane biaxial strain in Ge reported in the literature at room temperature is 1.9%, corresponding to a Raman shift of 8.1 $cm^{-1}$ [133]. This configuration suffers from shear stress at the edge of the stretching arms, which most likely is detrimental to carrier injection.

Indeed, shear stress affects the bandstructure and in particular the alignment between the L and Gamma valleys.

2.1.4.2 Landed membranes

Electrical contacts to suspended membranes could be very challenging. Specific strategies have therefore been developed to land on the substrate the entire under-etched structure after its initial release and the eventually obtained high strain levels. Strain redistribution can for instance be achieved by using a very thin sacrificial etch-layer. Jan Petykiewicz *et al.* adopted a 25 nm thick layer of $Al_2O_3$ to release the strain. The strained structure was, immediately after the etching, re-attached to the substrate that consisted of Si with a 10 nm thick conformal $Al_2O_3$ layer deposited by atomic layer deposition [134]. This thin layer can potentially passivate the Ge surface. Another strategy to land strained structures is by controlling the wet etching speed of the thick $SiO_2$ sacrificial layer (1 µm). Figure 18a shows cross-sectional schematics of a suspended strained structure using high speed HF under-etching (bottom part) and of a landed strained structure using a low speed HF under-etching (upper part). Micro-crosses fabricated in a GeOI substrate with 1 µm of BOX have been landed on silicon (Figure 18b) using a low speed HF etching recipe. The stretching arms can subsequently be partially removed, provided that the bonding to the Si substrate is strong enough, with an elastic energy threshold of around 3 J/m², typically. Such a value is of the same order of magnitude as bonding energies in wafer bonding experiments [135]. In that case, the strain of a landed micro-bridge is kept after etching away the stretching arms, Figure 18c and 18d.

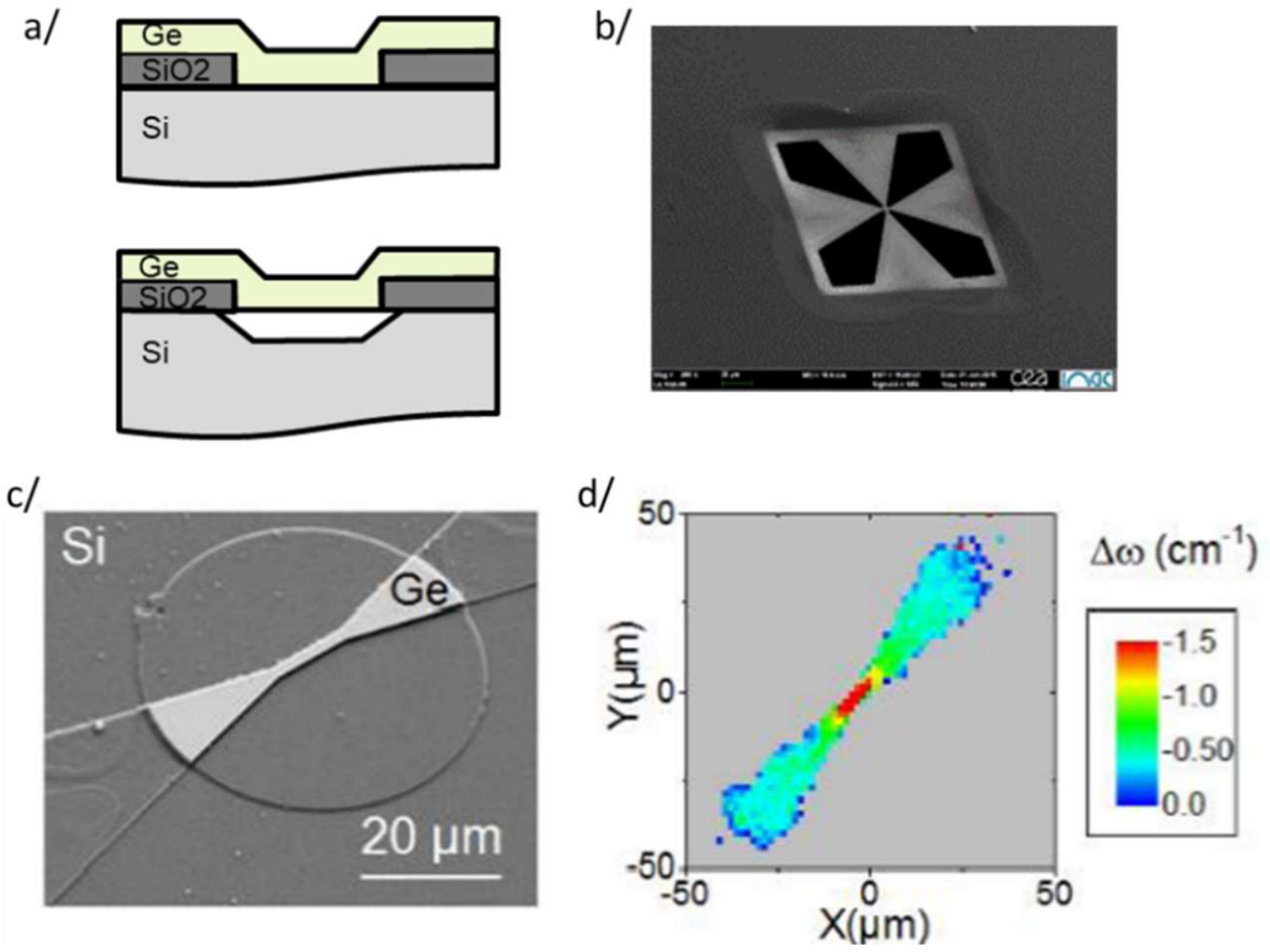

Figure 18: a/ Top (bottom) cross-section schematics of a landed (a suspended) strained structure obtained by slow (fast) etching. b/ Landed Ge micro-cross made from GeOI with a low speed HF etching recipe to remove $SiO_2$, c/ Landed Ge micro-bridge with a partial removal of stretching arms, d/ Raman spectroscopy mapping of the landed and strained Ge micro-bridge bonded on Si (figures from [135]).

### 2.1.5 Strain induced by external stress or by stressor layers

Tensile strain has also been applied to Ge using external force transponders [136], [137, 138, 139] and flexible polyimide which could be set under pressure and on which Ge layers were deposited [140,141].

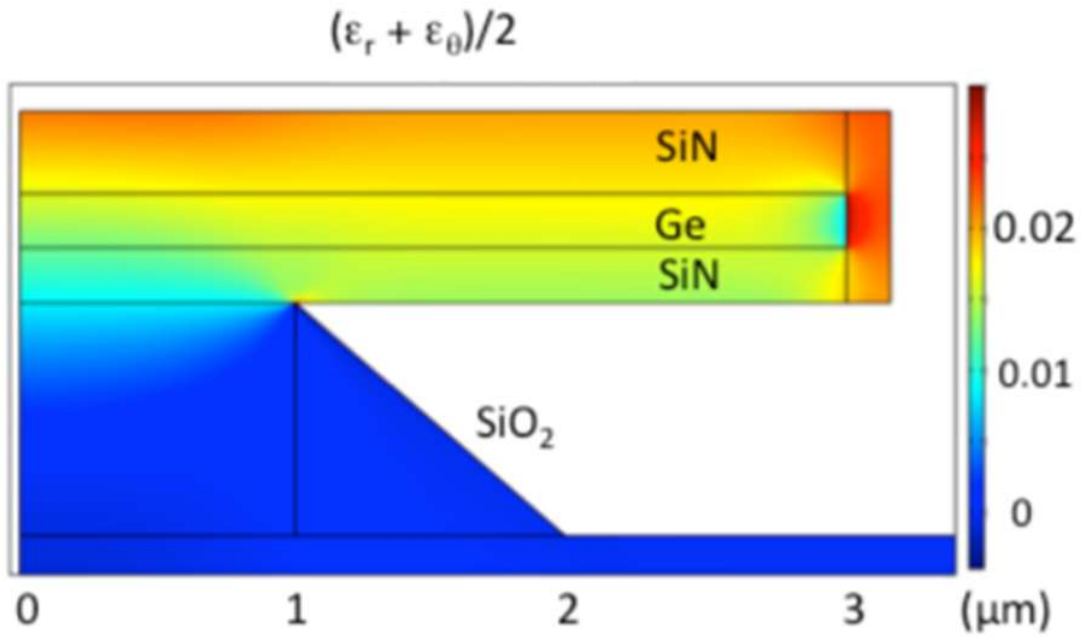

Figure 19: Strain distribution in a Ge micro-disk with SiN all around calculated by the finite elements method (figure from [164]).

Emission enhancement when reaching the Ge direct bandgap was observed for biaxial strain reaching 1.4 % in [141]. A more promising technology for applications are strain-tensors deposited around Ge and/or GeSn layers. Several stressors have been studied including SiN [142], Cr [143], SiGe [144,145], strained-Si [146] as well as GeSn [147,148]. Ge photodetectors have for instance been tensile strained using tungsten layers with residual compressive stresses higher than 4 Gpa [149]. However, SiN stressors are more often used, as their strain can be controlled from compressive to tensile through a proper selection of process parameters. In addition, SiN is transparent in the MIR and its electrical isolation properties as well as thermal properties are advantageous when combined with Ge or GeSn. Tensile strain has been induced using compressive SiN on Ge nanowires [150] and Ge membranes [134, 151, 152, 153,], as well as on straight Ge waveguides [154,155,156,157], nanowires [158], micro-disks [70], micro-rings [159], and micro-pillars [160,161].

| Straining method | Device form | Stress nature | Tensile strain (%) | Measurement | Ref. |
|---|---|---|---|---|---|
| Thermally induced strain | GeOI | | 0.16 | XRD, Raman, Micro-Laue | [128, 132] |
| | Ge on Si | Bi-axial (001) | 0.3 | XRD | [125] |
| | Patterned GeOI | | 0.4 | XRD | [126] |
| Strain by growth (pseudomorphic) | Ge on InGaAs | | 0.7 | XRD, Raman, PL | [111] |
| | Ge on InGaAs | Bi-axial (001) | 2.3 | XRD | [117] |
| | Ge quantum dots | | 2.4 | XRD, Raman, TEM | [118] |
| External stress | Ge membrane | Bi-axial (001) | 0.6 | Pressure sensor | [136] |
| | Ge membrane | Bi-axial (001) | 2.0 | Raman | [140] |
| | Ge wire | Uniaxial <111> | 2.5 | Raman | [139] |
| | Ge strip | Bi-axial (001) | 0.7 | Raman | [162] |
| | Ge membrane | Uniaxial < 100 > | 1 | Raman | [153] |
| | Ge pillar | Bi-axial (001) | 1.3 | Raman | [163] |
| Stressor layers | Ge nanowire | Uniaxial < 111 > | 1.5 | Raman | [72150] |
| | Ge micro-disk | Bi-axial (001) | 1.9 | Raman | [133] |
| | Ge micro-ring | Bi-axial (001) | 2 | Raman | [159] |
| | GeSn micro-disk | Bi-axial (001) | 1.4 | Raman | [3] |
| | Ge micro-bridge | Uniaxial <100> | 3.1 | Raman | [16] |
| Strain redistribution | GeOI micro-bridge | Uniaxial <100> | 4.9 | Raman, micro-Laue | [4] |
| | GeSn micro-bridge | Uniaxial <100> | 2.2 | Raman | [2] |
| | GeOI micro-cross | Bi-axial (001) | 1.9 | Raman, Laue | [133] |

Table 2: Methods and reported tensile strain values in Ge and GeSn. Methods include epitaxy as well as diverse routes based on a mechanical transfer of strain.

However, inhomogeneous tensile strain is achieved when the SiN layer is deposited asymmetrically [70] around the host. Stressor layers in most of the cases will thus lead to a poor overlap of, for instance, the optical mode with the strained media. To solve this issue, all-around Ge and GeSn structures were developed showing indeed a homogenous strain distribution in Ge and GeSn, Figure 19. A direct bandgap behaviour was observed in Ge with up to 1.75% of bi-axial tensile strain [164]. An all-around SiN strategy, with a SiN intrinsic stress of −1.9 GPa, was used to encapsulate GeSn, resulting in ultra-low lasing thresholds around 1 $kWcm^{-2}$ [3]. Table 2 summarizes the tensile strain induced in Ge and GeSn with the different approaches described previously.

## 2.2 Band structures and band alignment

### 2.2.1 Band structure of (Si)GeSn alloys

Ge is an indirect bandgap semiconductor, but the lowest conduction band valley, at the L point, is just ~140 meV below the Γ point. Si has its conduction band minimum near the X point, while the Γ-valley is much higher than X. Alloying Si and Ge thus does not result in a direct gap material. On the other hand, Sn has a negative bandgap at the Γ point. In GeSn alloys, the Γ-valley energy $E_\Gamma$ decreases faster than the L-valley energy $E_L$, as the Sn content increases, which may result in a positive band gap at Γ, and smaller than the indirect gap towards L, i.e. GeSn can be a direct gap material. The transition into a fundamental direct bandgap semiconductor occurs at a Sn concentration of about 8 at.%. Adding Si, to have a ternary SiGeSn alloy, slows down the transition into a direct gap semiconductor.

The band gap $E^{\Gamma,L,X}$ of an unstrained (Si)GeSn alloy can been calculated from the concentrations of the three elements $x_{Ge}$, $x_{Sn}$, $x_{Si}$ (where $x_{Ge}+x_{Sn}+x_{Si}=1$), as

$$E^{\Gamma,L,X} = E_{Ge}^{\Gamma,L,X}x_{Ge} + E_{Sn}^{\Gamma,L,X}x_{Sn} + E_{Si}^{\Gamma,L,X}x_{Si} + b_{SiGe}^{\Gamma,L,X}x_{Si}x_{Ge} + b_{GeSn}^{\Gamma,L,X}x_{Ge}x_{Sn} + b_{SiSn}^{\Gamma,L,X}x_{Si}x_{Sn}$$

where the first three terms are the weighted average (linear interpolation), and the other three terms give the correction, (i.e. represent the quadratic interpolation). The bowing parameters $b_{\Gamma,L,X}$ are either derived from atomistic band structure calculations or are found experimentally. While either of these methods can be used to obtain even higher order interpolation expressions, it is usually considered that the linear interpolation plus bowing is accurate enough for practical purposes. However, some recent publications cast doubt on whether the SiSn bowing for the Γ valley can be described by a constant parameter $b^{\Gamma}_{SiSn}$, for instance, because there is strong evidence that it depends on the Si and Sn concentration [61,165].

Another way to tune the band structure is by applying strain. The 6 components of the strain tensor ($\varepsilon_{xx}$, $\varepsilon_{yy}$, $\varepsilon_{zz}$, $\varepsilon_{xy}$, $\varepsilon_{xz}$, $\varepsilon_{yz}$) may generally be independent, and influence the band structure. However, the most interesting cases in real structures are the biaxial and uniaxial strains.

The biaxial strain is due to the lattice mismatch between the substrate and the overgrown thin layer of the active material, or is set by a stressor layer (usually SiN) into which the laser active layer is embedded. In conventional, [001] grown structures, it has two equal in-plane components ($\varepsilon_{xx}=\varepsilon_{yy}$), while $\varepsilon_{zz}$ amounts to $\varepsilon_{zz}=-2C_{12}/C_{11}\varepsilon_{xx}$ (where $C_{11}$ and $C_{12}$ are the stiffness constants of the layer material). Off-diagonal components in the stress tensor are zero. Biaxial tensile strain corresponds to $\varepsilon_{xx}>0$. If coming from the lattice mismatch, the biaxial strain is

$$\epsilon_{xx} = \epsilon_{yy} = \frac{a_{latt,subs} - a_{latt}}{a_{latt}}$$

where the lattice constant depends on the alloy composition in a similar way as the band gaps (weighted average plus bowing), i.e.

$$a_{latt} = a_{Ge}x_{Ge} + a_{Sn}x_{Sn} + a_{Si}x_{Si} + b_{SiGe}x_{Si}x_{Ge} + b_{GeSn}x_{Ge}x_{Sn} + b_{SiSn}x_{Si}x_{Sn}$$

The uniaxial strain usually appears in structures with external force applied to a thin rod-like semiconductor, e.g. in bridge-like structures. The value of one strain component, e.g. $\varepsilon_{zz}$, is there considered known, and the other two components are then found from $\varepsilon_{xx}=\varepsilon_{yy}=-C_{12}/(C_{11}+C_{12})\varepsilon_{zz}$, and the off-diagonal components are also zero. Uniaxial tensile strain corresponds to $\varepsilon_{zz}>0$. The effect of strain on the band edge energies is described via deformation potentials, and the shifts of band energies (in cases of biaxial or uniaxial strains, as considered here) can be found from the 8-band k.p model for bands at Γ [166], for a zero wave vector, which gives the shifts of the conduction band at Γ and L points:

$$\Delta E_{\Gamma} = a_c(2\varepsilon_{xx} + \varepsilon_{zz}) \qquad \Delta E_{\mathrm{L}} = a_L(2\varepsilon_{xx} + \varepsilon_{zz})$$

and for the valence bands (taking zero energy as the valence band top for the no-strain case)

$$\mathrm{P}_{\epsilon} = a_v(2\varepsilon_{xx} + \varepsilon_{zz}) \qquad \mathrm{Q}_{\epsilon} = -b_v(\varepsilon_{xx} - \varepsilon_{zz})$$

$$E_{\mathrm{HH}} = -P_{\epsilon} - Q_{\epsilon}$$

$$E_{\mathrm{LH,SO}} = -P_{\epsilon} + \frac{1}{2}\left(Q_{\epsilon} - \Delta_{SO} \pm \sqrt{\Delta_{SO}^2 + 2\Delta_{SO}Q_{\epsilon} + 9Q_{\epsilon}^2}\right)$$

where $\Delta_{SO}$ is the energy of the split-off band, below the heavy-hole and light-hole valence band (v.b.) degenerate top. The material parameters for Si, Ge and Sn, used in these calculations, and also their original sources, are given in [167].

Biaxial tensile strain decreases both the direct and indirect band gap (the direct bandgap decreases faster, resulting in a conversion into a direct gap material). It also splits the heavy-hole (HH) and light-hole (LH) bands, degenerate at zero strain, so that LH comes above HH. Compressive biaxial strain has the opposite effect. Uniaxial tensile strain also decreases the direct gap faster than indirect, leading to a direct gap material, but in this case the HH band is the highest, and LH is below it.

In these considerations, it is important to note that in these strain-split valence band states, the heavy and light holes are defined (in terms of their character – the wave functions) with respect to the z-axis described above (normal to the biaxial strain plane, or along the uniaxial strain). Within the simple parabolic approximation, the HH band effective mass in the z-direction is $m_{HH}^{z} = 1/(\gamma_1 - 2\gamma_2)$, while the in-plane mass is much smaller, $m_{HH}^{||} = 1/(\gamma_1 + \gamma_2)$. The LH band masses are (approximately, without corrections for strain) $m_{LH}^{z} = 1/(\gamma_1 + 2\gamma_2)$ and $m_{LH}^{||} = 1/(\gamma_1 - \gamma_2)$. Therefore, the effective-mass based labelling of HH and LH states applies only to their masses in the z-direction, if they are split by strain. This is important because the density of states of these bands is not the same as in unstrained bulk.

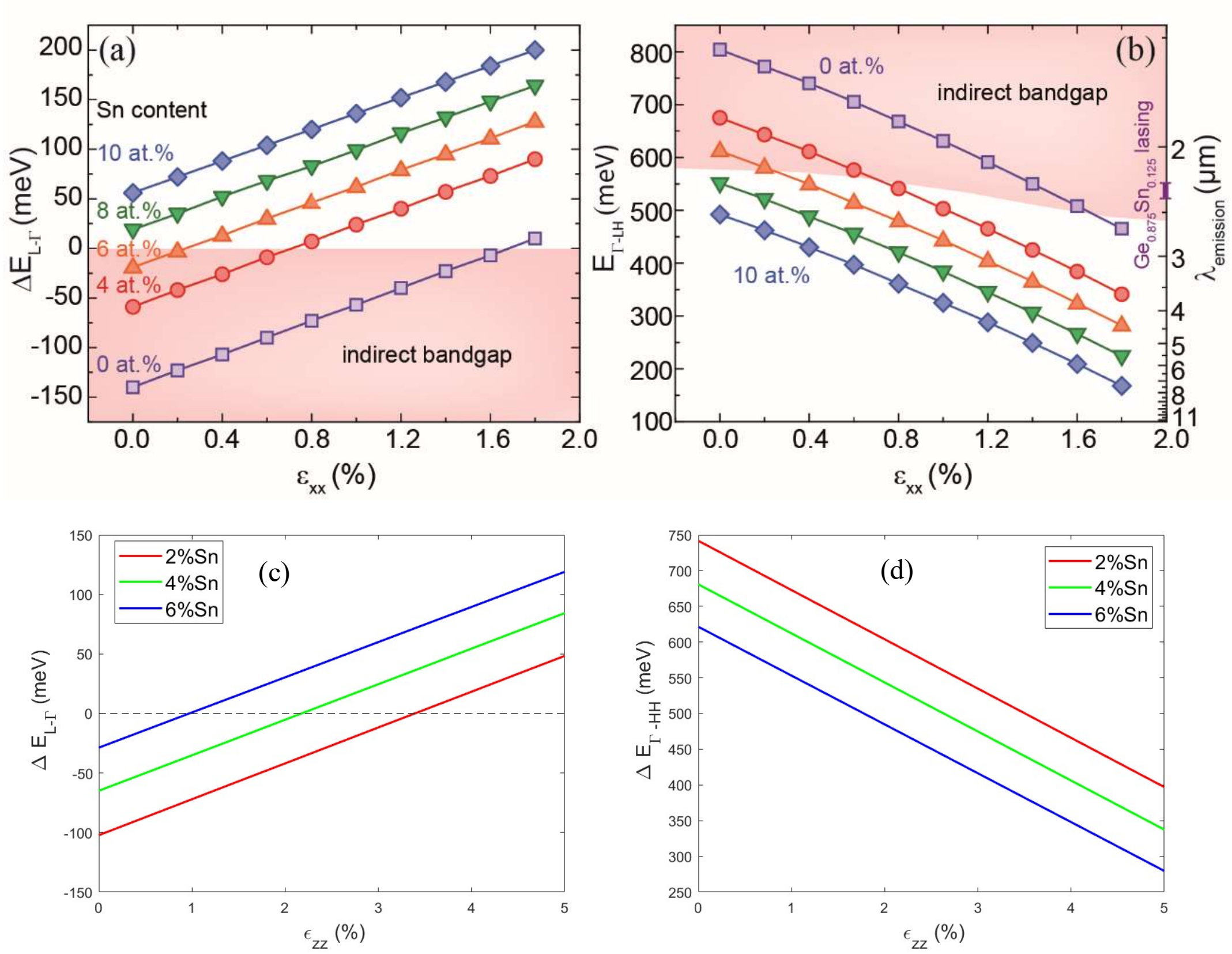

Figure 20: Directness $\Delta E_{L\text{-}\Gamma}$ (a), direct bandgap $E_{\Gamma\text{-}LH}$ and corresponding emission wavelengths (b) as functions of biaxial tensile strain for several Sn contents, at T=300 K. For comparison, the lasing region of bulk $Ge_{0.875}Sn_{0.125}$ from [5] is marked on the wavelength scale. (figure from [168]), (c) Directness $\Delta E_{L\text{-}\Gamma}$, and (d) direct bandgap $\Delta E_{\Gamma\text{-}HH}$ functions of uniaxial tensile strain for several Sn contents, at T=300 K. These results are obtained with parameters listed in [167], but some other methods, as discussed in [3], indicate that even larger strain is required for direct gap, e.g. 5-6% uniaxial strain for Ge.

Also, the optical transition matrix elements (as known from k.p modelling) between the top valence band and the conduction band are larger for z-polarization in biaxial tensile strained materials (implying that the light propagation is in the strained layer plane). In the case of a compressive biaxial strain, the xy-polarization is dominant. On the other hand, for uniaxial tensile strained material (with HH being the topmost valence band), transitions for xy-polarized light are much stronger, i.e. for light propagation along the strain direction (which is a practically favourable setup).

Combinations of tensile strain and Sn content yielding direct bandgap in GeSn alloys are shown in Figure 20. Figure 20(a) shows the difference of indirect and direct band gaps (“level of directness”), while Figure 20(b) shows the direct band gap, i.e. the approximate emission energy of a laser based on some material/strain combination. Given the much larger density of states in L-valleys than in the conduction band Γ-valleys, a directness of at least 100 meV is preferable for good lasing performance, where a good fraction of electrons will reside in the Γ-valley up to higher temperatures, c.f. our considerations given in section 2.3. While pure Ge requires a rather large biaxial strain (~1.8-2.1%) to get zero directness (with other aspects of that further discussed in Sec.3.2), alloys with over 4% Sn require much smaller strain levels. Figure 20(c,d) shows these two properties (directness and direct band gap) for uniaxial tensile strain. Lower Sn content GeSn alloys may be interesting, in combination with high enough levels of tensile strain, for lasing. Indeed, the number of crystalline defects in such alloys is likely lower than in higher Sn content GeSn layers (see section 1 of this chapter). Of course, alloys with large enough Sn contents can be direct even with some compressive strain, as was e.g. the case in the first demonstration of an operating GeSn laser [1], and also in many other research results which showed improved lasing parameters. An operating temperature of 270K was indeed achieved in GeSn 20% with a biaxial compressive strain of -0.5% in the active medium [52].

However, decreasing the compressive strain or, even better, introducing some tensile strain is generally beneficial. Ultralow threshold continuous wave and pulsed lasing was recently reported [3], using GeSn 5.4% with biaxial tensile strain of 1.4%. Lasing up to 273K was recently achieved in GeSn 16% suspended micro-bridges with 2% of uniaxial strain [2]. A large enough uniaxial strain of ~6% yielded lasing in pure Ge [4].

### 2.2.2 Band alignment in heterostructures

Using SiGeSn ternary alloys in GeSn-based lasers can be very attractive. While SiGeSn is not expected to be the laser active medium, it can be used for carrier confinement in GeSn, either in classical heterostructures or as the barrier material in quantum-well structures, where GeSn is the well material. It is therefore important to consider the band alignment at GeSn/SiGeSn heterojunctions, i.e. to find the band discontinuities for electrons and holes. To find these discontinuities, we started from the average valence band energy (average of HH, LH, and SO band energies), calculated according to the expression from [169]:

$$E_{v,av} = -0.48\ eV \cdot x_{Si} + 0.69\ eV \cdot x_{Sn}$$

which gives the average valence band energy in SiGeSn alloys, the reference zero being this energy in Ge. The SO band energy in the same material is obtained by weighted average of SO energies in Si, Ge, and Sn

$$E_{SO} = 0.297\ eV \cdot x_{Ge} + 0.044\ eV \cdot x_{Si} + 0.8\ eV \cdot x_{Sn}$$

which then gives the HH,LL band top in unstrained material

$$E_v = E_{v,av} + E_{SO}/3$$

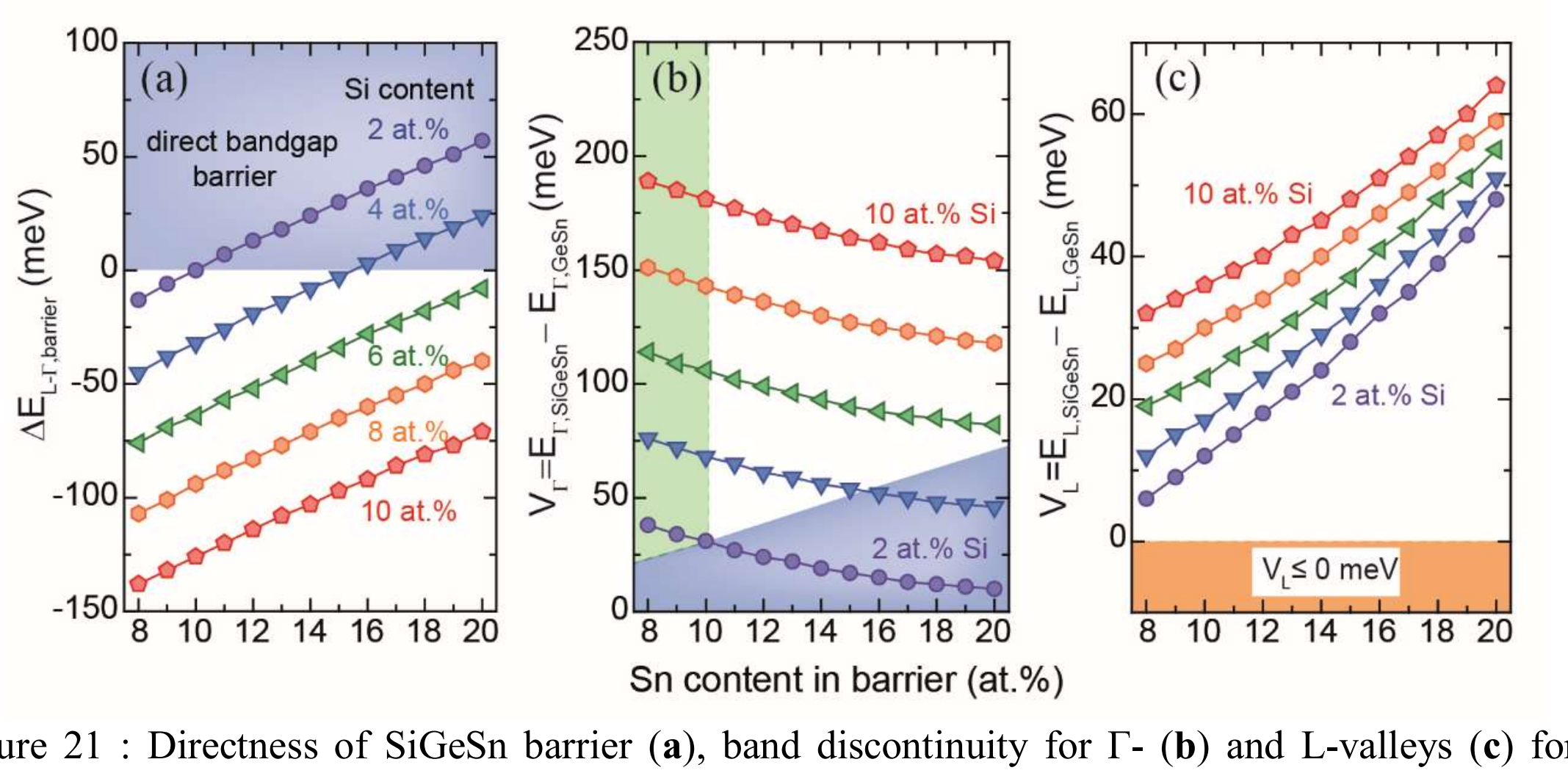

Figure 21 : Directness of SiGeSn barrier (**a**), band discontinuity for Γ- (**b**) and L-valleys (**c**) for a $Ge_{0.92}Sn_{0.08}$/SiGeSn MQW. The colored areas show regions where SiGeSn is a direct bandgap semiconductor (blue), regions where band discontinuities for L are negative (orange), and regions which correspond to realistically grown CVD layers (green). (figure from [167]).

After that, by adding appropriate band gaps, with applied strain, one can find energies of all bands of interest, all with the valence band top in Ge as the reference zero point. Calculating these for the two materials that make the heterojunction (with the strain in each one as exists in that heterojunction) will then deliver the discontinuity of any particular band at this heterojunction. There are also a couple of different expressions for the average valence band energies, instead of that in [169], e.g. [170,171], which give somewhat different values for band discontinuities. A more detailed experimental investigation is necessary before concluding about the accuracy of those calculations.

Band discontinuities in a GeSn/SiGeSn heterostructure, and also the level of directness of the barrier (which should preferentially be negative), are given as examples in Figure 21.

### 2.2.3 Quantised states in GeSn/SiGeSn quantum well structures

The full 8-band k.p model can be used for the calculation of quantum states in quantum well structures. Usually, a simpler method based on the effective-mass approximation will deliver reasonably accurate results. Here, we briefly describe the latter method. The effective-mass Schrodinger equation, with the potential found from the band alignment calculated as described above, and with the effective mass as the $m_{HH}^{z}, m_{LH}^{z}$ component for HH, LH, or the effective mass for electrons, is solved e.g. by the finite-difference, or some other method to get the states of the system. For L-valley states, the projection of the effective mass in the z-direction is used. It is given by $m_L^z = 3m_l m_t/(2m_l + m_t)$, where $m_l$ and $m_t$ are the longitudinal and transverse components for L-valley ellipsoids. The Luttinger parameters necessary for the calculation of hole effective masses depend on the alloy composition. They are calculated from the fitting expressions obtained by the pseudopotential method [172,173], which is more accurate than the simple weighted average of Luttinger parameters for elements. Within this effective-mass approximation, quantum states will have the density of states determined by the in-plane effective masses of the corresponding bands. This method in fact gives the same energies as k.p calculation does at subband edges (zero wave vector), and is reasonably accurate in cases where the quantum states of HH and LH are separated enough so that the HH-LH mixing is not large within the range of in-plane wave vectors of interest.
An example of bound states in a GeSn/SiGeSn Multi Quantum Well (MQW) structure, in which lasing was achieved, is shown in Figure 22.

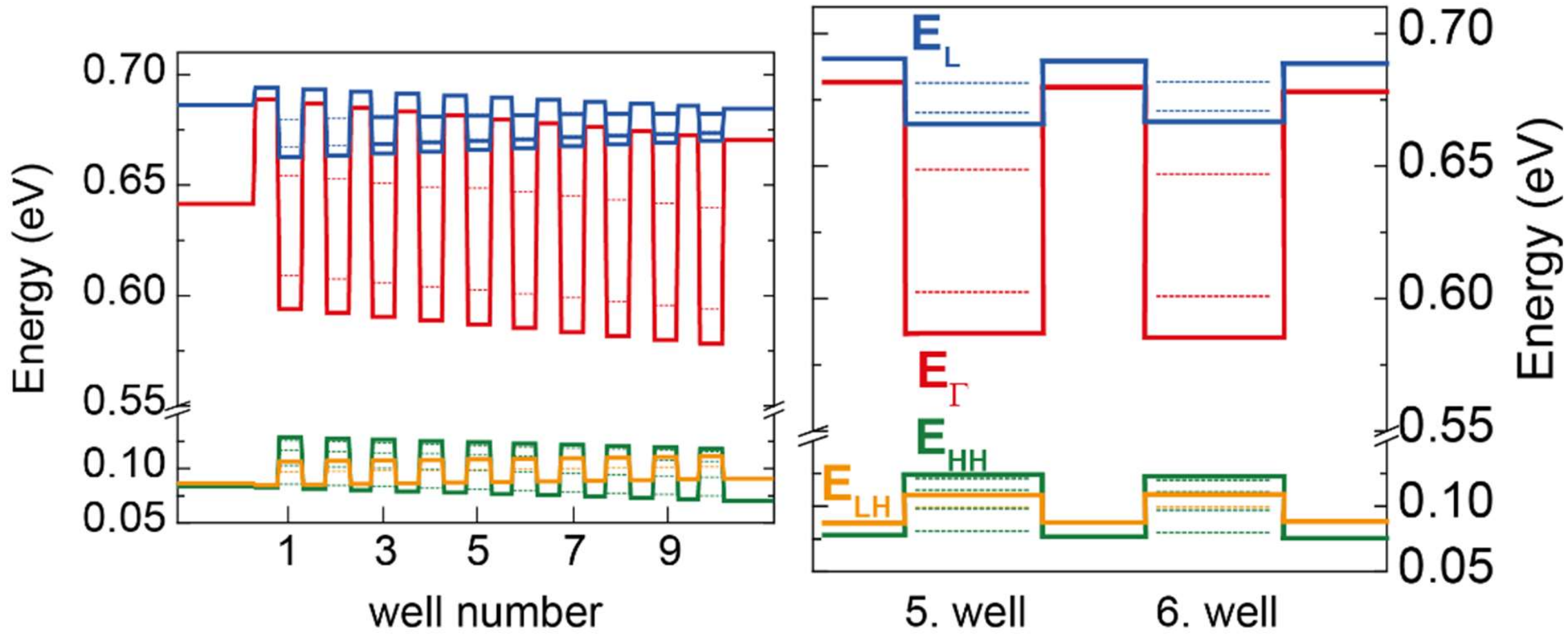

Figure 22: Band diagram of a MQW structure with 10 periods of $Si_{0.05}Ge_{0.82}Sn_{0.13}/Ge_{0.87}Sn_{0.13}$ with 22nm thickness for each layer, with a gradient of strain (from -0.05% to -0.2%), and confined states derived from the strain distribution for the whole stack (left) and for the two central QW’s (right). Inside the wells, the bulk band (solid) and quantized state energies (dotted lines) are shown for each of the valleys. (figure from [58]).

## 2.3 Gain calculation of strained Ge-based materials (Ge, GeSn, SiGeSn)

The gain calculation is based on the 8-band k.p model, using the electron and hole states coming from the diagonalization of the 8x8 k.p matrix for a range of wave vectors k. The gain is given by

$$g(\hbar\omega) = \frac{e^2}{8\pi^2 \omega n_r c \varepsilon_0 m_0^2} \sum_{c,v} |\hat{e} \cdot \boldsymbol{p}_{cv}|^2 \int \delta(E_c - E_v - \hbar\omega)\big(f_c(E_c(k)) - f_v(E_v(k))\big) d^3k$$

where $n_r$ is the refractive index, $\omega$ the radiation frequency and $|\hat{e} \cdot \boldsymbol{p}_{cv}|^2$ is the interband momentum matrix element. $f_{c,v}$ are the Fermi-Dirac functions for electrons in the conduction and valence bands, with the quasi-Fermi levels $F_{c,v}$, and $E_c(k)$ and $E_v(k)$ coming from the 8-band k·p Hamiltonian. The summation is here made only over bands around the Γ point. However, the Fermi level for electrons should be calculated by including L valleys. This equation does not account for any broadening of electron and hole states. This can be included by introducing the convolution of the gain spectrum calculated from the above formula with a Lorentzian function with appropriate linewidth. However, this is usually not essential for interband transitions in bulk, with very different dispersion of initial and final states.

Another important effect in semiconductor gain media is the inter-valence band absorption (IVBA), coming from transitions between different hole bands (HH, LH, SO). It is usually the transitions between HH,LH and SO bands which may result in considerable absorption at wavelengths close to interband transitions. This is calculated by an expression similar to that above for interband gain, except that both Fermi-Dirac functions are for the valence band states, and the summation should be done over all combinations of valence band states.

Finally, the calculated material gain has also to be corrected for free carrier absorption (FCA), which can be calculated using the second order perturbation model described in Ref. [174], or using the Drude-Lorentz model, as in [175]. The acoustic phonon scattering, deformation potential scattering (L-valley), intervalley scattering, ionized impurity scattering and alloy scattering all contribute to, and are included in the FCA calculation. After deducting the IVBA and FCA from the material gain, one gets the effective, net gain. Examples of the calculated gain achievable in biaxial tensile strained GeSn alloys and bandgap directness are given in Figure 23.

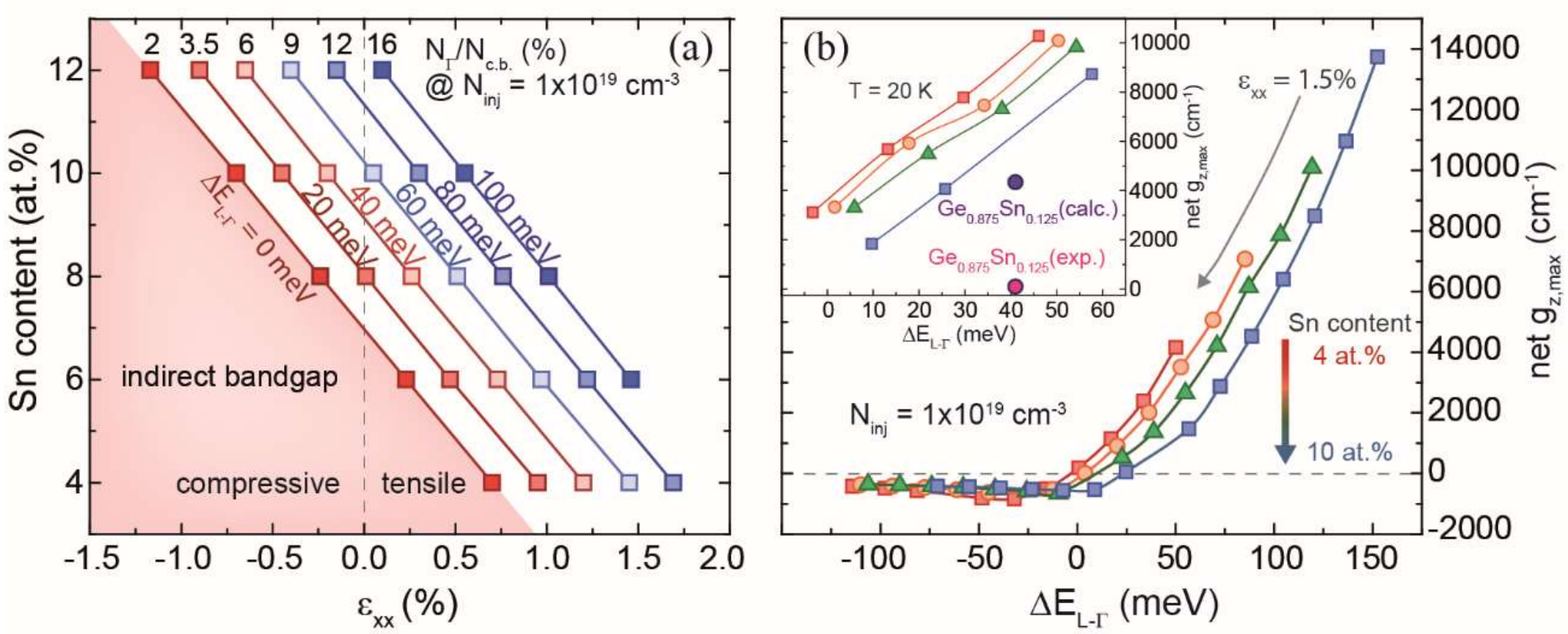

Figure 23: (a) Bandgap directness $\Delta E_{L\text{-}\Gamma}$ as a function of biaxial strain and Sn content. (b) out-of-plane net gain maximum $g_{z,max}$ for several Sn contents, as it depends on the directness, at an injection carrier density of $1\text{x}10^{19}$ $cm^{-3}$ at 300 K. Highest values of $g_{z,max}$ are achieved for strain values of 1.5 %. (figure from [168]). Clearly, the very high gain predicted for $1\text{x}10^{19}$ $cm^{-3}$ carrier density means that even much lower densities can provide practically useful gain.

In MQW structures, within the effective-mass approximation described above, used for state energies and dispersion, the gain is given [176]:

$$g(\hbar\omega) = \frac{\pi e^2}{n_r c \varepsilon_0 m_0^2 \omega} \sum_{n,m} |I_{hm}^{en}|^2 \int_0^\infty \rho_r^{2D} |\hat{e} \cdot \boldsymbol{p}_{cv}|^2 \frac{\gamma/\pi}{[E_{hm}^{en} + E_t - \hbar\omega]^2 + \gamma^2} \times [f_c^n(E_t) - f_v^m(E_t)] dE_t$$

where $|\hat{e} \cdot \boldsymbol{p}_{cv}|$ is the interband momentum matrix element, which depends on the in-plane wave vector (or $E_t$), as given in section 9.5 of [176], for transitions between conduction band-HH and conduction band-LH quantized states. This expression also includes the transitions broadening, described by a Lorentzian function with a Full Width at Half Maximum of $2\gamma$. The density of states $\rho_r^{2D}$ is defined using the reduced in-plane effective mass $m_r^*$ as:

$$\rho_r^{2D} = \frac{m_r^*}{\pi\hbar^2 L_z}$$

where $L_z$ is the well width, and $I_{hm}^{en}$ is the overlap integral of envelope functions of electron and hole states $n$ and $m$. The Fermi-Dirac distributions $f_{c,v}^n$ for electrons and holes with the quasi-Fermi levels $F_{c,v}$ are here defined as:

$$f_{c,v}^n(E_t) = \frac{1}{1 + exp\left[\left(E_g + E_{e,h\,n} + \frac{m_r^*}{m_{e,h}^*} E_t - F_{c,v}\right) / k_B T\right]}$$

Figure 24 shows the subbands in a MQW structure and the gain calculated by this method, as well as the influence of additional n-type doping on the net gain (this may improve the gain in a limited range of doping densities, but too large a doping degrades it).

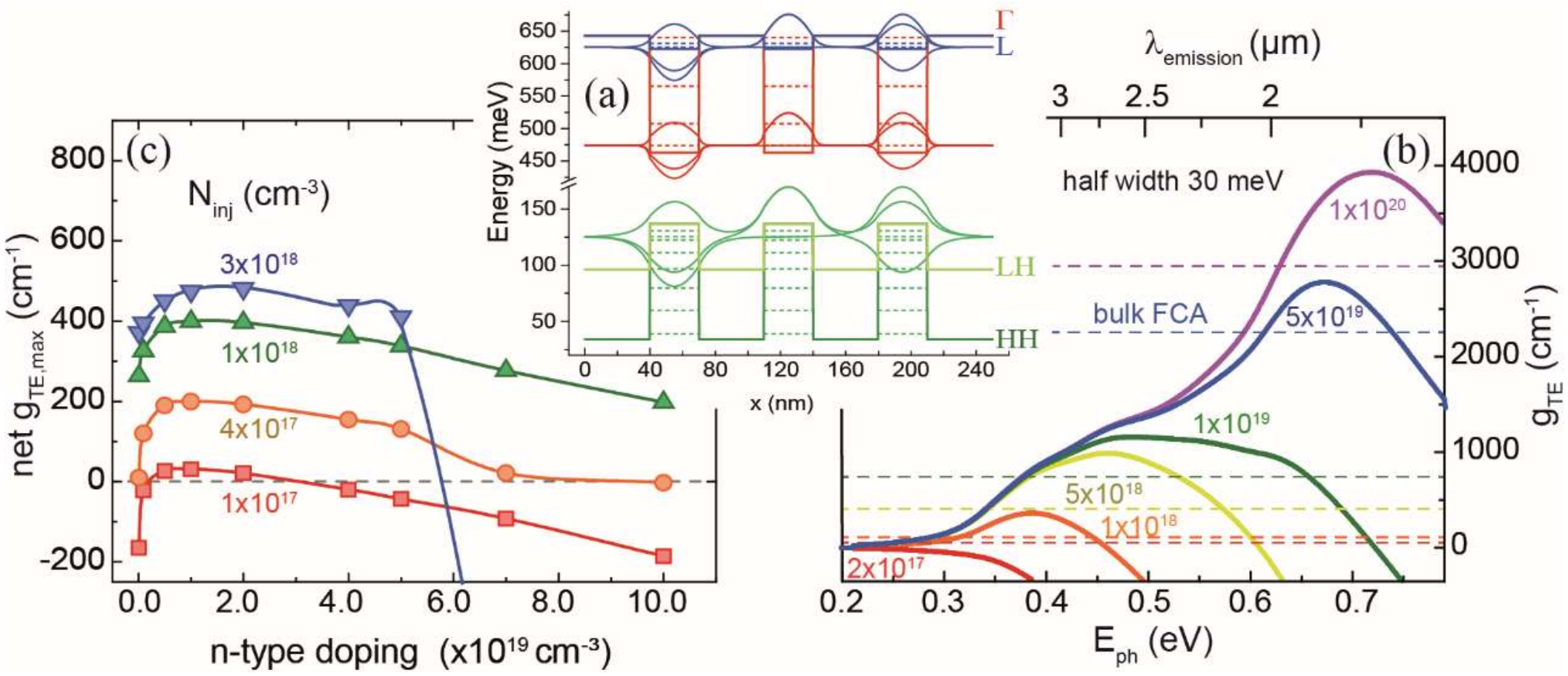

Figure 24: (a) Band diagram of an unstrained MQW with 3 x {30 nm $Ge_{0.84}Sn_{0.16}$/40 nm $Si_{0.1}Ge_{0.75}Sn_{0.15}$} stacks. (b) TE-mode interband gain at 300 K for different injected carrier densities. Dashed lines indicate FCA values for bulk $Ge_{0.84}Sn_{0.16}$. (c) Net gain dependence on n-doping density (with FCA subtracted) - (Figure from [167]).

We should note that the MQW gain calculations presented above are based on parabolic approximation within the k.p model for energy states, and that the SO band is not included. In some cases, this may be reasonably accurate, but in many other cases, the full k.p model should be used to include nonparabolic band dispersion and to calculate IVBA.

Finally, it is worth noting that other methods of band structure and gain calculations have recently been presented for this material system (in bulk form), e.g. the empirical pseudopotential method (EPM) [177], which also enables accurate calculation of both interband gain and IVBA. As shown there for Ge with large uniaxial strain, good accuracy of these calculations is important for finding parameters leading to useful net gain (i.e. with IVBA definitely smaller than interband gain).

# 3 Group-IV lasing

Lasing in silicon and Ge-based material – in addition to the impact it would make to traditional micro-electronics – is an interesting topic for fundamental research. In this section, we review recent laser results, elaborate on the optical cavity designs used and review material developments and corresponding laser devices. An overview over the impressive achievements made by the community in just 5 years is given below.

Since the first demonstration in 2015 of lasing in GeSn layers [1], progress has been vast and continuous, as shown in Table 3 summarizing the recent achievements in chronological order. For example, threshold densities were reduced by more than 2 orders of magnitude. Similarly, the maximal operating temperature ($T_{max}$) was at the beginning around the boiling temperature of nitrogen. It now approaches Room Temperature (RT). This – as we will detail below – is directly correlated with progresses made in the epitaxial growth of high Sn content GeSn layers. Defect management is another method to improve lasing. Record low thresholds were obtained when interface defects were removed in transferred materials and an external stressor was used [3]. These developments are detailed in subsection 3.3.2.

Two routes have been followed to obtain direct band gap material, one based on mechanically induced tensile strain in Ge [16,178,179], the other based on the epitaxial growth of GeSn. Recently, these two approaches began to merge [2,3]. Device structures and optical cavities are also different. In the case of suspended structures, the cavity design has to follow the strain enhancement rules. GeSn layers obtained by epitaxy on Ge SRBs allow the use of more traditional approaches such as Fabry-Perot waveguides or micro-disk cavities. Most of the lasing reports quantify threshold excitation densities and $T_{max}$. Gain and quantum efficiency are however rarely given, most likely because of experimental difficulties. To stimulate such investigations, which enable to quantify non-radiative recombination and optical loss from, for instance, intervalence band [180] or free carrier-absorption, a set of generic instructions how this could be done is presented in subsection 3.1.2.

| Systems | Strain | Substrate | Cavity | L-Threshold | $T_{max}$ (K) | Gain ($cm^{-1}$) | Q-eff (%) | Ref | Year |
|---|---|---|---|---|---|---|---|---|---|
| **GeSn (12.5%)** | -0.6% biaxial | Ge-VS | Ridge WG F-P | 300 kW/cm² | 90 | < 110 | ±1.5 | [1] | 2015 |
| **GeSn (12.5%)** | ± 0% | Ge-VS | MD | 220 kW/cm² | 130 | - | - | [67] | 2016 |
| **GeSn/SiGeSn multi QW (13%)** | ± 0% | Ge-VS | MD | 40 kW/cm² | 120 | - | - | [181] | 2018 |
| **GeSn DHS (16%)** | ± 0% | Ge-VS graded | MD | 134kW/cm² | 180 -230 | - | - | [68, 182] | 2017, 2018 |
| **GeSn (17.5%)** | ± 0% | Ge-VS | Ridge WG F-P | 170kW/cm² | 180 | - | - | [183] | 2018 |
| **Germanium microbridge** | 1.6% uniaxial | GeOI bonded | MC DBR F-P | - | 180 | - | - | [190] | 2018 |
| **GeSn (16%) Photonic crystal** | ± 0% | Ge-VS | PC membrane | 230kW/cm2 | < 90 | | | [39] | 2018 |
| **SiN stressor Germanium** | 1.7% biaxial | Ge (SiN) bonded | MD | - | 180 | - | - | [191] | 2018 |
| **GeSn (20%) step enhanced** | -0.5% biaxial | Ge-VS | Large area F-P | 50kW/cm$^2$ | 270 | - | - | [52] | 2019 |
| **Germanium microbridge** | +5.5% uniaxial | GeOI | MC CC F-P | 20kW/cm$^2$ | 90 | > 300 | >50 | [4] | 2019 |
| **GeSn DHS (±13%)** | +2% uniaxial | GeOI | MC CC F-P | 10KW/cm$^2$ | 273 | - | - | [2] | 2019 |
| **SiN stressor GeSn (6%)** | +1.5% biaxial | GeSn(SiN4) bonded | MD | 1.1kW/cm$^2$ | 100 | - | - | [3] | 2019 |
| **Electrically pumped GeSn p-i-n (11%)** | ± 0% | Ge-VS | Large area F-P | 600 A/cm$^2$ | 100 | - | 0.6 | [5] | 2020 |
| **GeSn (7-10.5%)** | ± 0% | Ge-VS | MD | 10KW/cm$^2$ | 90 | - | - | [184] | 2020 |

Table 3: Group IV lasing major milestones over the last 5 years. VS stands for Virtual Substrate, WG for Waveguide, F-P for Fabry-Pérot, DBR for Distributed Bragg Reflector, MD for Micro-Disk, MC for Micro-Cavity, CC for Corner-Cube, PC for Photonic Crystal.

## 3.1 Optical cavity design

The optical cavity or resonator is a major component of a laser; it provides the feedback of the laser light and thus eventually reinforces stimulated emission. The performance of a cavity is usually given by its quality factor (Q-factor), which is a measure of the number of oscillations cycles a photon takes before it decays, $Q = 2\pi\upsilon\tau_d$, where $\upsilon$ is the cavity resonance frequency and $\tau_d$ the photon decay time. Standard direct band gap semiconductors stand out due to large gain and therefore – in principle - do not necessarily require cavities with ultra-high Q-factors. We expect this to be true for direct bandgap group-IV materials, by analogy. Fabrication concepts suitable for standard III-V based semiconductor lasers should thus also suit group-IV lasers, as well as the traditional characterization methods.

### 3.1.1 Resonators

Resonator types used so far for group IV lasers include (i) Fabry-Perot (F-P) cavities with partially reflecting end-mirrors and single mode ridge waveguides (WG) or multi-mode large area WG, (ii) micro disc (MD) cavities, as well as (iii) F-P like micro cavities (MC) with end-mirrors made of distributed feedback (DFB) reflectors or total reflection corner cubes (CC). The F-P MCs comprise mainly suspended structures. The above list is completed by (iv) photonic crystal designs for membranes and waveguide slaps. Their main properties are summarized in Table 4, and detailed below.

Waveguide cavities are conveniently used with GeSn layers epitaxially grown on a bulk or virtual substrate. They are fabricated with a planar technology and are typically 1 mm long. Thanks to an under-etching of the waveguides, the mode overlap with the active laser medium is improved and part of the transversal strain is relaxed. Waveguide F-P cavities are convenient for gain measurements, for example by investigating the laser threshold or light emission dependence on cavity length over which the waveguide is excited, see below. The latter method is named the variable stripe length (VSL) method [185]. Both of these methods were used to characterize the first group IV lasing in GeSn [1]. The Hakki-Paoli or the Fourier transform methods [186], which can alternatively be used to measure the modal gain under subthreshold conditions, were on the other hand not used so far for group IV lasing. These two methods are based on the analysis of mode spectra taken at high resolution.

Micro discs (MD) are the most used cavities. By design, the compressive strain of typically 0.5% to 0.8 % in as-grown GeSn, will then elastically relax [67]. Through the relaxation, the directness of the band structure improves and so does the laser performances, as detailed in the theory section. The characteristically high Q-factor of MD was for most of the investigations of less importance because – as said – the material gain are expected to be rather high. And indeed, the above mentioned investigations performed on F-P cavities of GeSn layers with 12% of Sn [1] revealed gain values exceeding 100 $cm^{-1}$. MD are not only used with epitaxial GeSn layers but also in conjunction with GeSn layers which were transferred from the original epitaxial wafer and embedded in a SiN stressor layer to inject high amounts of tensile strain [3]. During the GeSn layer transfer, defects generated by plastic strain relaxation during the epitaxial growth on Ge SRBs were removed, making lasing possible under steady state excitation [3], as detailed in section 3.3.3.

FP Micro cavities around strained micro bridges were developed first with tensily strained Ge [4, 134, 187] and later on with GeSn grown on a Ge virtual substrate (VS) [2]. These designs have to comply with the specificities of strained bridges. The optical cavity should not, for instance, reduce the amount of strain injected by anchors or pads in the optically active structures. Such constraints exclude, for example, photonic crystals with etched holes in the strained membrane or slab. The cavity should also match the small micrometer size of the gain medium. Designs used so far are of the Fabry-Perot type, with "end-mirrors" made of (i) a pair of distributed Bragg reflectors (DBR) on either side of the constriction [134] or (ii) two corner cube shaped surfaces where the light is confined via total reflections [2], [4]. Reported Q-factors range between 1000 to 4000, which is not particularly high given that the gain medium fills only 1/3 of the overall cavity. The corresponding optical loss can then quickly exceed 100 $cm^{-1}$ [4]. The principle of such cavities with corresponding mode calculation is depicted in Figure 25.

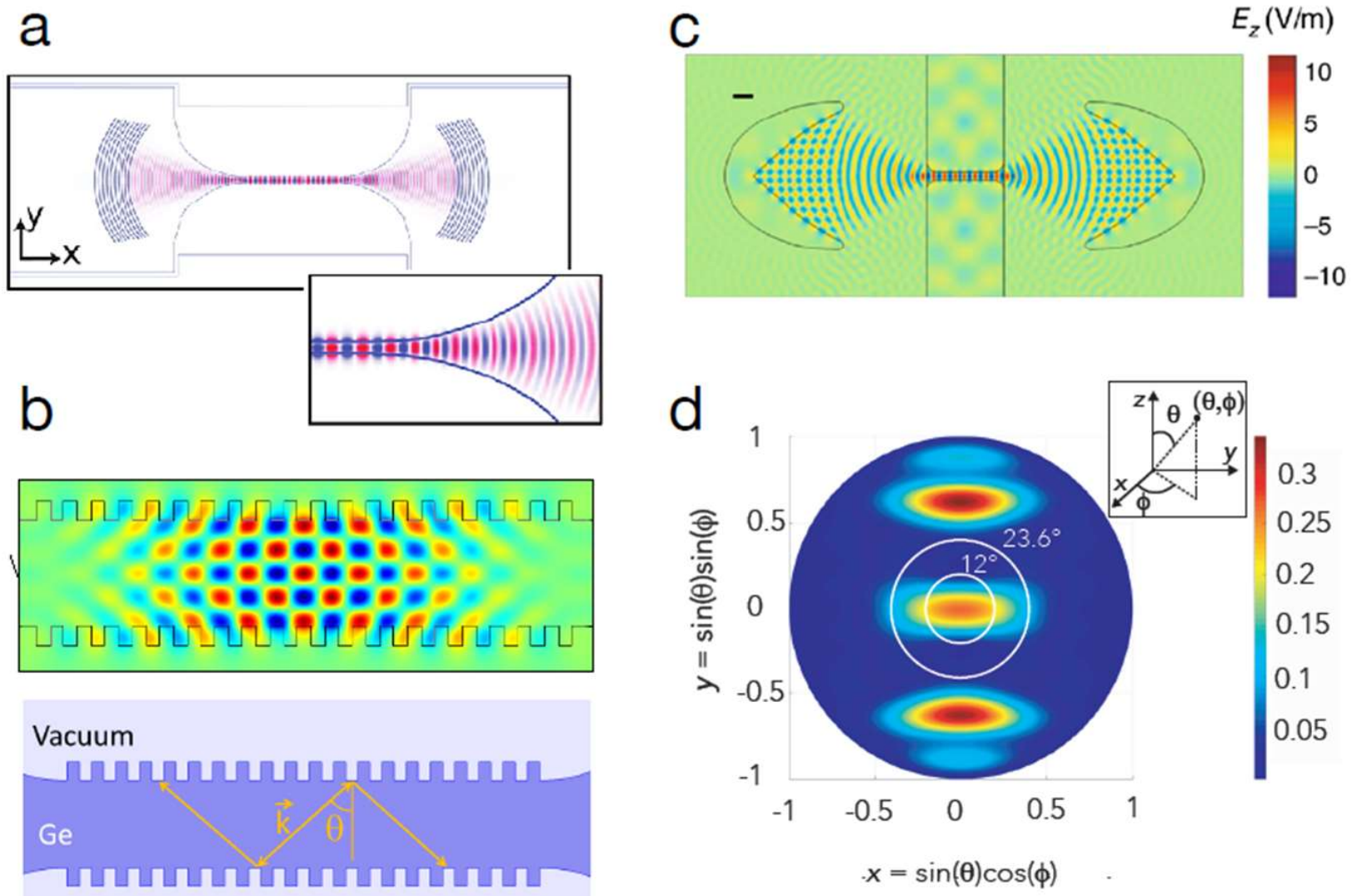

Figure 25: Microbrige laser cavities: (a) with DBR end mirror [134], (b) Design for a higher order mode DFB cavity. A high Q-factor (of > 10k) is achieved by the strong distributed feedback which can exceed 0.5 per reflection unit [188]. (c) Mode pattern of micro bridge cavity with parabolic end mirrors forming a corner cube [4]. (d) Upwards scattered light of cavity (c), from S.I. of [4].

Photonic crystals (PC) designs yield laser cavities with high Q-factors, which can operate in single mode. At the edge of the photonic Brillouin zone where the slope of the bands drops to zero, modes have a zero group velocity. These points of slow light are those where lasing can potentially take place. Such band-edge modes were obtained in a 2D hexagonal photonic crystals membrane cavity obtained by under etching a $Ge_{0.84}Sn_{0.16}$ layer deposited on a step-graded GeSn buffer [39]. The performance of the laser thus obtained did not outperform the MD laser from the same wafer. This was partly due to the resonator frequency, which was not optimized for the wavelength red-shifted emission at elevated temperatures [39]. Indeed, the lasing wavelength in micro-disks adapts itself to gain when the temperature rises, not so in PC-based designs, where the resonance frequency is fixed by the periodicity of the hole - or corrugation - pattern.
Band edge modes with high Q were also studied for use in suspended micro structures [188]. The laterally corrugated ridge structure, pictured in Figure 25b, was found to maintain the imposed strain and may have offered Q-factors exceeding $10^4$ with suitable mode filling. A sketch of its working principle is shown in Figure 25b, together with the mode pattern of a higher (5th-) order mode. In a semi-classical propagation picture, the strong localisation of higher order modes can be understood from resonant back reflection which may exceed 50% under certain angles. Key parameters for optimizations of the back reflection and thus the cavities Q-factor are the grating depth and period, the incidence angle - which is related to the mode order - and the duty cycle of the grating [188].

| Cavity type | | Strain | Size | Q-factor, empty cavity | Filling factor | Benefit | dis-advantage | [ref] |
|---|---|---|---|---|---|---|---|---|
| waveguide ridge | 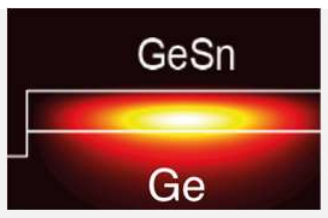 | Pre-served | ±1 mm | typ. 10'000 | ca. 60% | planar technology | strain/mode leakages | [1] |
| waveguide supported | 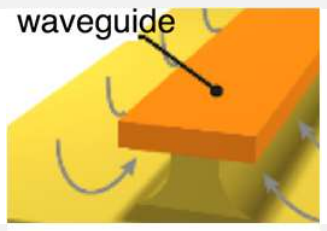 | Partially relaxed | ±1 mm | typ. 10'000 | up to 90% | planar technology | mode quality | [189] |
| Micro-disc supported | 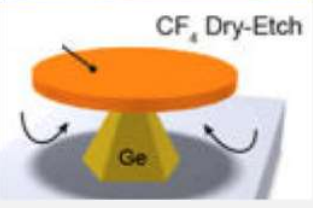 | Nearly fully relaxed | ø ±10 µm | > 10'000 | ca. 90% | strain relaxation | heat managemen t | [67] |
| Micro-bridges free standing | 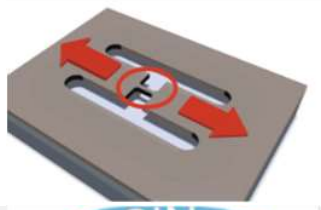 | Tensily strained | ±10 µm | < 2000 | ca. 20% | on wafer tunability | low mode filling factor | [4, 178,190] |
| Microcavity capped stressor | 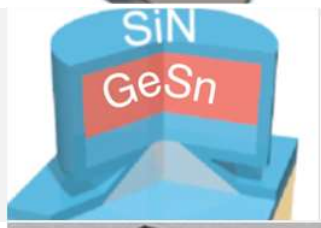 | Tensily strained | ø ±10 µm | > 10'000 | ca. 90% | defect removal | complex fabrication | [191,3] |
| Photonic crystal suspended | 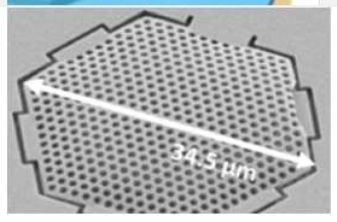 | Relaxed | ø ±35 µm | > 10'000 | ca. 40% | single mode | low fabrication tolerances | [39] |

Table 4 : Laser cavities used to demonstrate lasing in group-IV semiconductors.

### 3.1.2 Gain and Efficiency measurements

A direct method to measure the modal gain is based on luminescence collected from one of the edges of a long waveguide structure that is excited over a variable length L measured from that edge. If modal gain is present, the emission intensity increases exponential with the stripe length, L, and a substantial decrease in linewidth occurs. This is an indication of amplified spontaneous emission. The modal gain, g, is determined from :

$$I_{ASE} = I_o + \frac{I_{SP}}{g} \cdot (e^{gL} - 1) \, ,$$

where $I_{ASE}$ and $I_{SP}$ are the amplified and unamplified spontaneous emissions, respectively. $I_o$ contains higher-order modes as well as light scattered from the waveguide sidewalls. This method has successfully been used in [1], together with the threshold dependence on the cavity length. Similarly, the modal gain is obtained by following the lasing threshold change for a series of identically fabricated F-P waveguide cavities from the same material but with different lengths, l. The mirror – or extraction - loss R, expressed as gain needed to compensate this loss is :

$$\Delta g_{mirror} = -\frac{1}{2l} \ln(R_1 R_2)$$

where $R_j$ represent the reflection loss at the two waveguide edges. To give a simple example, by changing the length of a cavity from, for example, 300 µm to 600 µm, $\Delta g_{mirror}$ decreases from roughly 30 cm$^{-1}$ to 15 cm$^{-1}$ assuming $R_j = 0.4$. The thus obtained differential gain is obtained by dividing $\Delta g$ by the corresponding reduction of injection needed to reach threshold $\Delta I$. The latter is expressed in photon or current density for optical or electrical injection, respectively. This method has shown to give results in good agreement with the variable strip method [1]. Both methods gave a differential gain of approximately $\Delta g/\Delta I$ = 0.4 cm kW$^{-1}$.

A measurement of a device’s quantum efficiency is obtained from the measured (change of) optical output power normalized to the (change of) input power for conventional FP lasers with end-mirrors made of waveguide facets. In such conventional cavity geometries, the total optical loss of the cavity corresponds indeed, and to a good approximation, to the reflection loss $R_j$ per facet. In closed cavities, for instance in micro cavities around strained micro bridges, the optical loss is primarily due to scattered light (in and out of plane). Its emission profile is a priori not known and thus has to be calculated using a 3D solver. This solver must contain the cavity structure and its environment. The detection window may be defined by a corresponding field of view, conventionally parametrized with a numerical aperture (NA).

In the case of the microbridge cavity shown in Figure 25c, the light scatters out of the plane from two distinct places, (i) the intercept of the ellipsoids which form a discontinuity for the fields and (ii) the two filets of the microbridge. The amount of upwards scattered light from the latter depends on how narrow the bridge is and on how fast it narrows. The part which is collected within the NA is obtained by integrating the outwards scattered emission profile, Figure 25d, over the corresponding solid angle $d\Omega = \sin\theta \cdot d\theta \cdot d\Phi$ normalized to the total intensity of the outwards scattered light $i_{tot}^{s}$ :

$$\eta_{col}^{s} = i_{col}^{s} \Big/ i_{tot}^{s} = \int_{NA} i_{out}^{s} \, d\Omega \Big/ i_{tot}^{s}$$

$\eta_{col}^{s}$ represents thus the collection efficiency of the set-up. The external QE, $\eta^{QE}$ , is defined as the ratio between scattered photons ($p_{tot}^{s}$) and the injected carriers. The latter is related to the number of absorbed photons ($p^{abs}$) or the injected current.

Quantifying the amount of scattered light (in terms of photons or power) requires a proper calibration of the set-up which includes all the loss channels starting from the light collection to transmission losses, coupling losses to the spectrometer and so on. The detector signal has also to be converted to an equivalent in power or photon flux. Since we are interested in the overall efficiency of the measurement system, its performance is conveniently calibrated by placing a test emitter at the sample position. A good choice for such an emitter is a hot filament black body emitter of known emissivity and temperature. Taking the emission profile into account which follows Lambert's cosine law, the power collected from the test emitter defined as $p'_{col}$ is readily calculated. To account for the image blurring of the optical system, the size of the test emitter and laser device should be similar. The amount of scattered light from the lasing device follows from :

$$p_{tot}^{s} = S_{det} \cdot \left(\frac{p'_{col}}{S'_{det}}\right) \cdot (\eta_{col}^{s})^{-1}$$

where $S_{det}$ and $S'_{det}$ are the detectors signals as obtained from respectively, the laser and test device. The above method is used to estimate the quantum efficiency of highly strained microbridge lasers under near resonant excitation [4].

## 3.2 Lasing in Ge

In comparison to GeSn, achievements in the field of strained germanium lasing might seem slower. One may therefore wonder about the significance of Ge laser experiments given the rapid developments underway with GeSn. Before answering this question, let us first define what we mean by a strained Ge laser. After all, the term “Ge-laser” has already been used quite a bit, for instance for the inter-valence-band p-type Ge laser [192], the so-called pseudo direct Ge structures obtained by band folding [6], and confinement [7] and also for heavily n-type doped Ge [193].

In contrast to the examples given above, here, we mean by “strained-Ge” those systems whose band structure is fundamentally modified by applying a high enough amount of tensile strain. We therefore often call it "direct bandgap strained-Ge". This approach has the elegance of adding, through a careful tuning, the ability of pure Ge to lase to its already large application portfolio. One could for instance think of using its lasing capabilities for quantum technologies where germanium starts to play an important role [194]. Another good reason to continue this research is that strained Ge is the only semiconductor laser system that consists of only one element. Fundamental (ab initio) models - such as band structure calculations, phonon dispersion, charge carrier transport as well as fundamental laser modelling tools can simply be applied and tested on it.

However, the strained Ge approach inherently comes with disadvantages, notably when constructing a laser. Namely, semiconductor lasers are often built up of heterostructures, or quantum wells. Moreover, for electrical carrier injection, some layers must be doped. In essence, a good laser will not consist of only a homogeneous layer such as Ge. Thus, for the moment, strained Ge may just be a powerful testbed helping the development of the final group IV laser. Applications may nevertheless come, but in very specific areas.

In the following, we will highlight the impact research on Ge lasing has had so far. Firstly, we should mention the work performed by CEA-LETI on testing previously introduced band structure predictions, i.e. the deformation potential theory [195]. Similarly, the phonon dispersion under strain was investigated. For any strain above 3%, notably for uniaxial loading a pronounced bowing of the Raman strain tensor has been demonstrated [132]. For biaxial loading, non-linear dispersion corrections are found to be less pronounced. The bandstructure crossover from indirect to direct seems to match the deformation potential theory [195]. Obviously, this knowledge does have an impact on GeSn, in particular in a later stage when the fabrication and layer quality issues will have been resolved and accurate modelling will be needed for performance optimisation.

| | Cavity modes | Linewidth narrowing | Mode competition | Efficiency | [Ref] |
|---|---|---|---|---|---|
| **n-doped Ge laser** | no, only transversal | no, | no | low | [193] |
| **intermediate strained Ge** | y | y | no | low/not quantified, | [190] |
| **Stressor induced Biaxial strain** | y | y | no | low/not quantified | [191] |
| **Strained Ge laser** | y | y | y | > 50% | [4] |
| **GeSn (micro-disc)** | y | y | y | high/not quantified | [3,184] |

Table 5 : Experiments performed on strained germanium and on GeSn (for comparison). Examples given in the last two rows fulfil all four lasing criteria, longitudinal modes are resolved, linewidth narrowing, mode competition, and high efficiency.

What the lasing demonstration in strained Ge [4] has shown so far and what we can learn from it is the following:

(i) The direct band gap concept works. Group IV lasers of that kind behave very similarly to other semiconductor lasers.

(ii) Gain values as well as lasing efficiency are high, especially at low temperatures.

(iii) The latter is very prominent in gain clamping and mode competition, where one cavity mode prevails at the expense of all others.

(iv) A multi-mode spectrum - which is uniformly amplified as the excitation increases - does not mean lasing. Such a spectrum may well contain stimulated emission, but this can hardly be distinguished from transparency and is therefore not sufficient to evidence lasing. Instead, lasing with a high gain material such as Ge or GeSn will occur only on a few, possibly even only one mode(s), as Figure 26, taken from Ref [4], convincingly shows.

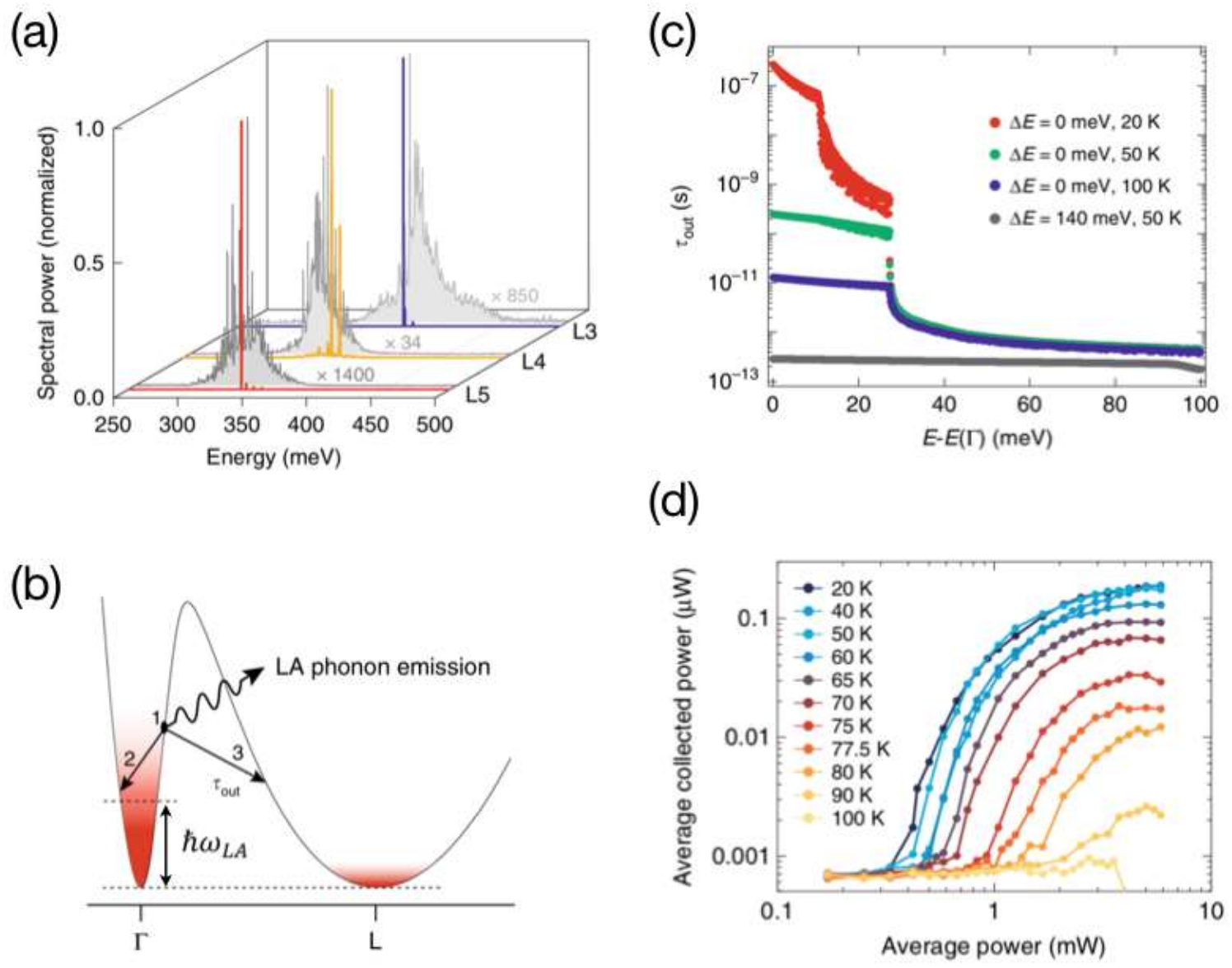

Figure 26 : (a) Few mode to single mode behaviour of the strained Ge laser, typical of high gain semiconductor lasers. (b) and (c) At low temperature the scattering of electrons from Γ to the L valley is blocked up to energies exceeding the energy of the zone boundary phonon $\hbar\omega_{\mathrm{LA}} \cong 28\mathrm{meV}$. (d) Light-in-light out curve as a function of different temperatures. Images are adapted from Ref [4].

The above given features help to categorize previously published work, c.f. Table 5. We also mention, for comparison purposes, the most recent works on GeSn. Comparing the lasing experiments performed on biaxial strained Ge [191] and GeSn [3] is particularly instructive because the experiments were performed on practically identical set-ups. In the germanium case [191], only 2 of the 4 laser criteria were fulfilled. For GeSn [3], however, all 4 criteria were fulfilled. Biaxial strained germanium did not seem, in Ref. [191], to have reached the laser threshold, most probably because the tension (1.7%) was not high enough, then, to close the offset between Γ and L. Surprisingly, this possibility was not considered in [191].

Additional understanding to be gained from lasing experiments on strained Ge [4] concerns the valley scattering from Γ to L. This scattering is very fast in natural germanium, where the offset is about 140 meV. The situation however changes when the offset is approaching zero under high strain, c.f. Figure 26b. The scattering then depends on the energy the electron carries with respect to the L-band edge. The electron will hardly scatter as long its excess energy is less than the energy of phonons at the Brillouin zone boundary [4], Figure 26c. Interestingly, the blocking of valley scattering as demonstrated here at low temperature reminds the original proposal by the MIT group [193] to block this scattering by filling up the L states at energy below the conduction band minimum at Γ through a heavy n-type doping of Ge with phosphorous. Seemingly, the blocking effect by doping is less effective and/or has drawbacks as level broadening.
For electrons with higher excess energy, valley scattering rates increases quickly. This effect is strongly dependent on temperature. Indeed, at elevated temperatures, (i) the probability electrons have of occupying states with energies definitely higher than the Fermi energy does increase, and so does (ii) the number of the zone boundary phonons. The latter may have the stronger impact, as is deduced from comparisons between experiments and model predictions, shown in Figure 26d. Time resolved laser excitation experiments on strained Ge layers but also GeSn may be the best way to study these processes in more details. From such studies, we expect to gain a quantitative understanding of the impact of reduced scattering by phonon blocking for group IV lasing. In Ref [4], with nearly direct band gap Ge and comparably short excitation, phonon blocking is indeed highly effective up to almost 100 K. For continuous wave excitation and under nanosecond long excitation, this effect is expected to be weaker, though not negligible. Indeed, in a vast amount of barely direct systems, lasing stops near $T < 100$ K [1, 3, 184], as shown in the following section. This peculiar conformity could come in part from the blocking effect vanishing at this temperature.

## 3.3 Lasing in GeSn

The fabrication of an efficient, electrically pumped laser operating at room temperature based on the epitaxy of group IV materials would have several advantages like an easy integration in the present day Si-photonics platform. In the Subsection just above, Ge lasing was presented. Here, we will discuss lasing from GeSn alloys and GeSn/SiGeSn heterostructures. The band structure engineering lessons learned from Ge can also be applied here. There will however be specific issues coming from the epitaxy, on Ge SRBs, of thick GeSn layers or heterostructures.

### 3.3.1 From bulk to heterostructures

As detailed above, incorporating Sn atoms into the Ge lattice modifies the electronic band structure. The energy of the conduction band at the Γ-valley (direct transition) shrinks faster than the energy of the conduction band at the L-valley (indirect transition), resulting in a cross-over to a fundamental direct bandgap semiconductor once a critical amount of Sn is reached. Photoluminescence (PL) experiments on a series of GeSn layers with Sn contents varying from 8 to 12.5 at% proved that as early as in 2015 [1]. A definite change of the evolution of the integrated PL intensity with the measurement temperature highlighted a fundamental change in the semiconductor bandgap. Using the conduction band offset ΔE (energy difference between the Γ-valley and the L- valley) as a modelling parameter, an indirect bandgap with ΔE=-80 meV was extracted for $Ge_{0.92}Sn_{0.08}$ and a direct bandgap with ΔE =+25meV for the $Ge_{0.875}Sn_{0.125}$ alloy. These values are in good agreement with k.p band structure calculations. The conduction band offset ΔE will be called, in the following, the bandgap "directness". It is an important parameter determining lasing properties. The above mentioned PL experiments, together with strain effects calculations, yielded an experimental Sn content of 8 at.% at which a fully relaxed and therefore cubic GeSn alloy became a direct bandgap semiconductor .

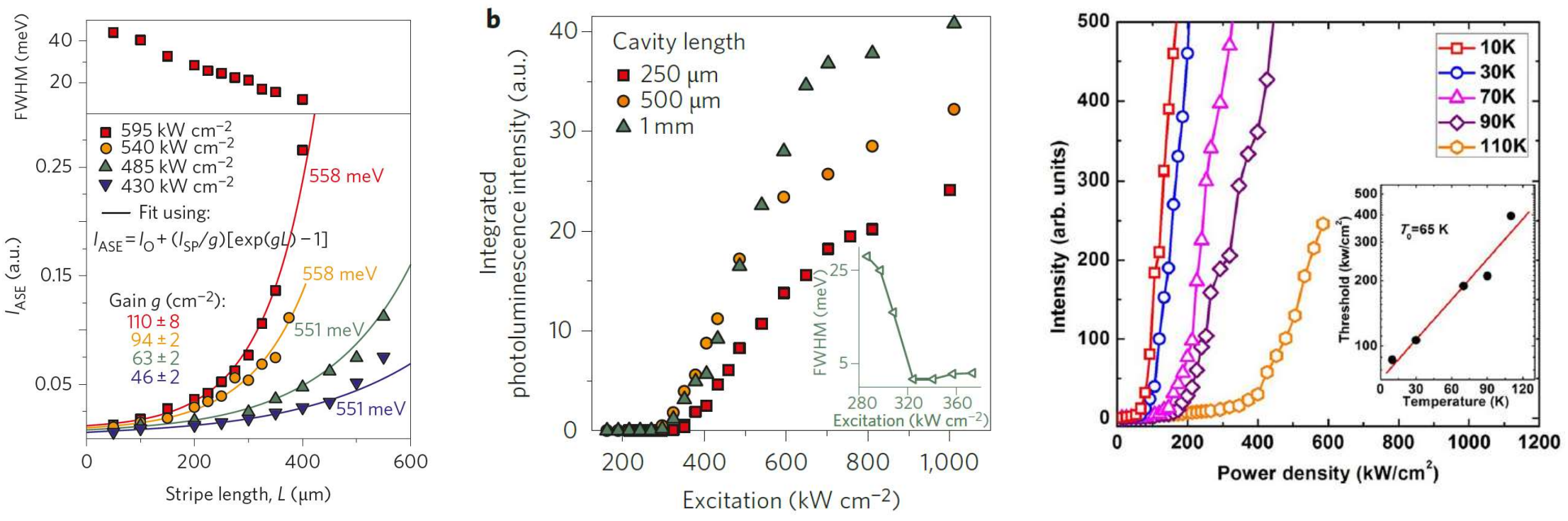

Figure 27: (a) Amplified spontaneous emission (ASE) spectra for different peak energies. The full line is the fit, using the equation given in the figure, to determine the modal gain [1], (b) Integrated photoluminescence intensity as a function of optical excitation for 250 µm, 500 µm and 1 mm long waveguides. Inset: Full Width at Half Maximum of the peak around the lasing threshold for the 1 mm-long GeSn waveguide [51], (c) Photoluminescence intensity of a 1100 µm-long waveguide between 10K and 110K. Each curve shows lasing threshold characteristics. The temperature-dependent thresholds varied from 87 to 396 kW/cm$^2$. Inset: laser threshold versus temperature [51].

As shown in the theory section, a direct bandgap is a necessary condition for optical gain and lasing, in case the optical gain can compensate cavity and material losses. The modal optical gain – the remaining gain of a mode after subtracting the material losses - can be extracted experimentally by the variable stripe length (VSL) method as explained in section 3.1.2. Using the VLS method, it was shown that a 500 nm thick $Ge_{0.875}Sn_{0.125}$ alloy yielded, at the maximum emission energy of 558 meV, a modal gain of 110 cm$^{-1}$ (Figure 27). Using the methods outlined in section 3.1.2., experiments shown in Figure 27 (a) and (b) gave a differential modal gain of approximately 0.4 kWcm$^{-1}$.

With a direct bandgap semiconductor and the technology to fabricate Fabry-Perot resonators, the demonstration of lasing under optical pumping naturally followed. By optically pumping a 1 mm - long waveguide, lasing was unambiguously proved, with a distinct threshold in output intensity, as shown in Figure 27. Despite the importance of being the first GeSn laser, performances were poor. The laser threshold at 20K was about 350 kW/cm$^2$ and increased up to almost 1000 kW/cm$^2$ at the maximum operation temperature of 90K.

Shortly after, these results were confirmed by the Univ. of Arkansas group using a 760 nm thick GeSn alloy with a Sn concentration of 11 at.% [51]. Lasing properties were similar, with a maximum lasing temperature of 110K. The poor laser performances were attributed to different causes:

i) a very low directness ΔE;
ii) a relatively high compressive strain in the layer;
iii) a high density of misfit defects at the GeSn/Ge-substrate interface.

These drawbacks, which are interconnected and not easy to separate, are discussed in the following. A cubic GeSn alloy has a larger crystalline lattice than elemental Ge. During epitaxy, a large in-plane compressive strain will build up in GeSn alloys grown on Ge SRBs. According to theoretical band structure calculations, biaxial compressive strain results in an increase of the energy of the Γ-valley and therefore a reduction of the energy separation between Γ-and L- valleys, or directness, ΔE. This is equivalent to a lower “effective Sn content” in the GeSn alloys. The influence of tensile and compressive strain on bandgap directness is the opposite. Tensile strain will indeed increase the bandgap directness. For valence bands, an in-plane compressive strain will lift up the heavy-hole (HH) energy band while a tensile strain will lift up the light-hole (LH) energy band. In the following, we will first discuss the impact of biaxial compressive strain. Later on, we will elaborate on the role of uniaxial and biaxial tensile strain. The calculated electronic band structure of a $Ge_{0.875}Sn_{0.125}$ alloys as a function of compressive strain is shown in Figure 28a. The solely strain based indirect-direct bandgap transition is clearly seen for a biaxial compressive strain of -1.05%. The directness, ΔE, of a $Ge_{0.875}Sn_{0.125}$ layer increases from 0 meV to 120 meV as the compressive strain decreases from -1.05% to 0%, respectively. As a consequence, the integrated PL intensity continuously increases as the compressive strain decreases [67].

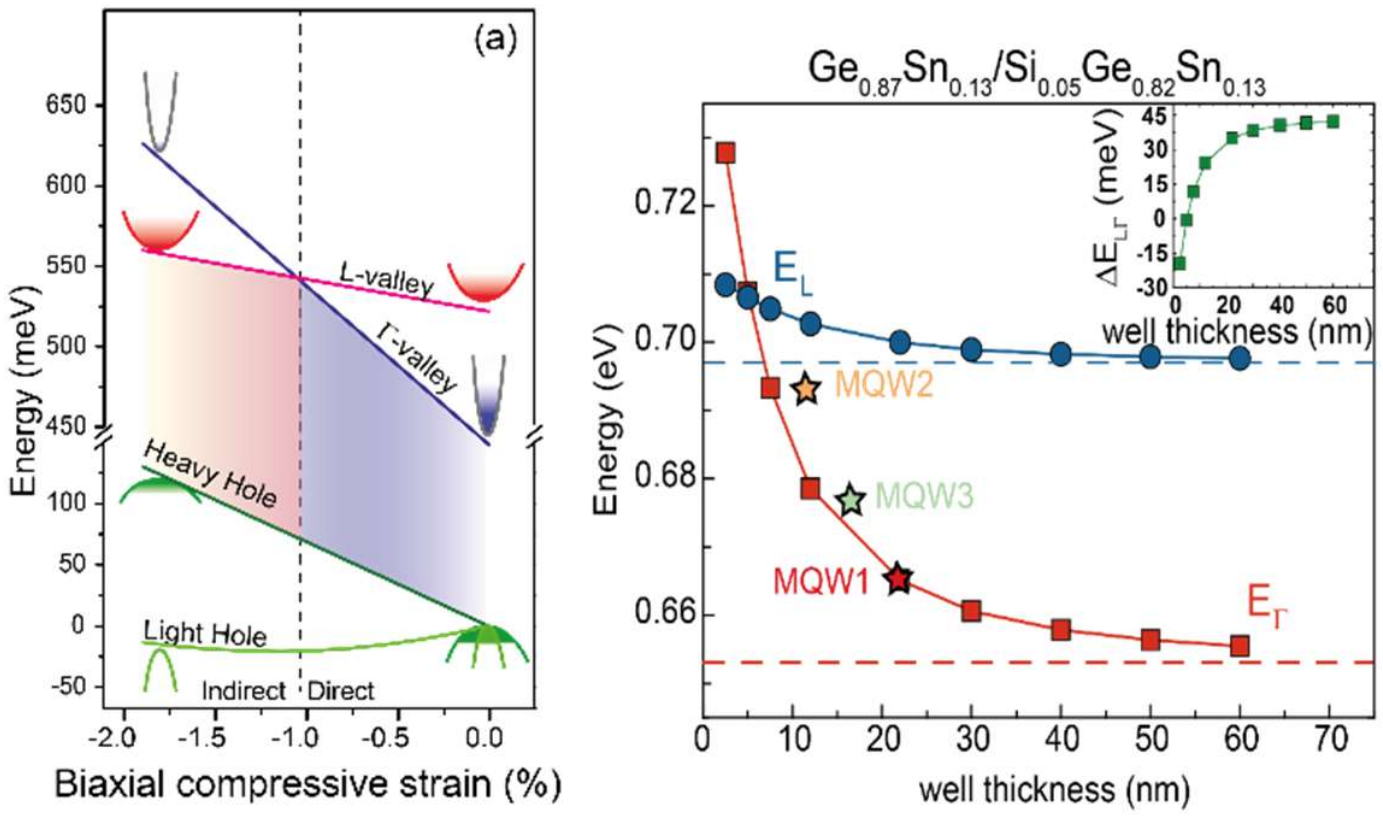

Figure 28: (a) Calculated electronic band structure as a function of compressive strain in $Ge_{0.875}Sn_{0.125}$ alloys. The semiconductor has an indirect bandgap for compressive strains larger than -1.05%. It becomes a direct bandgap semiconductor for strains lower than -1.05%. (b) Energy value of Γ- and L-valleys as a function of the quantum well thickness. Inset: directness versus quantum well thickness.

The biaxial compressive strain build-up during the epitaxial growth can be relaxed by depositing thick layers, well above the critical thickness for strain relaxation. However, it was shown, in the well-known SiGe on Si system, that 100% strain relaxation was hardly achievable in hundreds of nm thick layers [196]. In addition, a high density of crystalline defects density was present in those layers. Reduced crystallinity and point defects reduce the radiative emission by creating paths for non-radiative recombination.

Forming free standing regions where the lattice can elastically relax enabled to maximize relaxation in GeSn layers. Various strategies are described below. However, we would like first to discuss the possibility of improving laser performances thanks to the use of heterostructure and quantum wells into the active material. Growth aspects were discussed in a previous section.

Roughly an order of magnitude reduction in lasing threshold was achieved in III-V lasers each time the dimensionality of the gain material was reduced (from 3D to 2D to 1D and finally 0D, thanks to quantum wells, quantum wires, and quantum dots, respectively). In order to go along this path with GeSn, a second group IV semiconductor material with a larger bandgap than GeSn is required. The interest of using Ge in the barrier layers of p-i-n Ge/GeSn multi quantum wells (MQW) LEDs was evaluated. Unfortunately, no advantages over p-i-n GeSn homo-junction LEDs [42] were evidenced. In 2017, efficient electro-luminescent GeSn/SiGeSn MQW [197] p-i-n LEDs were fabricated, showing the interest of Sn-based heterostructures for lasing.

The effective bandgap in a two dimensional biaxial compressive GeSn QW corresponds to the energy differences between the first quantized levels of heavy holes (HH) and Γ-valley electrons. The quantum state energies are inversely proportional to the effective masses of the conduction and valence bands. This has an impact on the directness of the material, $\Delta E$, which is considerably reduced in thinner wells since the electron effective mass at Γ ($m^*_\Gamma$ [100] = $0.031m_e$) is ~2.7 times smaller than at L ($m^*_L = 0.117m_e$). In Figure 28, the energy of the first quantized levels of Γ- and L- valleys as a function of the quantum well thickness is given for a 10x{$Si_{0.05}Ge_{0.82}Sn_{0.13}/Ge_{0.77}Sn_{0.13}$} MQW grown on a partially relaxed $Ge_{1-y}Sn_y$ buffer, itself on a Ge SRB. There is, for $Ge_{0.77}Sn_{0.13}$ wells less than 10 nm thick, a transition back to an indirect semiconductor.

A major way to improve GeSn lasing is reduce the amount of structural defects that result in non-radiative recombinations. For instance, there is, in thick plastically relaxed GeSn layers which are the most used for optically pumped lasing, a misfit dislocation network at the interface with the Ge SRB, i.e. very close to the optical mode in GeSn. To address this problem GeSn buffers with a grading [51, 68], or a lower Sn content buffer are grown first [181]. The optical active medium is then grown lattice matched on this buffer. Interestingly, ternary SiGeSn alloys offers a balance of lattice strain and mismatch reducing additional interfaces relaxation. A similar strategy was also used in GeSn/SiGeSn MQW lasers.

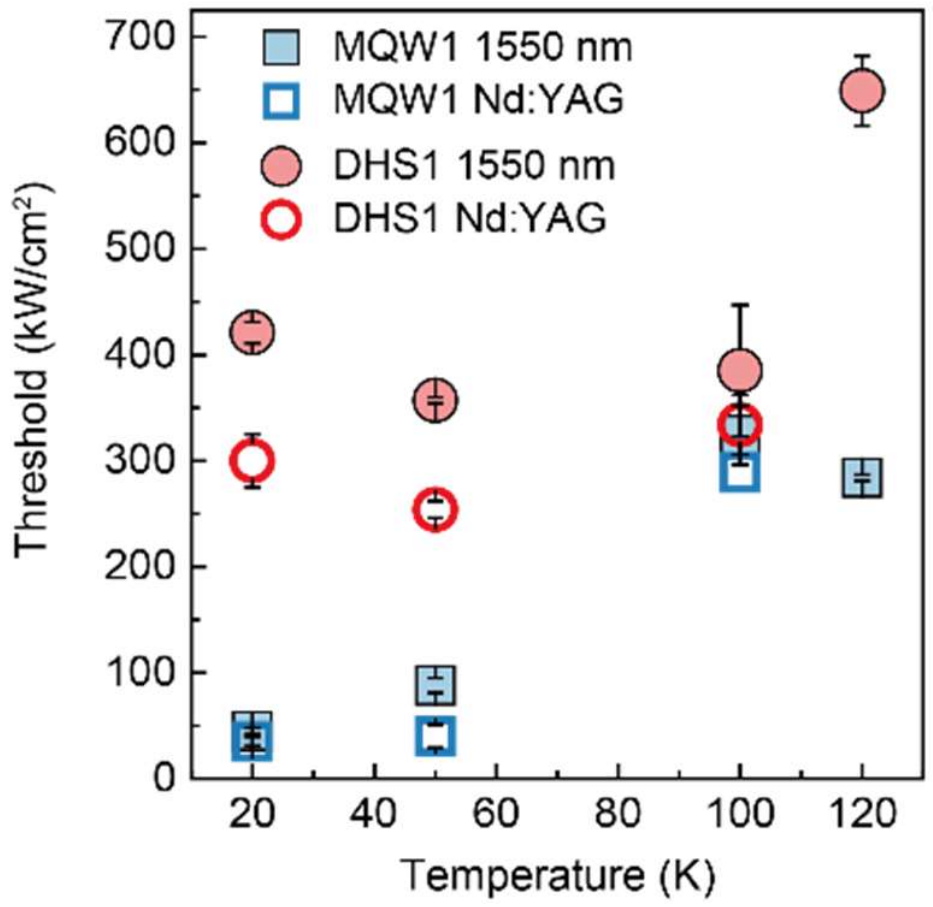

Figure 29: MQW and DHS laser threshold versus temperature for two different optical excitations.

A 200 nm thick strain relaxed $Ge_{0.9}Sn_{0.1}$ layer was first grown, as a separation layer, on a Ge SRB followed by a 10 periods {22 nm $Si_{0.05}Ge_{0.82}Sn_{0.13}$/22 nm $Ge_{0.77}Sn_{0.13}$} MQW (called MQW 1). Another MQW structure (called MQW2), with 12 nm thick wells and 16 nm thick barriers, as well as a double heterostructure (DHS) with a single 380 nm thick $Ge_{0.855}Sn_{0.145}$ active layer cladded with $Si_{0.055}Ge_{0.83}Sn_{0.115}$ were grown for comparison purposes. Details on the epitaxy of the heterostructures can be found in Ref. [181**Erreur ! Signet non défini.**].

Undercut microdisks were optically pumped using two different lasers: a Nd:YAG laser emitting at 1064 nm, with a pulse length of 5 ns and a repetition rate of 17 kHz and a fiber laser emitting at 1550 nm, with a pulse length of 800 ps and a repetition rate of 20 kHz [58]. Experimental conclusions, from comparing lasing data for these three heterostructures under 2 optical excitations, were the following:

- The maximum lasing temperatures were similar, for the DHS and the MQW1 heterostructures: 100K and 120 K under 1064 nm and 1550 nm excitation, respectively, although the DHS directness, 82 meV, was almost 2.5 times higher than in the MQW1 structure (32 meV only).
- The MQW2 structure did not lase under 1064 nm but only under 1550 nm excitation and at very high power density above 1000 $kW/cm^2$. This is attributed to the very low directness of only 8 meV and the very low active volume of only 120 nm (compared with 220 nm for MQW1 and 380 nm for DHS), with a gain not able to compensate the optical losses.
- The most interesting comparison concerns the lasing threshold (Figure 29). The laser threshold at 20K for the DHS structure was found to be ~300 ± 25 $kW/cm^2$ , to be compared with 40 ± 5 $kW/cm^2$ only for the MQW1 laser. However, the same laser threshold was obtained for temperatures above 50K.

Those experimental results support the initial assumptions. Reduced non-radiative recombinations due to a *spatial separation from the dislocated interface* with the Ge substrate by an intermediate strain relaxed buffer and an enhanced radiative recombination because of the *localization* of electrons and holes in multi-quantum wells result in the low directness MQW having a similar maximum lasing temperature than the larger directness DHS.

The lower threshold of MQW1 compared to DHS is mainly due to the reduced density of states (DOS), resulting in a decrease of the carrier concentration at transparency. Since the threshold reduction is more than 1 order of magnitude, it is reasonable to assume that the carrier lifetime is improved due to a localization of carriers in the QW. What is also striking is that, above 50K, lasing thresholds for DHS and MQW become similar. This is explained based on band structure calculations which indicate that the conduction band offset is about 50 meV at 20K and below 20 meV at room temperature. When pumped with high energy photons at high temperature, carriers easily can gain energy above the band offsets and escape the wells. Quantisation disappears and the structure behaves as a DHS, then.

In conclusion, quantum well structures are viable solutions to strongly reduce the lasing threshold. Experimental data and theory tell us that, in a MQW, the thickness of the well should be such that there is some carrier confinement and the quantization just high enough to reduce the DOS. Large quantization effects will strongly and negatively influence the GeSn directness. MQW improvements would result from the epitaxy, if feasible, of higher Si content SiGeSn barriers: the QW to barrier energy band offset would be higher, then, and carrier confinement still present at elevated temperatures.

Last but not least, optical excitation closer to the bandgap reduces the electron density and consequently the L-valley population, which contributes to losses by free carrier absorption. The effect was discussed, for the Ge laser, in a previous section.

### 3.3.2 GeSn optical cavity under-etching

By deeply underetching a GeSn epilayer grown on a Ge SRB, a significant fraction of the built-in strain will elastically relax. This compressive strain reduction will increase the bandgap directness and improve the gain. The increased refractive index contrast between air and semiconductor will also improve the mode filling. Last but not least, surface defects can be removed and/or passivated depending on the chemistry used. Those performance boosts are shown in the following discussion on undercut waveguide and micro-disk lasers.

The fabrication process is based on standard CMOS processing. Section 1 gives some details about the selective etching used. The control samples (not under-etched) and the under-etched samples were passivated with 10 nm of $Al_2O_3$ dielectrics in order to passivate the GeSn surface and reduce surface carrier recombination. Moreover, under-etching will remove part of the highly defective region at the GeSn layer / Ge SRB interface, reducing non-radiative Shockley-Read-Hall recombinations.

The Light in-Light out (L-L) characteristics at 20K of not-etched and under-etched, same dimensions $Ge_{0.875}Sn_{0.125}$ waveguides are shown in Figure 30. It is clear that the under-etching definitely improves the lasing characteristics, as expected, due to the increased GeSn directness via strain relaxation (Figure 30a).

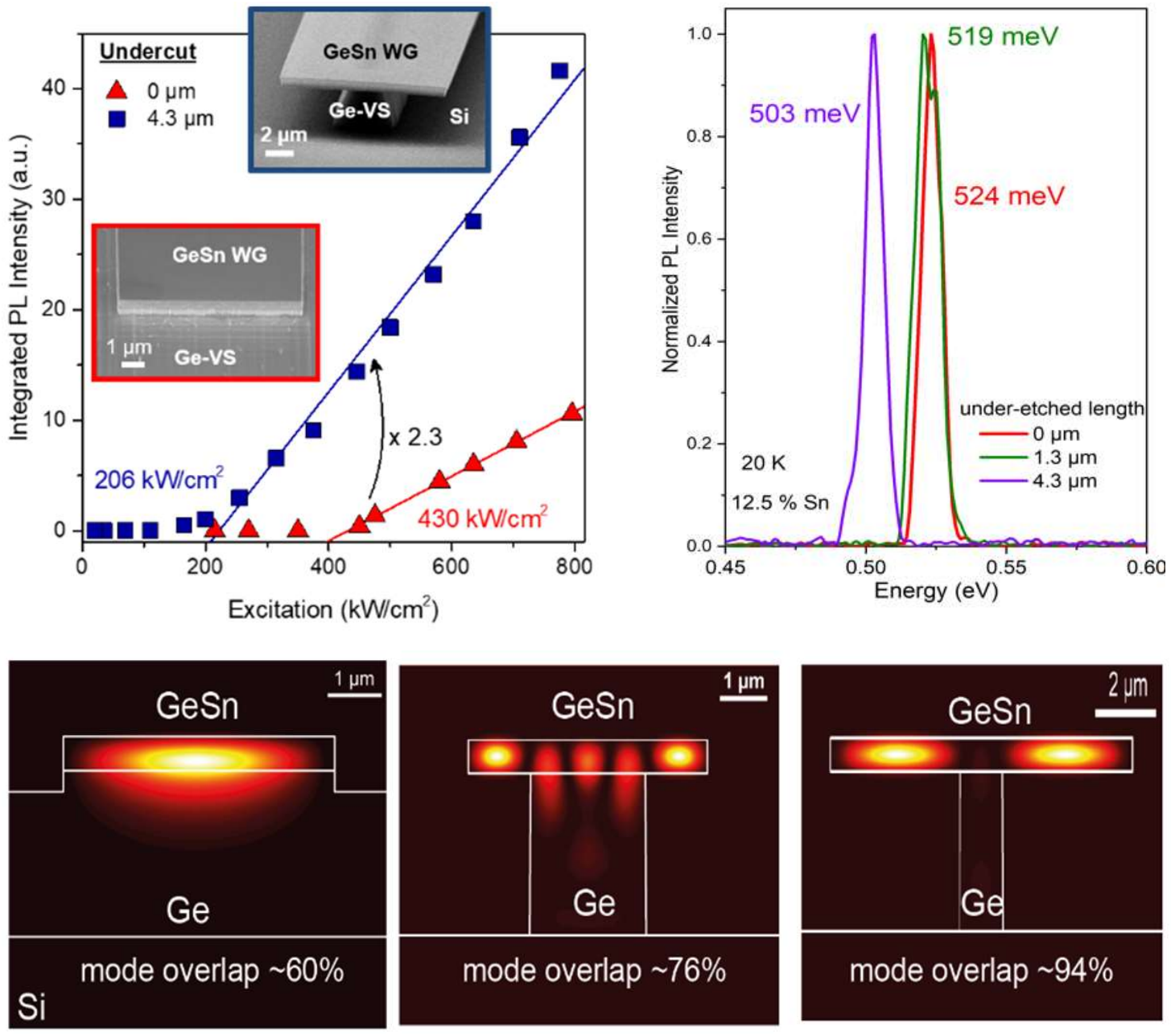

Figure 30: (a) L - L characteristics of non-under-etched (red plot) and 4.3 µm under-etched (blue plot) GeSn waveguides. SEM images of the two GeSn laser cavities are shown in the insets. (b) Laser emission from GeSn waveguides with different under-etch lengths. The shift of the laser emission energy is due to strain relaxation induced bandgap change. (c) Optical mode distribution in the GeSn waveguides with different under-etch lengths.

The lasing threshold decreases from 430kW/cm$^2$ down to 206 kW/cm$^2$, while the slope of amplified emission increases by a factor of 2.3. In the bulk state, the 560 nm thick $Ge_{0.875}Sn_{0.125}$ layer has a residual compressive strain of -0.6%, which translates into a $\Delta E$=25 meV. According to strain relaxation simulations data used for band structure calculations, a medium directness of about $\Delta E$=60 meV at the edges of the GeSn "wings" is obtained in the under-etched structure. The bandgap change can be seen by the energy shift of the laser emission for waveguides with different under-etching lengths, as shown in Figure 30b. The increased directness will result in an increase of the Γ-valley population, with less optical pumping required to reach the lasing threshold, as seen experimentally. Moreover, the Γ-to L-valley scattering is reduced for a higher bandgap directness, resulting in a higher lasing temperature. Indeed, the 4.3 µm under-etched waveguide lases up to 135K, a temperature 45K above that of the non–under-etched structure.

The additional advantage of suspended cavity designs is coming from a higher (GeSn/air) refractive index difference in the free standing regions, increasing the optical mode confinement in the gain material and reducing the bending losses. In the present samples, the optical mode overlap is 64% in the bulk case (no under-etching) and increases up to 76% and 94% for 1.3 µm and 4.3 µm undercuts, respectively (See Figure 30c). Such an effect is highly beneficial in order to increase the operating temperature and reduce the lasing threshold.

By fabricating such under-etched waveguides, it was possible for the first time to demonstrate multiple laser actions in the mid-IR region, or, in other words, tune the laser emission by changing the Sn atomic content. Using roughly 800 nm thick layers with a large strain relaxation in the as-grown state together with under-etching waveguide cavities, laser emission from GeSn layers with Sn contents of 8.5 at.%, 10 at.%, 12.5 at.% and 14 at.% was demonstrated. The laser peak energies and the L-L characteristics at 20K for such optically pumped lasers are shown in Figure 31.

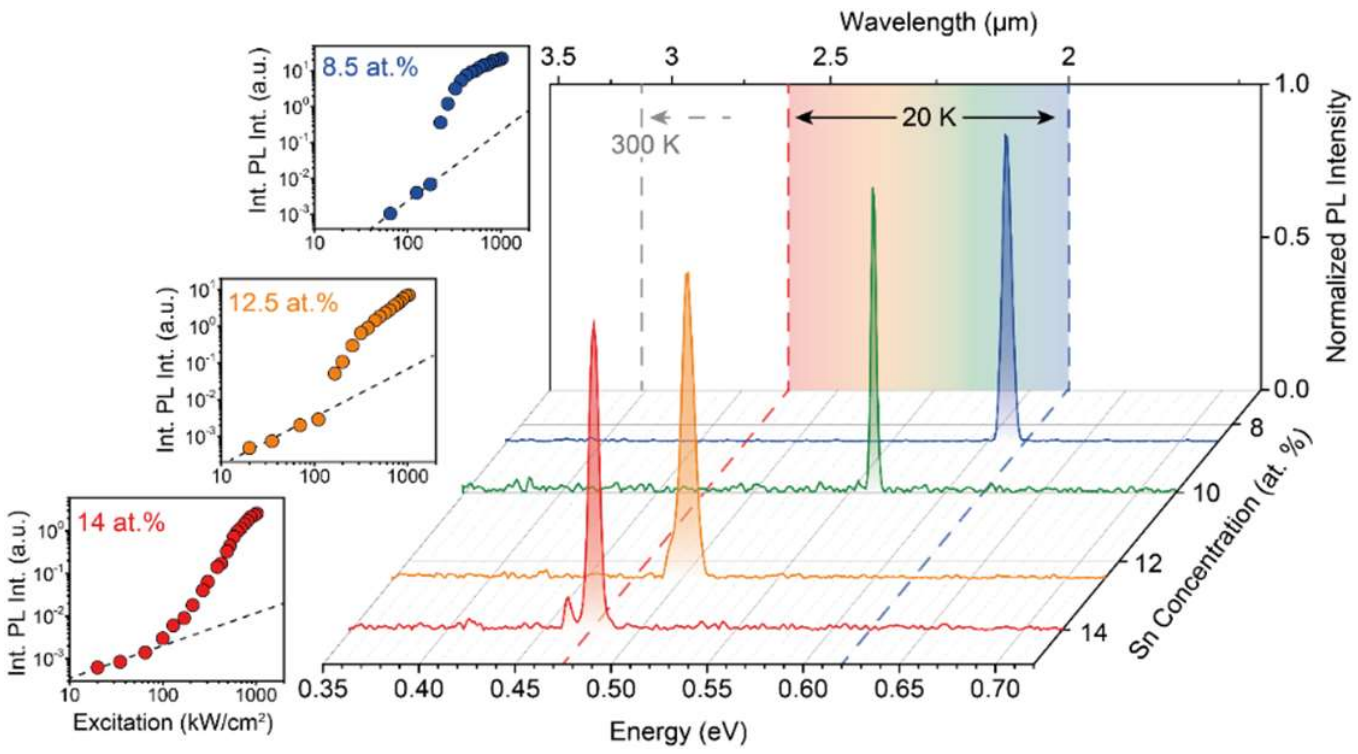

Figure 31: (Right side) Laser emissions of 8.5 at.%, 10 at.%, 12.5 at.% and 14 at.% GeSn Fabry-Perot lasers with 4.3 µm under-cuts. Bandgap changes induced by Sn contents which varied resulted in shifts of the laser emission energy. (left side) L - L characteristics at 20K for 8.5%, 12.5% and 14% GeSn Fabry-Perot lasers.

The cavity fabrication technology described above was also used to fabricate micro-cavities with micrometers round trip lengths (microdisk resonators and micro Fabry-Perrot corner cube cavities). The advantage of such cavities is a simpler experimental set-up, with a homogeneous pumping of the complete cavity which is more difficult for long and wide waveguides. The importance of defects management on GeSn lasing characteristics is critical and can result, if done properly, in drastically reduced lasing thresholds in microdisk laser cavities [184].

### 3.3.3 Toward higher operation temperature

GeSn growth improvements, with for instance the use of GeSn graded buffers to gradually accommodate the lattice mismatch between GeSn and Ge, yielded good optical quality GeSn layers with Sn contents up to 16 at.%. [18]. Compared to the GeSn 12% layer discussed above, with a bandgap directness of only ΔE~50 meV, the use of a graded GeSn buffer resulted in a large strain relaxation in the $Ge_{0.84}Sn_{0.16}$ as-grown layer on top, with defects confined in the bottom layers. The latter after the microdisk undercut became completely stress relaxed. The bandgap directness then increased to the 146 meV theoretical value.

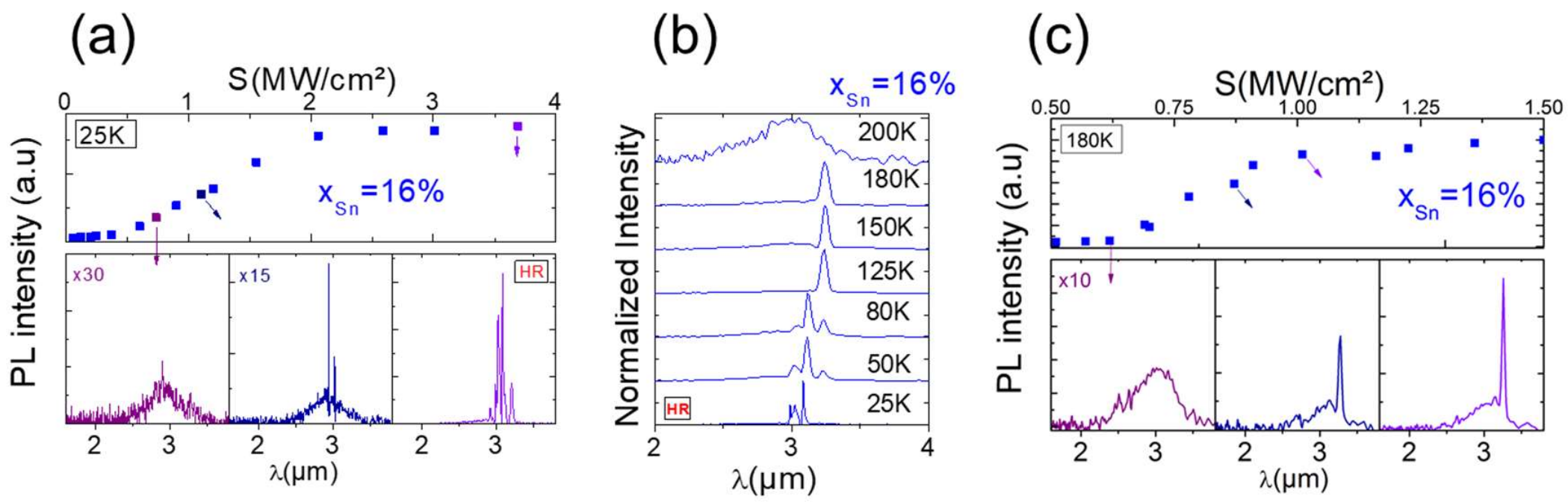

Figure 32: (a) L-L curve of a 8 μm micro-disk fabricated using a $Ge_{0.84}Sn_{0.16}$ layer grown on a GeSn step-graded buffer at 25 K (b) Evolution of the laser spectra as a function of temperature. (c) L-L curve at 180 K of the GeSn 16% microdisk [68].

Figure 32a shows the L-L characteristics at 25 K of a 8 $\mu$m diameter $Ge_{0.84}Sn_{0.16}$ microdisk together with emission spectra at three different pumping levels. The lasing threshold was 377 kW/cm$^2$ at 25 K. Figure 32b shows the evolution of the lasing spectrum as a function of temperature, with a mode switching as the temperature increased from 50 K up to 125 K. The optical gain shifted rapidly with the temperature, while the cavity modes evolved at a much slower rate. At 200 K, a clear broadening of the emitted spectrum was observed compared to lower temperature spectra, indicating no laser action anymore. Spectra were measured at 4.3 MW/cm$^2$. L-L curves at 180 K are shown in Figure 32c. The increase of the lasing temperature, from 135K up to 180K, was attributed to the strongly increased bandgap directness, from 50 meV for $Ge_{0.875}Sn_{0.125}$ up to 146 meV for $Ge_{0.84}Sn_{0.16}$ samples. Increasing the Sn content however degraded the layer quality. This is believed to be responsible for the increased laser threshold at 25K, from 220 kW/cm$^2$ for $Ge_{0.875}Sn_{0.125}$ sample up to 377 kW/cm$^2$ for $Ge_{0.84}Sn_{0.16}$.

With a thicker $Ge_{0.84}Sn_{0.16}$ active zone (418 nm) and a capping with a higher bandgap $Ge_{0.87}Sn_{0.13}$ layer, a maximum lasing temperature of 230K was reached. There was also a decrease of the threshold, down to 134 kW/cm$^2$ at 15K, because of a better overlap between the material gain and the optical modes [182].

Some conclusions can, at that point, be drawn regarding the evolution of the GeSn laser performance with the Sn content. Experiments described above show that an increase of the conduction band directness thanks to thick, higher Sn content GeSn alloys result in an increase of the maximum lasing temperature. However, a Sn content increase also seems to result in a lower crystalline quality of the material and thus to larger laser thresholds at high temperatures, where phonon scattering becomes important. For instance, the lasing threshold, for a Sn content of about 16 at.%, increases from approximately 100 kW/cm$^2$ at 20K up to more than 3400 kW/cm$^2$ at 270K [52]. A tentative solution to decrease the threshold while keeping high lasing temperatures could then be to grow SiGeSn/GeSn multi-quantum wells with high Sn contents in the wells. A decrease by one order of magnitude was indeed evidenced for a medium Sn content of 13% in the wells compared to bulk structures. It is then clear that the combination of two approaches seems the most reasonable way to address efficient room temperature GeSn based laser.

All the GeSn lasing researches presented so far were based on the epitaxial growth of GeSn layers on Ge buffers, with very large thicknesses and/or under-etching to reduce the residual compressive strain. The main direct transition then took place between the conduction band Γ- valley and the heavy hole (HH) as the top valence sub-band. Theoretical estimations indicate that if some biaxial tensile strain is induced in GeSn, the light hole (LH) becomes the top valence band. Due to splitting of HH- and LH-bands, the majority of injected holes will be in the LH which is beneficial for lasing. The calculated maximum net gain, $g_{max}$, function of the GeSn directness for different Sn content GeSn alloys is shown in Figure 33a. It can be seen that, for the same bandgap directness, the net gain is larger in low Sn content layers. This is why the optical transition matrix element for z-polarized radiation for the LH-G transition (z-polarized) is 30% larger than the matrix element for the HH-G (x-polarized) transition. The same directness can be obtained by injecting tensile strain in the GeSn material. Theoretical and experimental data [168] indicate that, under a biaxial tensile strain of +1.5%, a GeSn layer with 5at. % of Sn has the same Γ- to L- valley energy difference as a 12.5 at.% GeSn layer under a biaxial compressive strain of -0.6% (as discussed above). The Sn content influences the conduction band energy separation but *not* the LH and HH valence band separation. The tensile biaxial strain positively impacts both the conduction band directness and the LH-HH splitting. Biaxial tensile strain is the only type of strain that lifts the LH band above the HH band. The laser transition will then occur between two low DOS bands (Γ- and LH), with a large LH-HH energy separation. This should result in lower laser thresholds.

Based on the above considerations, a $Ge_{0.95}Sn_{0.05}$ layer was used as the optically active medium for laser emission. A $SiN_x$ stressor layer was deposited on the grown GeSn layer and subsequently transferred by wafer bonding onto a Si substrate covered with a thick Al layer [3]. The bottom part of the as-grown GeSn layer, which was defective with the presence of a regular array of edge dislocations, was at that stage on top of the stack. It was removed thanks by chemical etching. This step was crucial, as it eliminated one of the main sources of non-radiative recombination. This effect was also demonstrated in Ref [184], where defects were removed at the bottom of microdisks during the under-etching of 7 to 10.5% tin content alloys. Low temperature lasing thresholds were then reduced to 10kW/cm$^2$ upon optical pumping at 1.5µm [25MHz, 3.5ns]. After microdisk patterning and under-etching, the GeSn optical cavity was supported by a large Al metallic post that acted as a heat sink. The final structure was covered with a second stressor $SiN_x$ layer, resulting in a GeSn layer embedded in a $SiN_x$ stressor layer and suspended on a large Al pillar. The strain injected into the nearly 300 nm thick GeSn layer was 1.45%. It was high enough to transform the initially indirect bandgap $Ge_{0.95}Sn_{0.05}$ layer, with a directness of ΔE=−60 meV, into a direct bandgap one, with ΔE=70 meV. The valence band splitting was $E_{LH}-E_{HH}$=170 meV.

Lasing experiments were performed with a 1550 nm wavelength continuous wave pump laser focused on the sample surface into a 12µm diameter spot. PL emission spectra from a 9µm diameter disk, collected at various incident pump powers at 25K, are shown in Figure 33b. At low excitation levels, microdisks exhibited a broad spontaneous emission background, attributed to Γ - LH direct transitions. For higher CW pump powers, whispering gallery modes (WGM) appeared. Higher excitation resulted in an exponential intensity increase, typical of amplified emission, of the main optical mode at 485meV. At 2.3mW pump power, the lasing emission was four orders of magnitude stronger than the background, as shown by the high resolution spectrum shown in Figure 33.

The emission intensity at 25K of the lasing mode at 485meV, corresponding to 2.55µm wavelength, in the threshold region, is shown as a function of the incident pump power by the linear light-in light-out (L-L) characteristic in Figure 33-inset. The lasing threshold, as obtained by linear extrapolation the LL curve shown in the inset of Figure 33 (b) is 1.3 kW/cm$^2$ only. The laser emission is also characterized by a linewidth narrowing below the instrumental resolution of 58 µeV. This GeSn device was the first capable of continuous wave laser operation and had the smallest threshold and linewidth reported so far of any group-IV semiconductor lasers.

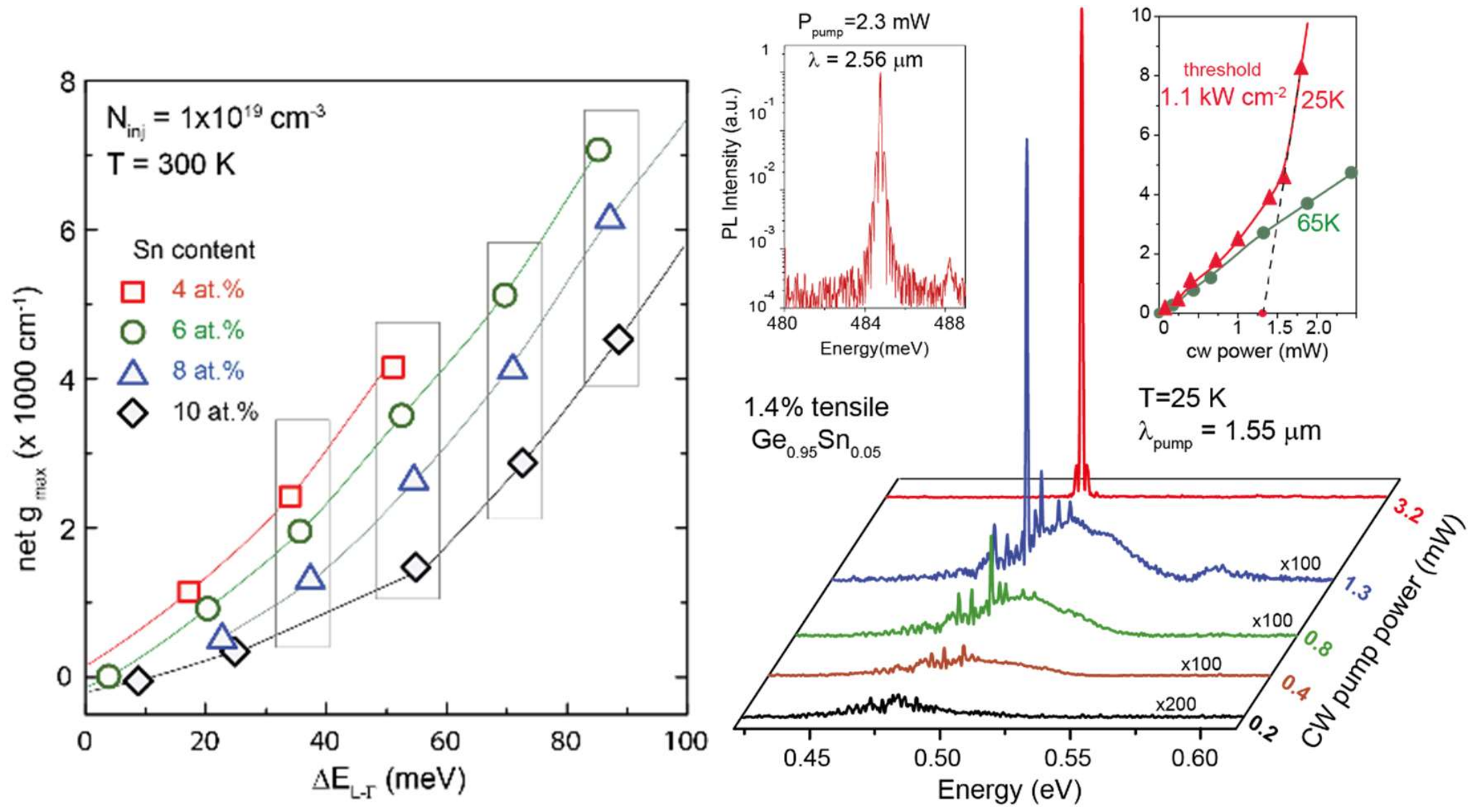

Figure 33 : (a) Calculated net $g_{max}$ function of the GeSn directness for different Sn content GeSn alloys. A large directness is obtained by injecting high amounts of in-plane tensile strain [167]. (b) PL spectra of tensile strained $Ge_{0.95}Sn_{0.05}$ under CW excitation, insets: 2.55 μm emission peak intensity at 25K and L-L curve detail taken close to the threshold.

Finally we focus on a device described in a previous section (Figure 16), where an uniaxial stress was injected into a microbridge with a 400 nm thick $Ge_{0.84}Sn_{0.16}$ layer on top of a step-grading [2]. The cavity was defined by two corner cube mirrors, as shown in Figure 34. The epitaxy of the step graded stack occurred on a 5 μm thick Ge SRB, itself on a silicon on insulator wafer. In contrast to pure germanium suspended membranes, the GeSn stack was in a compressive state. Etching of the membrane enabled to reverse the strain state thanks to pulling by the germanium arms, as discussed in section 2.1.3.2. The overall length L of the Ge pulling arms governed the amount of tensile strain in the central part of the structure. At low temperature, a maximum of 2.2 % uniaxial tensile strain was reached for L = 250 μm.

(a) (b)

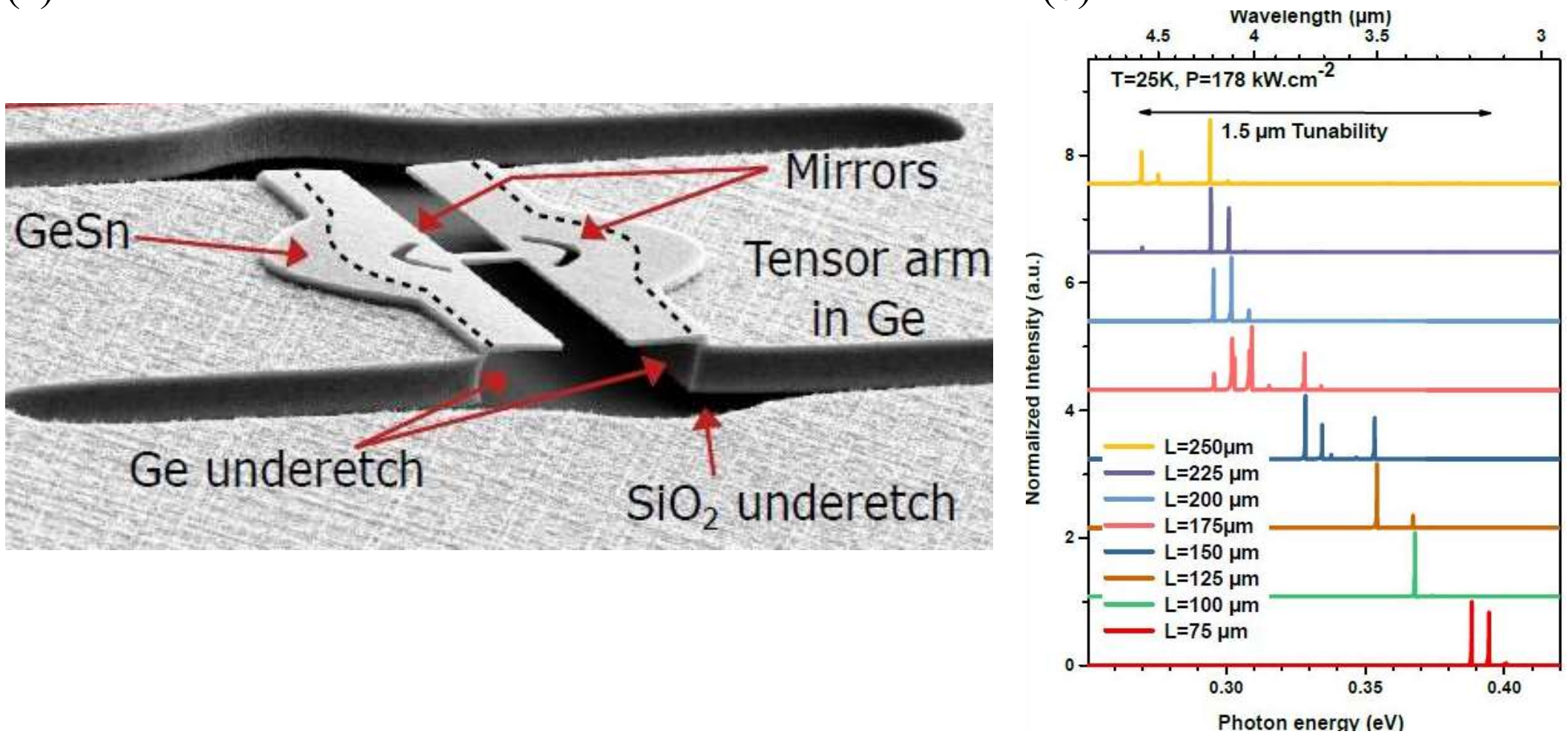

Figure 34 : a) Uniaxially tensile-strained $Ge_{0.84}Sn_{0.16}$ step graded layer laser cavity, (b) Laser emission spectra at 25 K versus arm length (the higher the arm length, the higher the uniaxial strain).

The uniaxial strain was evaluated from the photoluminescence spectral shift of the layer. Figure 34b shows the laser emission spectra are presented as a function of the arm's overall length (i.e. the tensile strain). A wide tunability of the laser wavelength, from 3.1µm up to 4.6µm, is obtained for a fixed 16% tin content in the active zone, at 25K. Thresholds are around 10kW/cm$^2$ for a pumping wavelength of 2.6µm (82MHz, 0.100ns). As the built-in strain is created by thermal expansion coefficients differences, the strain decreases for higher temperature, and thus also the directness of the gain material. Laser emission is still observed up to 273K, with a 2000kW/cm$^2$ threshold for a pumping wavelength of 1.06 µm [50kHz, 0.6ns]. At 273K, the strain is around 0.3%. So far, 273K is the highest temperature reported for any GeSn laser operation. The race to reach room temperature lasing operation is still ongoing.

# 4 Optoelectronic devices

A formidable interest has been building up for the last 10 years to develop and use photonic components operating in the mid infra-red, ranging from integrated biomedical and chemical sensors to short-range optical interconnections. A major roadblock is the current lack of a cost-efficient and complete MIR photonic platform with light sources and photodetectors fully compatible with the existing complementary metal–oxide–semiconductor (CMOS) platform. Current mid-IR technologies rely on bulky components and stand-alone laser sources operating in free space. Being able to monolithically integrate efficient and tunable MIR light sources and efficient group-IV photodetectors on Si would immediately give birth to new applications.

The continuous development of non-equilibrium growth schemes in industry-compatible reduced-pressure chemical vapor deposition (CVD) systems have enabled the epitaxy of high-quality GeSn binary [46, 53, 198, 199] and SiGeSn ternary [50, 61, 200, 201] materials. Despite their metastable nature, alloys are thermally stable up to at least 300°C (and possibly, depending on the Sn content and internal material strain, even above 600°C) [202], which is high enough for back-end processing. Germanium and Germanium-tin (GeSn) have indeed shown great promises for the direct integration of group IV semiconductor lasers and photodetectors. Light emission and photo-detection will occur in the mid infrared, thanks to the small bandgap of GeSn and tensile germanium. In the following, we will discuss the different strategies adopted in recent years to develop the required components to build an integrated circuit platform.

## 4.1 Photodetectors

Silicon and germanium, semiconductors from the group-IV column of the periodic table, are known for their pivotal role in microelectronics. The comparatively large elemental bandgaps of 1.12 eV (Si) and 0.80 eV (Ge, direct bandgap) at room temperature prevents their use as infrared detector materials, however. α-Sn is actually a group-IV semimetal with a negative bandgap of -0.41 eV. Alloying Ge with α-Sn and forming a binary with a diamond cubic crystal structure, enables one to reach longer wavelengths in the infrared region. Even more control in designing material properties can be achieved thanks to SiGeSn ternaries. The incorporation of Si atoms increases the L-Γ-energy difference in the conduction band and might even reverse the Sn-induced direct bandgap transition. This may lead to the unique opportunity of simultaneously having a narrow and yet indirect bandgap material, not existing in III-V materials. The generally larger carrier lifetimes in indirect bandgap materials may result in larger generated photocurrents, which will increase the performance of photodiodes.

The wavelength tunability of Si-GeSn alloys has been experimentally proved via transmission/ reflection measurements, as shown in Figure 35. Reflection measurements on 700-1000 nm thick CVD-grown GeSn samples showed a strong absorption edge – linked to the direct bandgap transition in the Brillouin zone center – which shifted into the Short-Wave Infrared as the Sn content increased (Figure 35a). Being able to deposit SiGeSn ternaries yielded more freedom in designing the electronic structure. Individual tuning of incorporated Si atoms (Figure 35b) and Sn atoms (Figure 35c) led to decreased/increased absorption wavelengths.

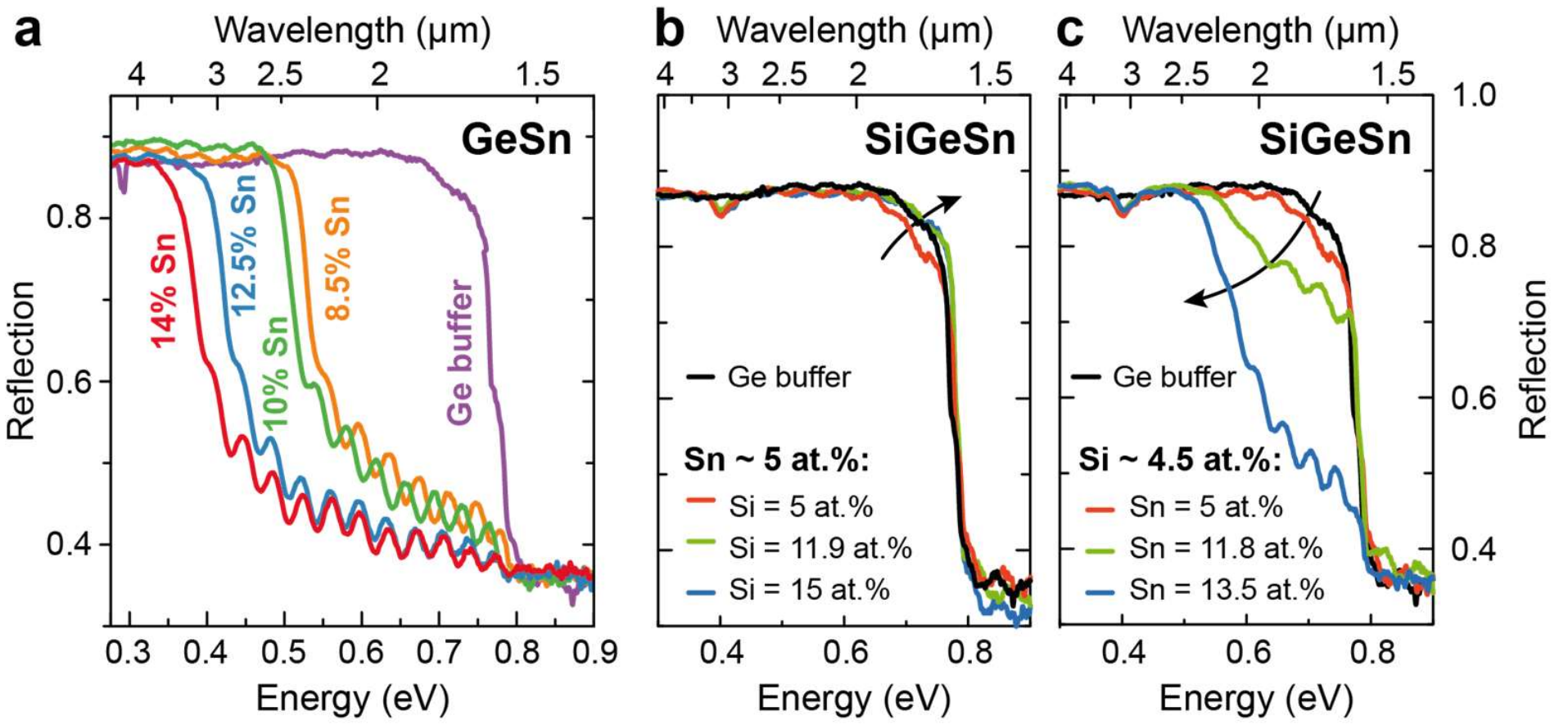

Figure 35: a) Room temperature reflection spectra of 700-1000 nm thick GeSn alloys, from reference [189]. SiGeSn ternaries enable to tune the absorption by changing the Si content b) and also Sn content c). Both from reference [61].

In the following, a brief overview of the history of GeSn-based photodetectors is given. Several important works from that field and key figure-of-merits are summarized in Table 6. We will show how previous works could help to improve the performance of present day detectors and stimulate new developments in this field.

Different types of GeSn-based photodetectors were developed since 2009, each with their specific advantages and disadvantages. Photoconductive devices made from GeSn, for example, rely on enhanced electrical conductivity due to generated photocurrents. They can be fabricated from intrinsic material without the need for any doping. In photodiodes (PD), the most common detector type, photo-generated carriers are separated by the internal electric field across a reverse-biased p-(i-)n junction. These structures can be combined with internal amplification mechanisms, e.g. through avalanche multiplication [203] or in a phototransistor geometry [204], to enable high-sensitivity photo-detection. Metal-semiconductor-metal (MSM) structures were also explored for high-speed photo-detection [205].

Ge photodetectors were already mature in 2009, with a monolithic integratation into CMOS processes. They delivered acceptable performance up to the middle of the telecom C-band (roughly 1550 nm), where their direct absorption edge was positioned [206]. At higher wavelengths, however, the material's responsivity was low as only indirect bandgap absorption was available. Reducing the direct absorption edge by incorporating small amounts of Sn (~2 at.%) into Ge thus became an interesting option to enhance the responsivity. It also extended the spectral range of Ge photodetectors to span the complete telecommunication wavelength band between 1260 nm and 1675 nm [207]. Continuous progress in epitaxial growth techniques yielded a steady increase of Sn content in such devices, opening new optical I/O communication windows around a 2 µm wavelength [208]. A Sn content of 11 at.% led to an extension of the long-wavelength cutoff to roughly 2.6 µm (room temperature) in state-of-the art GeSn photodetectors [209]. Even more freedom in shifting the wavelength regime could be obtained by integrating SiGeSn into SWIR photodetectors. However, only a limited amount of work has been devoted to that field up to now [210,211].

Another important figure-of-merit for photodetectors is the dark current. In photodiodes, which are driven under reverse bias, this results in a serious source of noise, limiting the sensitivity of the device. Intrinsic factors, including the (thermal) generation of carriers in the depletion region, can partly be tackled by improving the crystalline quality of the material. In contrast to silicon, GeSn alloys do not have a single surface oxide, but, instead, several germanium and tin sub-oxides. The latter may even show some conductive behavior. Leakage current from an insufficient surface passivation is therefore another important dark current source, which has previously been mitigated using low-temperature Si surface passivation [212] or ozone surface treatments [213].

Large material volumes are required to have high responsivities. However, the large lattice mismatch between GeSn and Ge/Si results in the formation of dislocations even in rather thin films. The critical thickness for strain relaxation of a $Ge_{0.9}Sn_{0.1}$ thin film grown on Ge is, for instance, around 90 nm [214]. Using GeSn wells separated by SiGeSn or Ge barriers with a lattice parameter close to or equal to that of the Ge SRBs, is an attractive way of working around the critical thickness for plastic relaxation issue and benefit from larger active volumes without the formation of lattice defects. Another way to mitigate the thickness limit in a vertically illuminated diode is to insert the absorbing layers in a resonant cavity. The responsivity of the detector can be enhanced by allowing not-absorbed photons to cross the active region multiple times [215]. Ultimately, the responsivity can completely be decoupled from the absorber thickness using a waveguide geometry. Light is then guided laterally into the detector material, yielding high-speed operation and high responsivity at the same time [216,217].

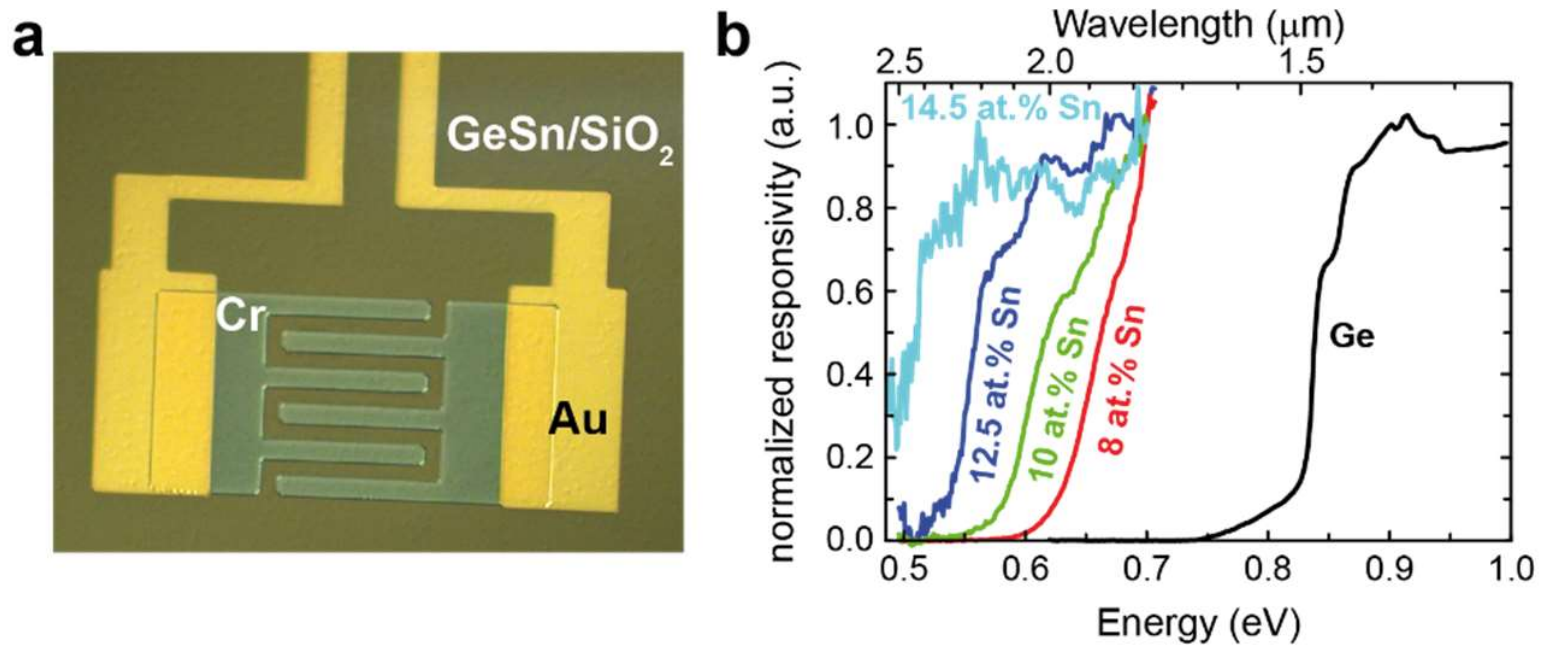

Figure 36: a) GeSn-based photoconductive detector with interdigitated electrodes. b) The normalized photo-response of several devices with different Sn contents shifted towards smaller energy/higher wavelengths as the Sn content increased.

Another advantage of heterojunction photodetectors is a decoupling of the active and contact material. Using a larger bandgap material for the top contact minimizes absorption losses and the deleterious impact of surface recombination. Furthermore, it acts as a low-pass filter to remove high energy photons and enhance the device's sensitivity in the targeted wavelength regime. While Ge has already been evaluated for such a purpose [40, 213, 217, 218], the use of SiGeSn ternaries, with similar or larger bandgaps, as shown in Figure 35b, has not yet been explored.

The first steps towards a co-integration with electronics were undertaken by fabricating photodetectors on a GeSn-on-insulator platform [205, 219] or bonding a number of photodetectors to a customized readout circuitry for focal-plane arrays [220].

We have also evaluated the properties of photoconductive detectors, made from GeSn alloys with different Sn contents. Devices, shown in Figure 36a, were built using a geometry with interdigitated electrodes, which helped to maximize the device area (minimize the RC delay) and keep the carrier drift region as short as possible. Using higher Sn contents shifted the photo-response towards higher wavelengths, as shown in Figure 36b.

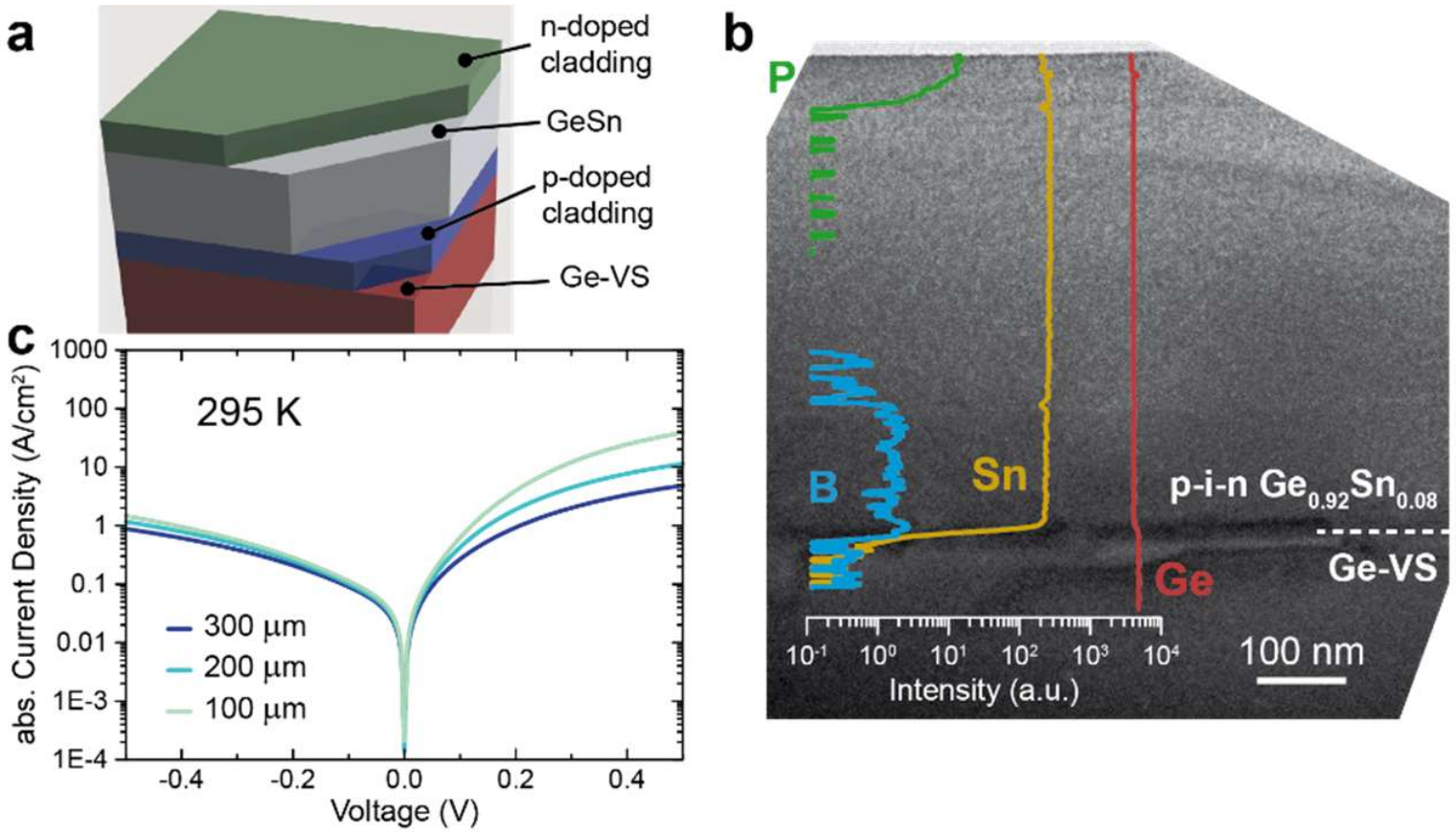

Figure 37: a) A schematic of epitaxially grown GeSn p-i-n photodiode. (b) cross-sectional TEM picture and SIMS depth profile of B, P, Sn and Ge in such photodetectors. Room temperature I-V curves for three fabricated devices with different mesa sizes. Adapted from reference [61].

Besides intrinsic GeSn thin films, epitaxy of different types of photodiodes was performed [42, 61, 197]. Homo-junction p-i-n diodes were epitaxially grown on top of Ge-buffered Si(001) wafers, as schematically shown in Figure 37a and Figure 37b. In-situ *p*- and *n*-type doped regions were separated by a 300 nm thick intrinsic GeSn layer, which acted as the absorption region, as indicated by the SIMS depth profiles. Room temperature I-V-curves in Figure 37c show distinct forward- and reverse bias regimes for different mesa structures. The dark current density at -0.5 V is roughly 1 A/cm$^2$, without any optimized surface passivation, however.

The combination of different group IV materials for GeSn-based heterostructures has been also evaluated, including undoped material stacks [58, 181] and heterostructure diodes [42, 61]. Germanium was used as barrier material in GeSn/Ge MQW diodes, as shown by the TEM micrograph in Figure 38a. When aiming for larger wavelengths – therefore higher Sn concentrations – Ge is not suitable anymore, since tensile strain from coherent growth on GeSn prevents them from being efficient barriers [42].

SiGeSn barriers can then be implemented in heterostructures, as shown by an Atom Probe Tomography (APT) line scan of Si, Ge and Sn in a GeSn/SiGeSn MQW sample in Figure 38b [181]. Furthermore, these material stacks were inserted in p-i-n light emitting diodes to fabricate room temperature Short-Wave InfraRed light emitters, as shown in Figure 38c. Measured I-V characteristics, shown in the inset of Figure 38c, demonstrate that the diodes are generally also suited for applications under reverse bias, such as photo-detection. To cancel absorption in contact layers, highly doped larger-bandgap SiGeSn layers should be used as top layers. Low resistive p- and n-type SiGeSn layers were thus grown and evaluated, as shown by the resistivity measurements in Figure 38d.

Additionally, the use of a double heterostructure design enables an electronic separation of the absorption region from the defective GeSn/(Ge/)Si interface. These defects are inherent to the GeSn material system and are due to the large lattice mismatch between GeSn and Ge/Si.

Table 6 provides the reader with an overview of various GeSn-based photo-detectors.

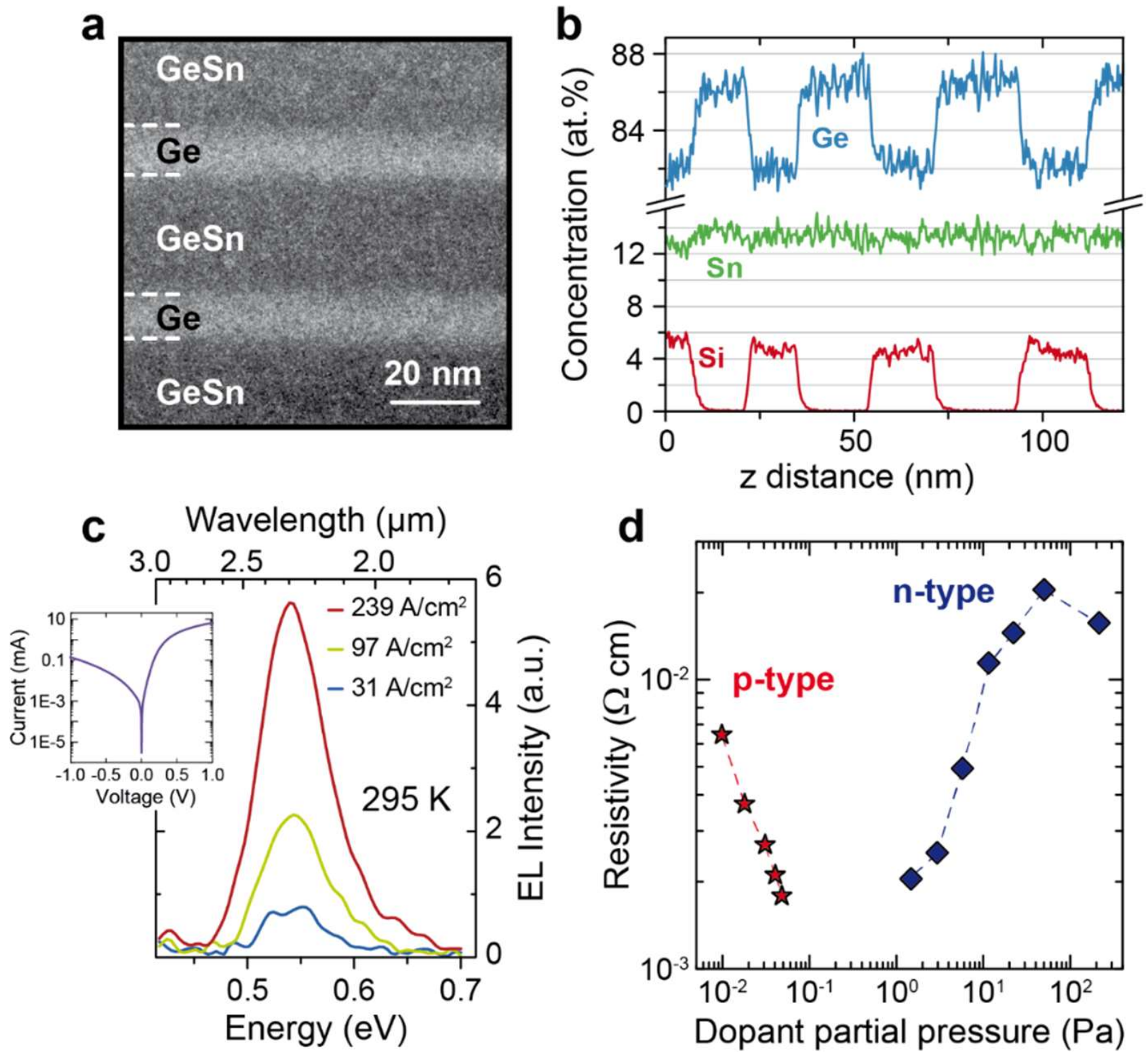

Figure 38: GeSn-based multi quantum well heterostructures, with a) Ge (TEM micrograph) or b) SiGeSn barriers (APT line scan) [181]. c) A GeSn/SiGeSn MQW diode shows room temperature electroluminescence and distinct forward and reverse bias regimes (inset). d) n- and p-type doping of SiGeSn material yields low-resistive contact material.

| Year | Sn (at.%) | Structure[a] | Responsivity (1.55 µm) @ -1V (A/W) | Responsivity (>1.55 µm) @ -1V (A/W) | $I_{Dark}$ @ -1 V (A/cm²) | Group |
|---|---|---|---|---|---|---|
| 2009 | 2 | PD | 0.05 @ -0.16 V | - | 1 | ASU [207] |
| 2011 | 3 | PD | 0.23 | 0.12 @ 1.64 µm | 1.8 | CAS [221] |
| 2012 | 9 | QW PC | 1 @ -5 V | 0.1 @ 2.2 µm | 5e-4 | Ghent [40] |
| 2013 | 3.9 | PD | 0.27 | 0.165 @ 1.6 µm | 0.4 | Taiwan [222] |
| 2013 | 3.6 | PD | - | 0.71 @1.8 µm (-3 V) | 6e-3 | CAS [223] |
| 2013 | 8 | MSM | 0.015 | - | 3e-6 | Stanford [205] |
| 2014 | 1.8 | WG PD | 0.18 @ 0 V | - | 5e-2 | Taiwan [216] |
| 2014 | 4.2 | PD | 0.22 @ 0 V | - | 8.9e-2 | Stuttgart [218] |
| 2014 | 7 | QW PD | 0.13 | - | 1.5e-1 | Stuttgart [224] |
| 2014 | 12 | PD | - | 0.03 @ 2 µm (-0.5 V,100 K) | 5e-2 (-0.5 V) | Delaware [225] |
| 2015 | 5 | APD | 1.21 @ -9.8 V | 0.95 @ -9.8 V | 196 (-9.8 V) | NUS [203] |
| 2015 | 5 | PD | 0.18 | 0.06 @ 1.63 µm | 7.3e-2 | NUS [212] |
| 2016 | 10 | PC | 0.11 | - | - | UA [226] |
| 2016 | 2.5 | PD | 0.45 @ -0.1 V | 0.2/0.045 @ 1.71 µm, -0.1 V (front/backside illum.) | 1e-4 (-0.1 V) | NTU [220] |
| 2016 | 8 | PD | 0.07 @ 0 V | 0.093 @ 2 µm | 3 | CAS [227] |
| 2017 | 2.8 | MQW WG PD | 0.06 @ 0 V | - | 5.9e-2 | Taiwan [217] |
| 2017 | 3 | MQW PD | 0.07 @ 0 V | - | 4e-3 | Stanford [213] |
| 2017 | 10 | MQW PD | 0.2 | 0.023 @ 2 µm | 3e-2 | NUS [228] |
| 2017 | 6.5 | HPT | 1.8 | 0.043 @ 2 µm | 1.5e-1 | NUS [204] |
| 2018 | 2.5 | PD | 0.38 | - | 0.7 | Taiwan [215] |
| 2018 | 7 | Lateral PD | - | 0.016 @ 2 µm | - | NUS [219] |
| 2018 | 11 | PD | - | 0.25 @ 2 µm | 1.5 | UA [209] |

[a] (A)PD – *(Avalanche) Photodiode*; PC – *Photoconductor; MSM – metal-semiconductor-metal;* (M)QW – *(multi) quantum well*; WG – *waveguide*; HPT – *heterojunction phototransistor*

Table 6: A brief overview of GeSn-based photodetectors history together with key figures-of-merit such as the responsivity at different wavelengths and the dark current.

## 4.2 Electrically pumped devices

Interest has been growing in recent years to develop light sources for the mid-infrared. Such devices could be used for light detection and ranging [229], biomedical sensing [230], gas sensing [231] and MIR optical communication [232]. GeSn devices are good candidates, as the GeSn bandgap can be tuned by strain and the Sn content in the Ge matrix and because of their compatibility with existing CMOS platform. High speed and efficient GeSn detectors have lately been demonstrated (see the photodetectors part of this chapter). We will review advances on GeSn light emitting diodes (LEDs) but start with earlier attempts using strained germanium and will conclude with the recent demonstration of an optically pumped GeSn laser.

### 4.2.1. Light emitting diodes

We saw previously that the energy offset between the Γ and the L valley was reduced by injecting tensile strain in Ge. The electron occupancy was then increased in the Γ valley, improving photon absorption and emission [233].

As early as 2012, an approach using external stressors technique was used to fabricate highly strained Ge LEDs. A tungsten stressor with a 4 GPa compressive stress was deposited at the bottom of the device (Figure 39a) to take advantage of a biaxial tensile strain close to 0.75% in the active region of a vertical Ge p-n junction. I-V characteristic of the devices was improved at room temperature under stress by a factor of 3 in the forward current mode as compared to the as-grown device with 0.2% of strain. A red shift of 100 nm was measured for the strained LED with a light emission up to 1.7 µm [152].

Strain has been as well successfully transferred into Ge micro-disks using SiN stressors. Biaxially tensile strain up to 0.72% was locally achieved at room temperature in Ge LEDs [70]. A 250 nm n-doped Ge thick layer was grown on n-type GaAs substrates. The latter yielded electron injection and carrier confinement in the Ge layer thanks to the higher GaAs band gap. A compressive SiN layer transferred locally tensile strain to Ge thanks to its mechanical relaxation (Figure 39b). The electroluminescence maximal emission peak was shifted from 1580 nm for an unstrained LED up to 1680 nm for a LED with a 0.72% tensile strain, locally. A suspended Ge LED was bi-encapsulated in compressive SiN using GeOI wafers (Figure 39c). A strain up to 1.92% in Ge was claimed, resulting in a 274 nm redshift of the emission spectrum [234]. Lateral p-i-n junction micro-bridges were used to reach 1.76% strain along the ⟨100⟩ direction [235]. The RPCVD grown Ge devices were fabricated on SOI substrates with doped tensile arms made by implantation and activated by rapid thermal annealing (Figure 39d). Electroluminescence up to 1885 nm was achieved at the centre of the bridge with a homogenous strain distribution. Strain in the Ge micro-bridges was limited due to dislocations between Ge and Si. A very different approach was explored using Ge-ion implated Ge quantum dots (QDs) in a defect-free Si matrix. Extraordinary properties of those defect-enhanced Ge quantum dots were measured showing a short carrier lifetime and minor thermal quenching in the PL emission up to room temperature. Readers can refer to this work in the other chapter of this book or to references [9, 236].

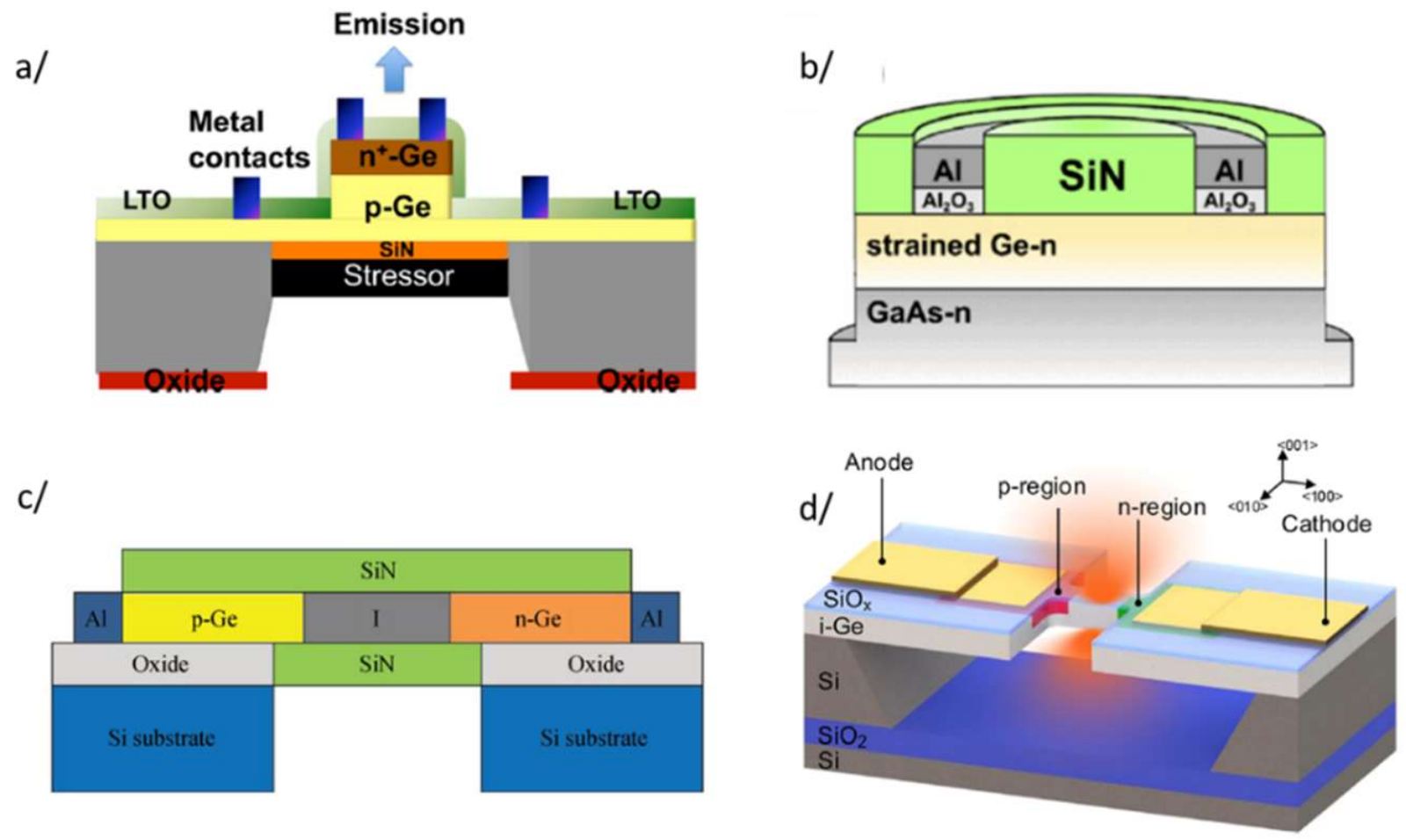

Figure 39: Schematic cross-section of a/ a strained Ge LED with a tungsten stressor at the bottom of the device (figure from [149]), b/ a strained Ge LED with a silicon nitride stressor deposited on top of the Ge layer (figure from [70]), c/ a strained Ge LED with a silicon nitride stressor deposited on both sides of the Ge layer (figure from [234]), d/ Schematic of a strained Ge micro-bridge LED with a lateral p-i-n configuration fabricated on a SOI substrate (figure from [235]).

Bulk GeSn LEDs have been reported, with light emission between 1.8 and 2.2 µm. Low Sn content GeSn active layers were epitaxied on p-type Ge SRB and capped with n-type material. As expected, they did not show any direct bandgap behaviour [237, 238,239, 240]. Fabry-Perot modes were observed when such GeSn stacks were epitaxied on SOI substrates [238]. Electroluminescence (EL) emission was reached with direct bandgap GeSn materials up to 3.4 µm for Sn contents up to 16% [241, 242]. Carrier confinement at room temperature seemed however to be insufficient to have an efficient light emission. Multiple quantum wells structures are known to improve the luminescence performances of LEDs. GeSn/Ge multiple quantum wells were investigated [243]. They showed a five time stronger electroluminescence emission compared to n-type Ge LEDs [244]. However, GeSn/Ge heterostructures seemed to poorly confine carriers at room temperature. SiGeSn carrier barriers have been introduced to overcome this limitation. These ternary alloys give more freedom to adjust the crystal lattice of the stack during the epitaxy. A type-I band alignment was demonstrated with binary GeSn alloys, with efficient light emission from direct bandgap layers in almost strain-free stacks [197].

The barrier height between the active layer and the barrier layer is critical to have efficient light emission at room temperature. In the last cited work, 10.5% of Si and 10.9% of Sn were incorporated in the SiGeSn barrier layers. The barrier height should increase as the Si/Sn ratio increases. Meanwhile, adding Sn results in an energy bandgap reduction. However, the exact band alignment between SiGeSn and GeSn is not yet know. Multiple SiGeSn/GeSn quantum well LEDs were investigated, with 10% of Si and 5% of Sn in the SiGeSn layers [245]. The EL intensity increased super-linearly with the applied voltage in those LEDs and was higher than in conventional GeSn/Ge heterojunction LEDS [246], showing that radiative recombination was efficient in such devices. Table 7 gives some of the specificities, at room temperature, of the GeSn-based LEDs discussed previously.

| LED structures | GeSn epitaxy | GeSn composition (%) | Biaxial strain | Wavelength emission | Ref. |
|---|---|---|---|---|---|
| Ge/GeSn/Ge heterostructures | RPCVD | 9.2 | - 0.48 % | 2.2 | [237] |
| | MBE | 4.5 | - 0.66 % | 1.9 | [238] |
| | MBE | 7.8 | ---------- | 2.2 | [239] |
| | MBE | 9.7 | ---------- | 2.25 | [240] |
| | RPCVD | 8 | ---------- | 2.06 | [246] |
| SiGeSn/GeSn/SiGeSn heterostructures | RPCVD | 16 | - 0.66 % | 3.4 | [241-242] |
| | RPCVD | 13 | ~ - 0.5 % | 2.8 | [242] |
| SiGeSn/GeSn/SiGeSn Multiple Quantum Wells | RPCVD | 9.1 | - 0.1 % | 2.2 | [197] |
| | MBE | 8.5 | - 1.6 % | 2.06 | [245] |

Table 7 : GeSn-based LED characteristics at room temperature. (RPCVD: Reduced Pressure - Chemical Vapor Deposition, MBE: Molecular Beam Epitaxy).

### 4.2.2 Lasers

Since the first claim of a laser effect in highly phosphorous doped Ge [193, 247, 248], there is a growing effort to fabricate an efficient CMOS-compatible Ge-based laser. Yet, no consensus has been reached on the physical interpretation of lasing by doping, and the impact of doping on the intervalence band absorption as shown in Ref. [180, 249]. To improve the performances of n-doped Ge layers for lasing, direct electron injection into the Γ conduction band of Ge was investigated in [250]. The radiative recombination of electrons injected by a forward biased p-i-n Zener diode with the holes in the indirect semiconductor Ge appeared to be more efficient than from a p-n junction with an n-doped active layer [250]. For both approaches, no further activities in the direction of lasing has been reported.

While the above-mentioned results did not result in any follow-up work, as far as we know, research on group IV lasing continued, mostly on the development of direct bandgap GeSn. Such R&D efforts very recently resulted in an electrically pumped GeSn laser [5]. The there applied electrically-pumping scheme is shown in Figure 40a. A threshold close to 0.6 kA/cm$^2$ was obtained at 10K and lasing was observed up to 100 K (Figure 40c). An external quantum efficiency of 0.3 % and a peak power of 2.7 mW per laser facet were reported. Since the threshold was measured on a series of waveguides with various lengths, we may use, on data provided in Table SII of the supplementary information of Ref. [5], the differential gain analysis protocol described in section 3.1.2. From a threshold that increases from $I$ = 0.77 kA/cm$^2$ up to 1.2kA/cm$^2$ for 0.6 mm and 0.3 mm long waveguides, respectively, $\Delta g/\Delta I$ for the lasing device operated at 10 K is approximately 35 cm $kA^{-1}$. By comparison, the differential gain observed during the optical pumping of Ref [1] devices was 0.4 cm $kW^{-1}$ only. The efficiency of the electrical injection thus seems to be orders of magnitude higher than the optical pumping, independent on the particular normalization one may apply to convert the photon injection to the injection by a current. This distinct difference may relate to the near resonance condition in electrical pumping. As shown for the strained Ge case, resonant pumping can indeed be very efficient, near unity as shown in [4] and detailed above in 3.2.

Further lasing details are given in the upcoming report by the same group available on arXiv [251]. There, electrically pumped lasing is investigated for various thickness and Sn content optically active GeSn layers, as well as various contents in the SiGeSn barriers and indium doping for the electrical injection layers. The thickness of the capping layer above the gain region was shown to have a remarkable impact on lasing threshold. This was due to the mode overlap with the GeSn active layer (Figure 40b). Performances as shown in the first report [5] were not exceeded, however. Subsequent developments may thus have to involve active layers with higher Sn contents and/or elastic-strain relaxation or p-i-n structures with tensile strained active layers. Single mode waveguide structures as well as the use of shorter than 700 ns long pulses to avoid excessive heating of the device may also boost performances. Howsoever, these promising preliminary results will encourage continuation of the research along that track and give good hope that group-IV semiconductor lasers once will perform on par with those of integrated group III-V devices.

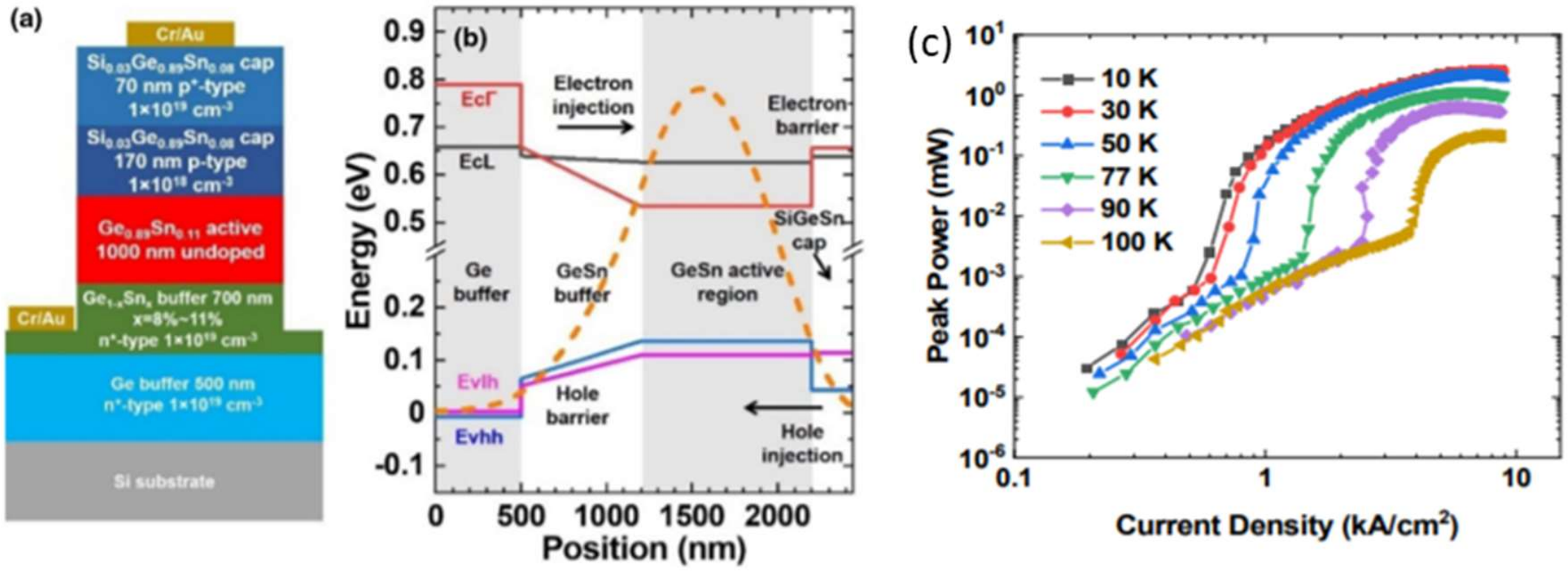

Figure 40 : (a) Cross-section of the electrically pumped laser stack, (b) Band structures and optical mode profile of the type II GeSn heterostructure, (c) Light output versus injection current of F-P cavity from 10 K to 100 K. Figure from [5].

# 5. Outlook and conclusion

As in the entirety of this book, the starting point for this topic is the technological compatibility with silicon photonics. Silicon is a poor light emitter. Bringing light into a silicon photonic chip involves either a coupling to external light sources or a cumbersome integration of better-emitting semiconductors on the chip. For such integration, both direct epitaxy and bonding of III-V chips are intensively explored today.

The idea of fabricating the entire Si photonic circuit in a standard CMOS foundry is obviously appealing. The only function lacking today is efficient and reliable light emission from group-IV semiconductors. A natural, straightforward approach would be to use a direct band gap group-IV semiconductor. A by-product of such development of light emitters would be efficient photodiodes and modulators, all based on direct-band-gap absorption by the same group-IV semiconductors.

To make a light emitter, particular attention was paid to Ge and derivatives for the following reasons. (i) Although it is still indirect, Ge is much closer to a direct bandgap semiconductor than Si. (ii) Processing Ge in microelectronics fabrication lines is already an industrial reality; (iii) Ge photodiodes are already mature devices in Si photonic transceivers.

As shown, a global tensile stressing of pure Ge layers converts Ge into a direct bandgap semiconductor, resulting recently in optically pumped Ge mid-IR lasers, only up to mild cryogenic temperatures, however. Although the crystalline quality of Ge is high and the straining technology advanced, intrinsic material properties, such as intra-valence band absorption and very high strain level requirements, make things difficult. Though technologically challenging, epitaxial as well as laterally defined ion implantation schemes for electrical injection are already available. Strained bulk Ge layers enable in-depth fundamental investigations of light emission from group-IV semiconductors, as shown above in this chapter.

Modifying the properties of Ge by alloying it with Sn is the second approach yielding a direct bandgap semiconductor. The good news for silicon photonics is that state-of-the-art GeSn layers and devices have been obtained with industry CMOS epitaxy tools and standard cleanroom technologies. This definitely favors GeSn alloys in material roadmaps. Still, many problems remain: GeSn has a large lattice mismatch with the Ge strain-relaxed buffers (on Si) typically used as templates. During the growth, the lattice mismatch is relaxed via a large number of interfacial defects. Moreover, an out-of-thermodynamic-equilibrium, low temperature growth typically results in a large number of bulk point defects, which result in a strong residual p-type doping and degrade the carrier mobility. Those crystalline defects will have to be dealt with. Many questions are still to be answered, as well: the band alignment at Ge/(Si)GeSn hetero-interfaces, predictive simulations of the band structure, etc…

However, even with an incomplete knowledge of the material parameters, by taking the best of the two approaches, external strain and alloying with Sn, several spectacular milestones were reached for GeSn lasers: low-threshold optically pumped strain- and dislocations-managed lasers [3], strain-wavelength-tuned optically pumped GeSn lasers operating at 273K [2] and the first electrically-pumped laser [5]. GeSn lasers’ characteristics are improving on a month-to-month basis. Outstanding challenges are not only to achieve continuous wave electrical pumping, but also to combine low lasing threshold and increased temperature lasing, as well as reaching reasonable wall-plug efficiencies.

The true goal, a room temperature electrically pumped laser, is not yet reached, but comes close as never before. Serious work is still to be done: optimizing the growth (residual doping, n-type and p-type doping, threading dislocations and point defect management), heterojunction band alignment within the (Si)GeSn quantum wells, designing optimized type-I heterostructures at room temperature, and finding strain-controlling cavities compatible with electrical injection.

While Ge is already a legacy detecting material for datacom applications at 1.55 µm, strained Ge and/or GeSn may become equally relevant for applications in the mid-IR range, typically from 2 to 4.5 µm. LEDs and photo detectors emitting or detecting in the mid-IR range fabricated with CMOS-relevant technologies will open to Si photonics new application fields such as gas sensing [242] and mid-IR imaging [252].

As mentioned in the introduction, other approaches probed to have light emission from group-IV materials, in addition to those described here, include diamond under mechanical strain [253], defect levels in the Ge quantum dots [10], tunnel injection of electrons directly in the Γ-valley of Ge [250], hexagonal Ge and SiGe alloys [8], and others such as quantum cascade lasers for the THz [254]. Those approaches are recent, and their progress is certainly worth observing in the near future.

Since the demonstration of the first room temperature semiconductor laser in 1971, our understanding of semiconductors and nanostructures has dramatically improved and an exceptionally large set of tools became available: ab-initio and quantum simulations, multiple epitaxial growth methods, nanofabrication and analysis of fine atomistic structure. Today, using a wide palette of such tools, we can tackle difficult solid-state physics problems, like modifying the band structure of an indirect semiconductor and converting it into a direct one.

## Acknowledgment:

We would like to acknowledge our scientific collaborators we have been working with, in particular, Francesco T. Armand Pilon, Lara Casiez, Jérémie Chrétien, Marvin Frauenrath, Andrea Quintero, Quang M. Thai, Joris Aubin, Rami Khazaka, Alban Gassenq, Kevin Guilloy, Mathieu Bertrand, Richard Geiger, Thomas Zabel, Stephan Wirths, Daniela Stange, Yann-Michel Niquet, Patrice Gergaud, Julie Widiez, Moustapha El Kurdi, Simone Assali, Oussama Moutanabbir, Guo-En Chang, Shui-Qing Yu and Jerome Faist, for our fruitful collaboration on group-IV lasers.
Some of the authors’ work was partially supported by the CEA Phare Photonics project, the Swiss National Science Fundation (SNF), the Elegante ANR Project and the CEA Gelato Carnot Project.